\documentclass[preprint,12pt]{elsarticle} 

\usepackage{amssymb}

\usepackage[colorlinks=true,linkcolor=black, citecolor=blue]{hyperref}

\usepackage[usenames, dvipsnames]{color}

\usepackage{ulem}

\newcommand{\etal}{\textit{et al.}}

\newcommand\figwidthhalf{67} 
\newcommand\figwidththird{44.7} 
\newcommand{\Z}{\mathbb{Z}}
\def \dirin {./figures}

\journal{none}

\begin{document}

\begin{frontmatter}


\cortext[cor1]{Corresponding author. \ead{henrygohkh@hotmail.com}}

\title{Continuous Fourier Transform: A practical approach for truncated signals and suggestions for improvements in thermography}


\author{K. H. H. Goh\corref{cor1}}

%

\begin{abstract}

The fundamentals of Fourier Transform are presented, with analytical solutions derived for Continuous Fourier Transform (CFT) of truncated signals, to benchmark against Fast Fourier Transform (FFT). Certain artifacts from FFT were identified for decay curves. An existing method for Infrared Thermography, Pulse Phase Thermography (PPT), was benchmarked against a proposed method using polynomial fitting with CFT, to analyse cooling curves for defect identification in Non-Destructive Testing (NDT). Existing FFT methods used in PPT were shown to be dependent on sampling rates, with inherent artifacts and inconsistencies in both amplitude and phase. It was shown that the proposed method produced consistent amplitude and phase, with no artifacts, as long as the start of the cooling curves are sufficiently represented. It is hoped that a collaborative approach will be adopted to unify data in Thermography for machine learning models to thrive, in order to facilitate automated geometry and defect recognition and move the field forward.
\end{abstract}


\end{frontmatter}


\section{Background}
\label{sect:bkgrd}

The aim of this work is to provide a framework for signal analyses and more precise understanding of scientific phenomena. It is done in hope that experimental scientists can use a more unified framework for the presentation of useful data, especially measured signals of interest. This will also provide the essential data required for Artificial Intelligence and Machine Learning algorithms to produce useful insight for the advancement of science. As some of the findings may be offensive to some who have to defend their research areas, I decided to write it as an individual, to insulate institutions from any potential backlash.

Over the years I have also provided advice to "technical experts" to avoid certain pitfalls in their supposed domains of expertise, to prevent the propagation of errors to the scientific community, as well as industry. One of the reasons for the generous advice to supposed internal competitors was to protect institutions of interest from any controversy arising from the use of specific methods. It was in 2016, when I was teaching a student about Fourier Transform, that I saw existing problems in certain areas of research. This data was not released in consideration of possible repercussions from certain individuals. I was told by a "technical expert" not to publish this, though his real intention will never be known. After much thought, the decision was made to organise the concepts and data into an article, to benefit the scientific community, as well as individuals and organisations that require the use of signal processing for their work.

The current version is a draft, which serves as a milestone for the work done to date. It is done in hope that more advanced algorithms will be developed in the realm of understanding physical phenomena, via a more precise representation in the frequency space. It would be great if readers can provide feedback to improve the suggested techniques. Collaborators are certainly welcome.

\section{Introduction}
\label{sect:intro}

Fourier Transform has a long history as a fundamental technique for understanding the information in signals, in the frequency space. There are a large number of applications such as in image compression, signal analyses, understanding of energy spectra in turbulence, Fourier Transform Infrared Spectroscopy (FTIR) for materials, and Thermography.

Fourier Transform of a function$f(t)$ in time $t$ is $F(k)$, defined to be \cite{FT2019Mathworld}:

\begin{equation}
F \left( k \right) \equiv \int_{-\infty}^{\infty}{f(t)e^{-2 i \pi k t}}dt
\end{equation}

where $k$ is the frequency in Hz. Inverse Fourier Transform is of the form:

\begin{equation}
f \left( t \right) \equiv \int_{-\infty}^{\infty}{F(k)e^{2 i \pi k t}}dk
\end{equation}

Most experimental measurements are conducted with limited resolution, either in spatial domain or temporal domain. As such, the common methods used would be Discrete Fourier Transform (DFT). For practical applications, it is common to use the Fast Fourier Transform (FFT) to compute the results in frequency space, because well established algorithms produce results faster than direct computations.

However, there exists certain applications where the DFT or FFT may produce undesirable characteristics. From the equations for Fourier Transform, it can be observed that these integrals represent areas under the curve of a function, split into real and imaginary space. Using FFT or DFT will result in uncertainties, as these methods represent said area as discrete rectangular blocks. Imagine if the measured signals can be represented accurately by different classes of functions. If the Fourier Transform of these functions have analytical solutions , the measured signals can be analysed in the frequency space via Continuous Fourier Transform (CFT), which is close to the exact representation of measured phenomena. The dependence on measurement resolution will also be reduced significantly.

In practice, signals and phenomena are measured in a finite amount of time or space. These results could also be non-cyclic in nature, which leads to leakage in DFT and FFT. These results are essentially measurements that span infinite time or space, truncated via rectangular windows. It is interesting to note that the mere existence of these rectangular windows allows some of these integrals to be mathematically tractable, and relatively easy to implement in computational algorithms.

Some simple functions will be used in the current work. The nature of the Fourier Transform allows for these simple functions to be added independently to represent signals or phenomena that look more complex e.g. a measured signal that can be represented by a polynomial and a sine function. However, it is important to ensure that the measured data is continuous in nature. All the solutions will involve truncation of the functions over the region of interest, to obtain the equivalent CFT for benchmarking with FFT. This region of interest is standardised as $[0,t_{s}]$ as most measurements are done in the time domain, with reference to time 0. It is relatively trivial to derive the equivalent for other regions of interests by working from the point where the limits of the integrals are implemented. The solutions for the CFT of simple functions shall be presented below.

For functions that are constants,

\begin{eqnarray}
f(t)  & = & c \nonumber\\
F(k) & = & c\delta (k), \delta (k) \equiv \int_{-\infty}^{\infty}{e^{-2 i \pi k t}}dt
\end{eqnarray}

Truncate the signal in range $[0,t_{s}]$, we get

\begin{eqnarray}
F(k) & = & \frac{c}{2 \pi k} \left[ sin(2 \pi k t_{s}) + i \left( cos(2 \pi k t_{s}) - 1 \right) \right]
\end{eqnarray}

This allows the theoretical computation of a truncated signal. For functions that are polynomial terms of the form $a_{n}t^{n}$, where $n$ is a non negative integer and $a_{n}$ represents the corresponding coefficient,

\begin{eqnarray}
f(t)  & = & a_{n}t^{n} \nonumber\\
F(k) & = & \sum_{m=1}^{n} \left( \frac{a_{n}t^{m}}{\left( -2 i \pi k \right)^{n-m+1}} e^{-2 i \pi k t} \left( -1 \right)^{n-m} \frac{\left( n+1 \right)! }{m! \left( n+1 \right)} \right)_{-\infty}^{\infty} \nonumber\\
        &    & + \left( -1 \right)^{n} \int_{-\infty}^{\infty}{\frac{a_{n} n!}{\left( -2 i \pi k \right)^{n}} e^{-2 i \pi k t}}dt
\end{eqnarray}

Truncate the signal in range $[0,t_{s}]$, we get

\begin{eqnarray}
F(k) & = & \sum_{m=0}^{n} \left( \frac{a_{n}t_{s}^{m}}{\left( -2 i \pi k \right)^{n-m+1}} e^{-2 i \pi k t_{s}} \left( -1 \right)^{n-m} \frac{ n! }{m! } \right) + {\frac{ \left( -1 \right)^{n+1} a_{n} n!}{\left( -2 i \pi k \right)^{n+1}}}
\end{eqnarray}

It is expected that this term reduces to the equivalent for constants when n is zero. For functions that are exponential with constants $a$ and $b$,

\begin{eqnarray}
f(t)  & = & e^{a t + b} \nonumber\\
F(k) & = & \frac{e^{b}(a +2 i \pi k)}{a^{2} + (2 \pi k)^{2}} \left( e^{(a-2 i \pi k) t } \right)_{-\infty}^{\infty} \nonumber\\
\end{eqnarray}

Truncate the signal in range $[0,t_{s}]$, we get

\begin{eqnarray}
F(k) & = & \frac{e^{b}}{a^{2} + (2 \pi k)^{2}} [  a \left( e^{at_{s}}cos(2 \pi k t_{s}) - 1 \right) + 2 \pi k e^{at_{s}}sin(2 \pi k t_{s}) \nonumber\\
        &  & + i \left( 2 \pi k \left(e^{at_{s}}cos(2 \pi k t_{s}) - 1 \right) -  a e^{at_{s}}sin(2 \pi k t_{s}) \right ) ]
\end{eqnarray}

For functions that are of the form $sin(2 \pi f t + \phi)$ where $\phi$ is a constant phase shift,

\begin{eqnarray}
f(t)  & = & sin(2 \pi f t + \phi) \nonumber\\
F(k) & = & \frac{i}{2} \left[ e^{-i \phi}\delta(k+f) - e^{i \phi}\delta(k-f) \right]
\end{eqnarray}

Truncate the signal in range $[0,t_{s}]$, we get

\begin{eqnarray}
F(k) & = &  \frac{1}{4 \pi} [ \frac{-cos(2 \pi (k + f) t_{s} + \phi) + cos \phi }{(k+f)} + i\frac{sin(2 \pi (k + f) t_{s} + \phi) - sin \phi}{(k+f)} \nonumber\\
        &     & +\frac{cos(2 \pi (k - f) t_{s} - \phi) + cos \phi}{(k-f)}+i\frac{sin(2 \pi (k - f) t_{s} - \phi) + sin \phi}{(k-f)} ]
\end{eqnarray}

For functions that are of the form $sin(2 \pi f t + \phi)$ with the special case of $f=k$, for a truncated signal in the range $[0,t_{s}]$, we get,

\begin{eqnarray}
F(k) & = &  \frac{1}{8 \pi f}[-cos(4 \pi  f t_{s} + \phi)  + cos \phi + 4 \pi f t_{s} sin(\phi) \nonumber\\
        &     & +i sin(4 \pi f t_{s} + \phi) - i sin \phi -i 4 \pi f t_{s} cos(\phi) ]
\end{eqnarray}

Note that the above solutions for truncated signals can be implemented readily via computer algorithms, for analyses of measured phenomena. Having introduced some analytical solutions for CFT, we move to examples of a practical applications e.g. using FFT on practical signals. As my current expertise is in Infrared Thermography, an example of a technique shall be presented henceforth to highlight the potential pitfalls of data presentation and understanding of phenomena resulting from the choice of FFT over CFT. In Thermography, one of the established techniques is Pulse Phase Thermography (PPT)~\cite{Maldague1996JAP}. It is essentially the heating of samples on one surface, and using thermal cameras to measure the surface temperatures over time, after the heating stops. These signals are then processed per pixel, using FFT, to observe derived phase data in frequency space, for the detection of defects and defect depths in structures. An aliasing problem was identified and presented by Galmiche~\etal~\cite{Galmiche2001ISOP}, which is essentially the dependence of  phase data on data sampling rates (image frame rates in this case).

Assuming that the temperature-time decay curves can be represented by exponential functions as presented before, it is possible to process the data using both FFT and CFT to make exact comparisons, while exploring the relative effects of different variables, without the other effects e.g. experimental error. In PPT, the variable of interest is the phase in frequency space, which is essentially the angle of a vector in the complex plane at a particular frequency, as the Fourier Transform results in complex numbers derived at every frequency. For a typical measurement done on 3D printed polymer samples, the measurements span 200 seconds, with 1 frame captured per second on infrared cameras. Assume a temperature decay curve of the form

\begin{eqnarray}
T & = & e^{at+b} + c \nonumber\\
   & = & e^{-0.025t+4.0} + 25.0
\end{eqnarray}

\begin{figure} 
\begin{center}
\includegraphics[width=\figwidththird mm]{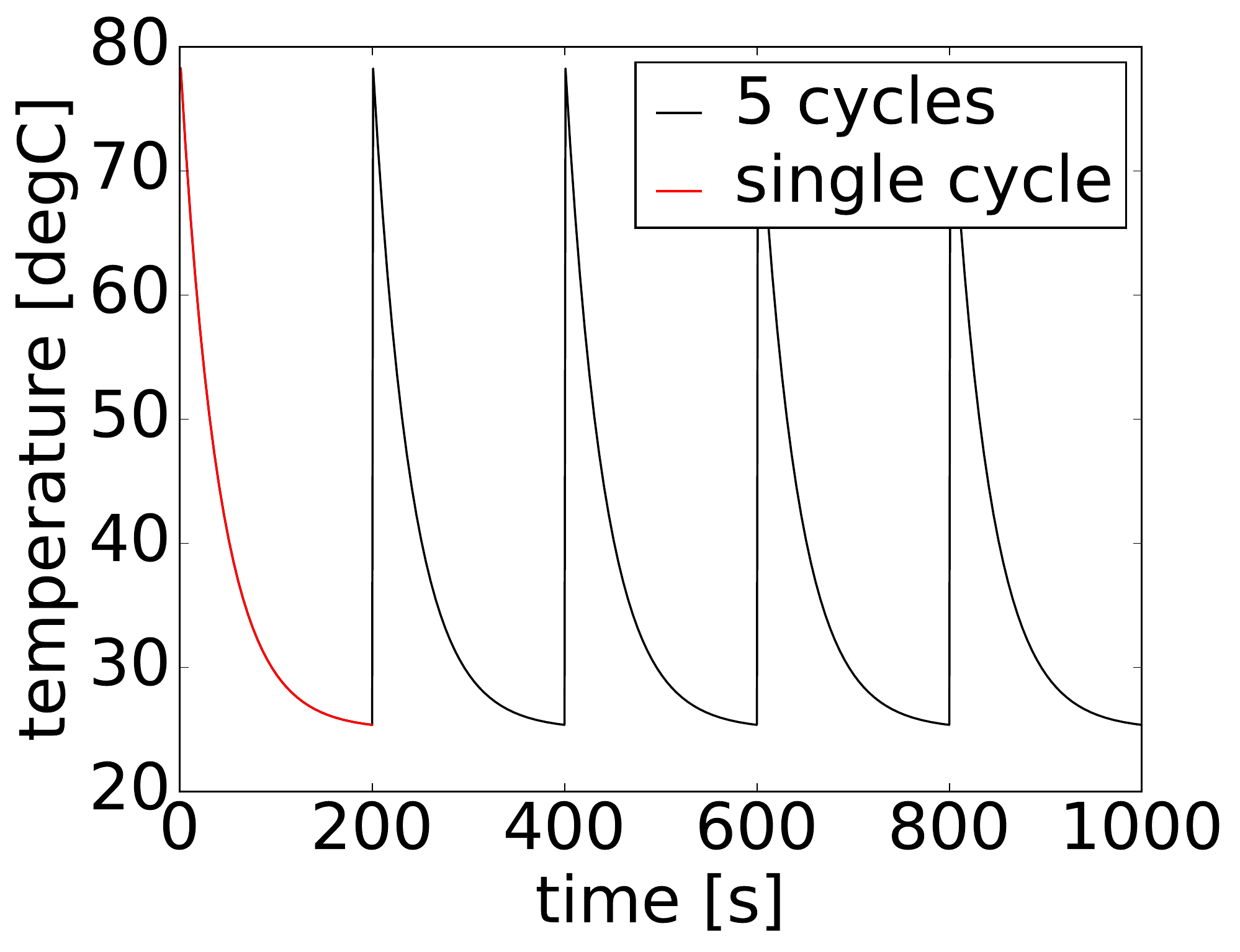}
\includegraphics[width=\figwidththird mm]{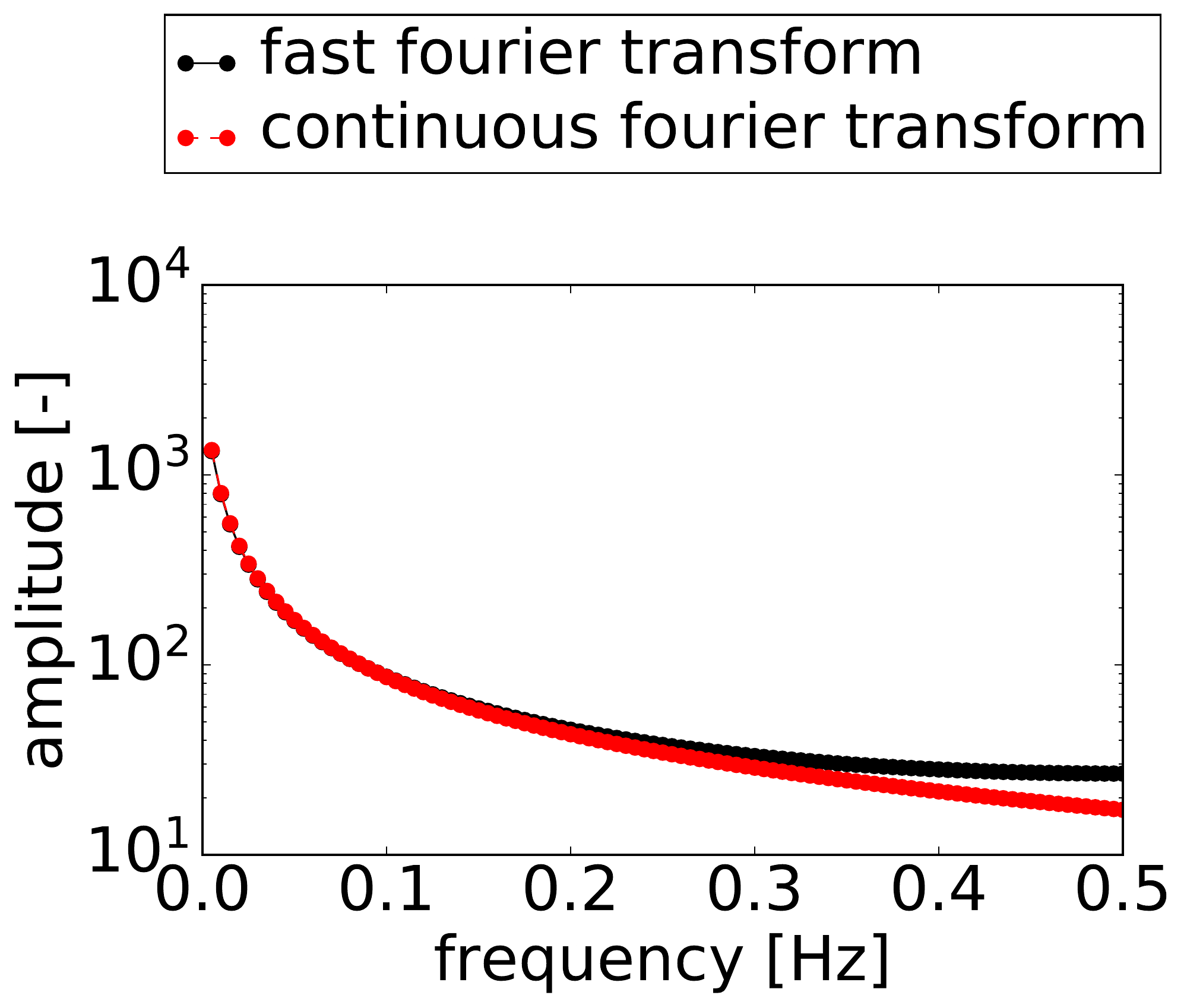}
\includegraphics[width=\figwidththird mm]{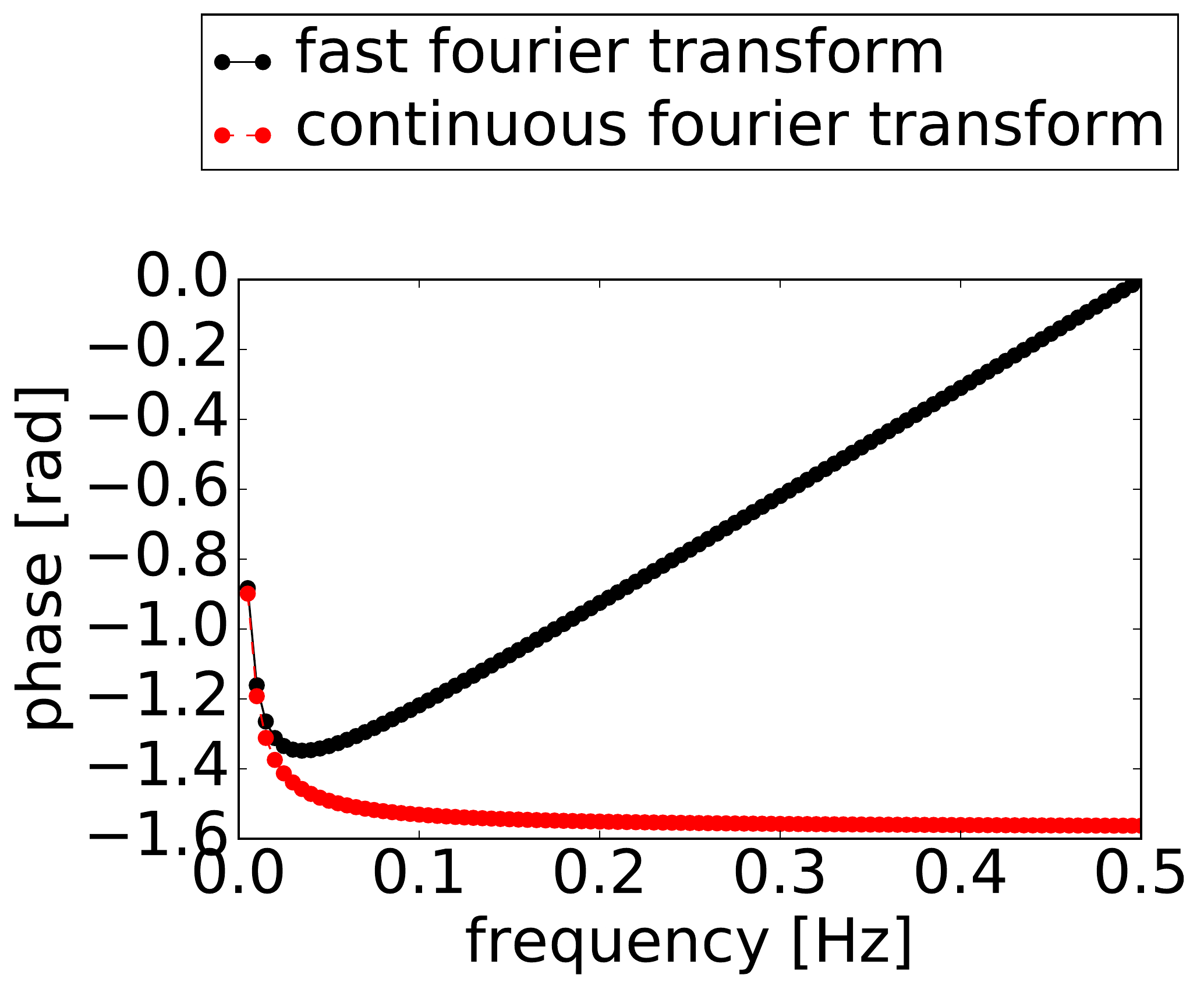}
\includegraphics[width=\figwidthhalf mm]{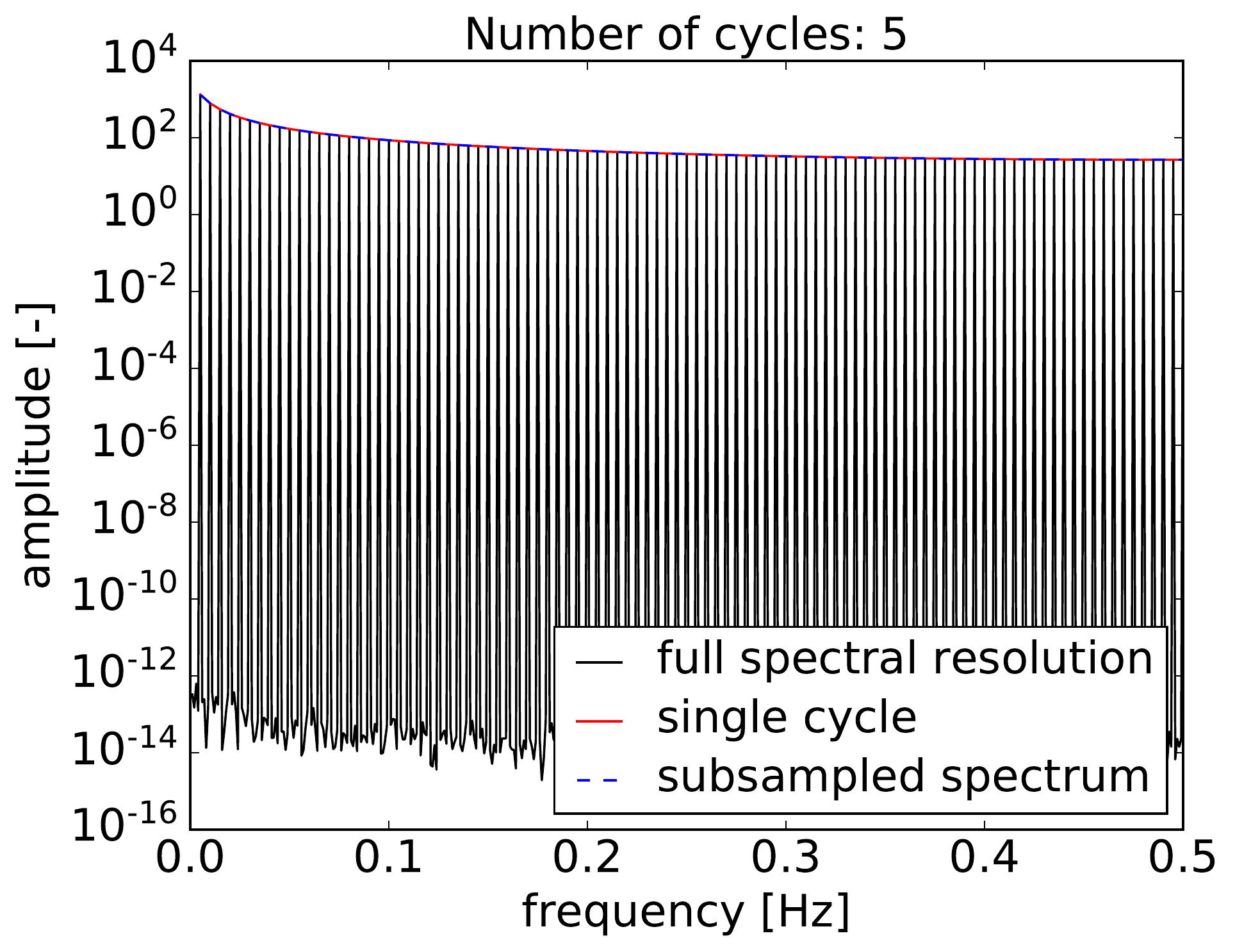}
\includegraphics[width=\figwidthhalf mm]{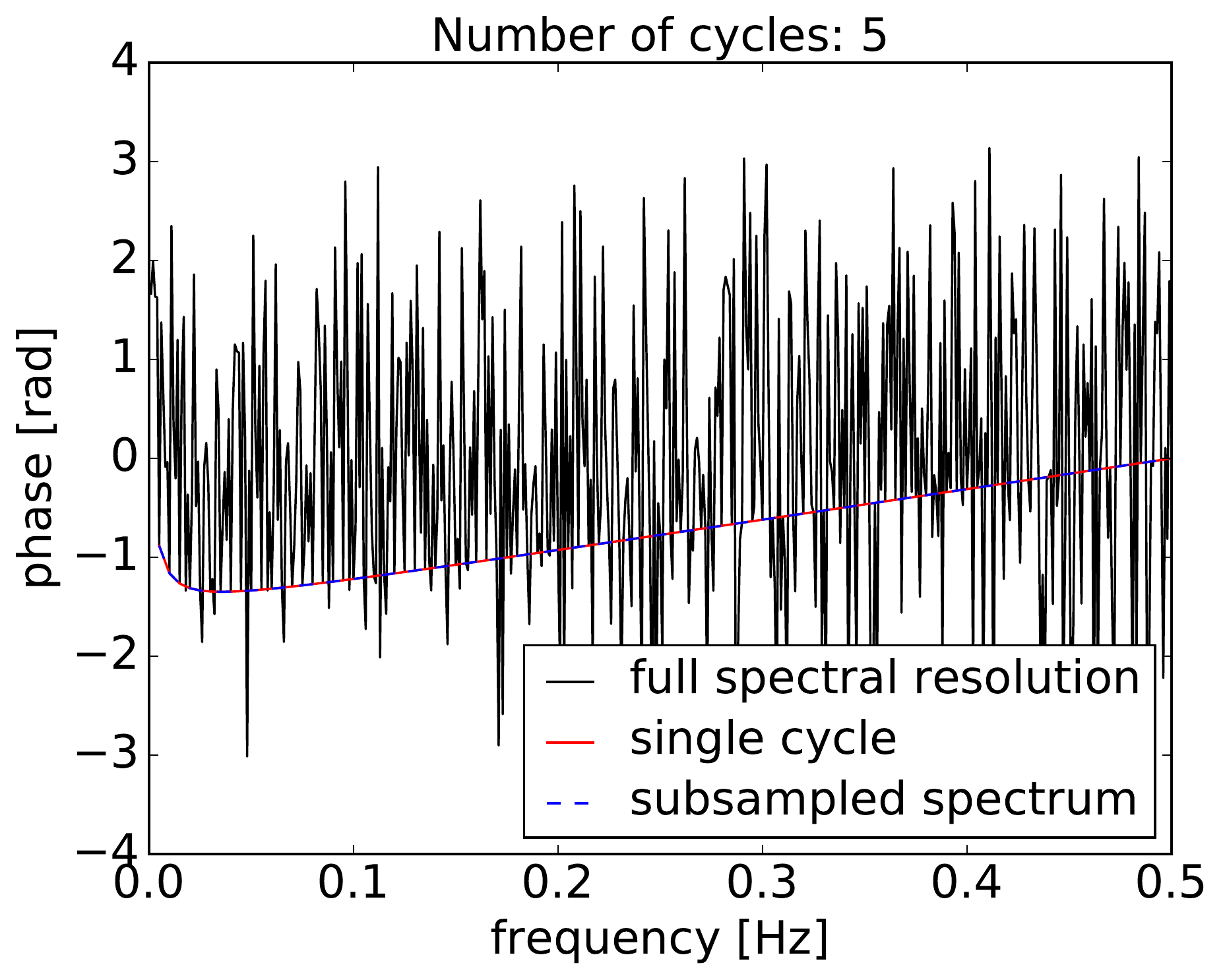}
\includegraphics[width=\figwidthhalf mm]{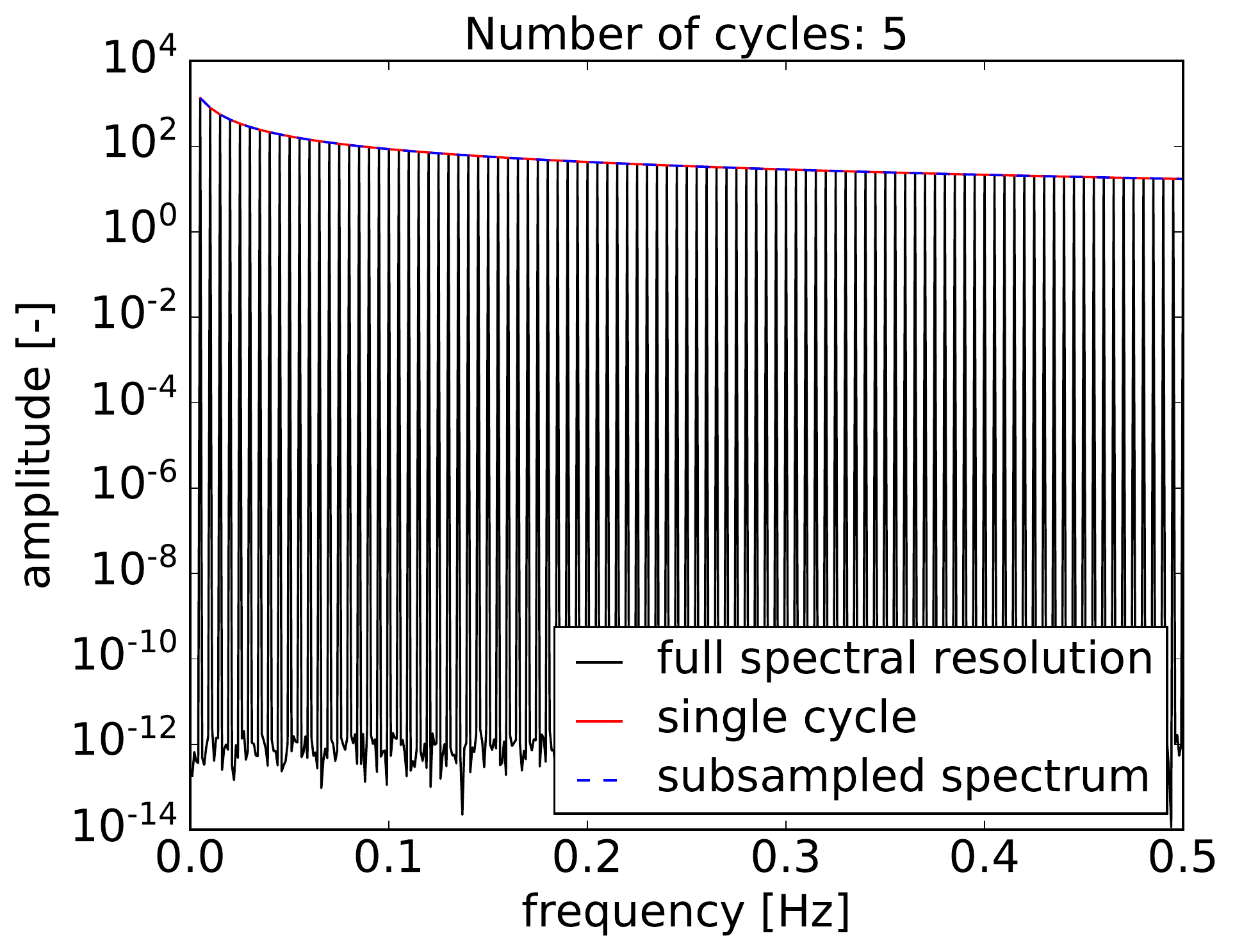}
\includegraphics[width=\figwidthhalf mm]{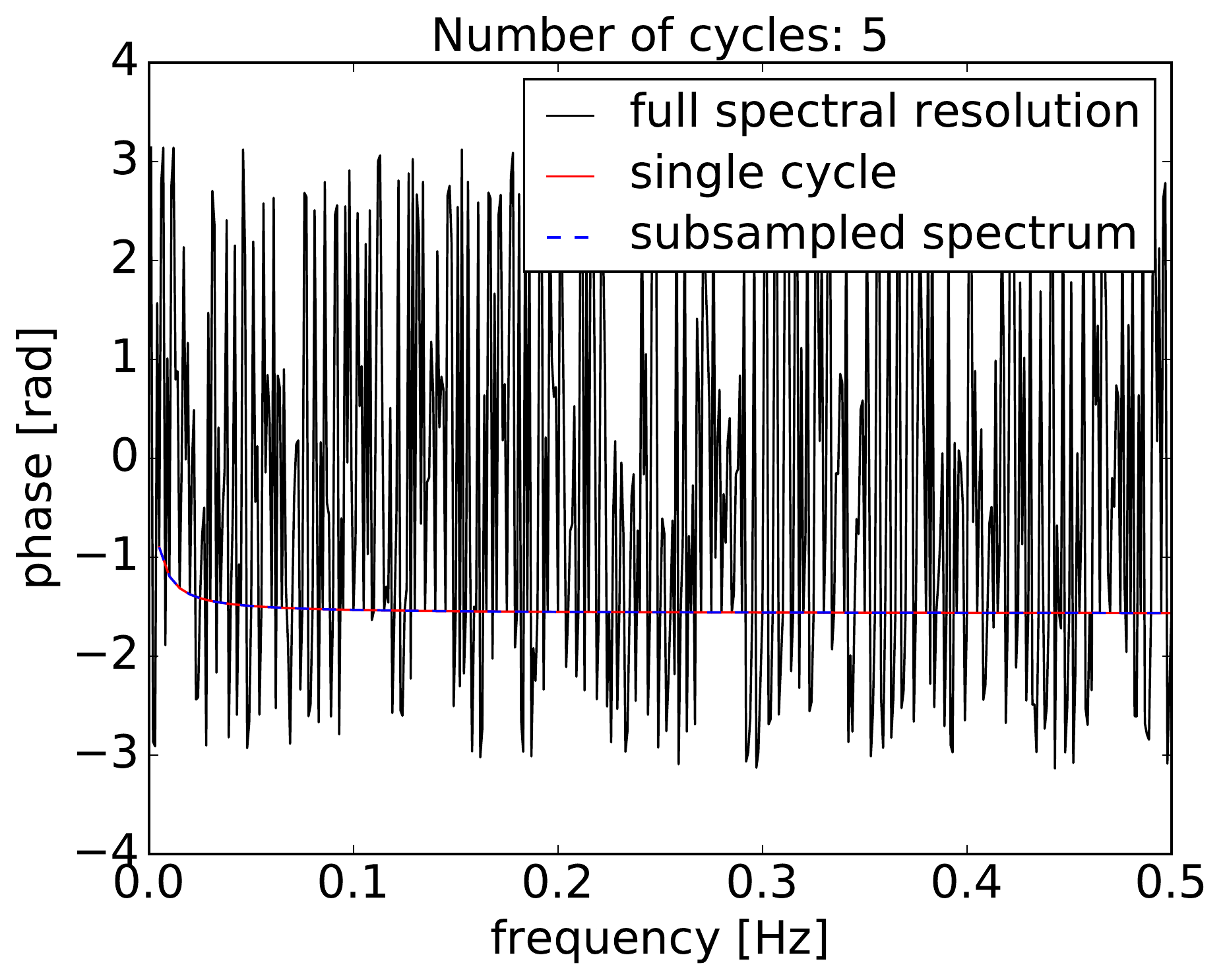}
\end{center}
\caption{Effect of number of cycles on CFT and FFT, and CFT versus FFT. Top row: Exponential function repeated over 5 cycles (left). Differences between CFT and FFT, for amplitude (middle) and phase (right). Middle row: Comparisons between FFT of single cycle, 5 cycles and subsampled spectra, for amplitude and phase. Bottom row: Comparisons between CFT of single cycle, 5 cycles and subsampled spectra, for amplitude and phase. All data processed from exponential signal at 1~Hz sampling rate.
\label{fig:fft_vs_cftexp_freq}}
\end{figure}

Using the given solutions for CFT, we can benchmark and compare with FFT directly, by truncating the signal with a rectangular window. The equivalent results for the amplitude and phase in frequency space can be seen in Fig.~\ref{fig:fft_vs_cftexp_freq}, in the top row, for the given exponential decay function with sampling rate of 1~Hz. It is obvious that there are clear discrepencies between CFT and FFT for both amplitude and phase, so the results from FFT are not exact. It was identified by Galmiche~\etal~\cite{Galmiche2001ISOP} that the values of phase will shift towards zero at higher frequencies, regardless of sampling rates, which is also reflected in the current work. However, this is not observed for CFT. The discrepencies between FFT and CFT increase with frequency values, for both amplitude and phase. This ties in with the previous point about FFT being an approximation. This is because the temperature values change rapidly at the start of the signal trace, so approximations via discrete methods will result in undesirable artifacts due to uncertainties. It is in no way a physical representation of the true phenomena of interest.

It was also identified by Galmiche~\etal~\cite{Galmiche2001ISOP} that for the CFT and FFT to be equivalent, the signal needs to be cyclic. Hence, by taking this example, and duplicating it for N cycles, from 1 to 16, it can be observed that when there is more than 1 cycle used to calculate the spectra, both cft and fft produced significant noise. However, when the results were subsampled by N, where N is the number of cycles, the results were similar to that achieved using a single cycle. These are shown in Fig.~\ref{fig:fft_vs_cftexp_freq} for 5 cycles. Similar results were observed when the sampling frequencies were set to 0.5, 1.0, 2, 5, and 10~Hz. This shows that the useful values of frequency $k$ to sample for a non-cyclic signal sampled at infinite frequency, are defined by the sampling time, via

\begin{eqnarray}
k & = & \frac{i}{t_{s}}, i \in \Z^{+}
\end{eqnarray}

\begin{figure} 
\begin{center}
\includegraphics[width=\figwidthhalf mm]{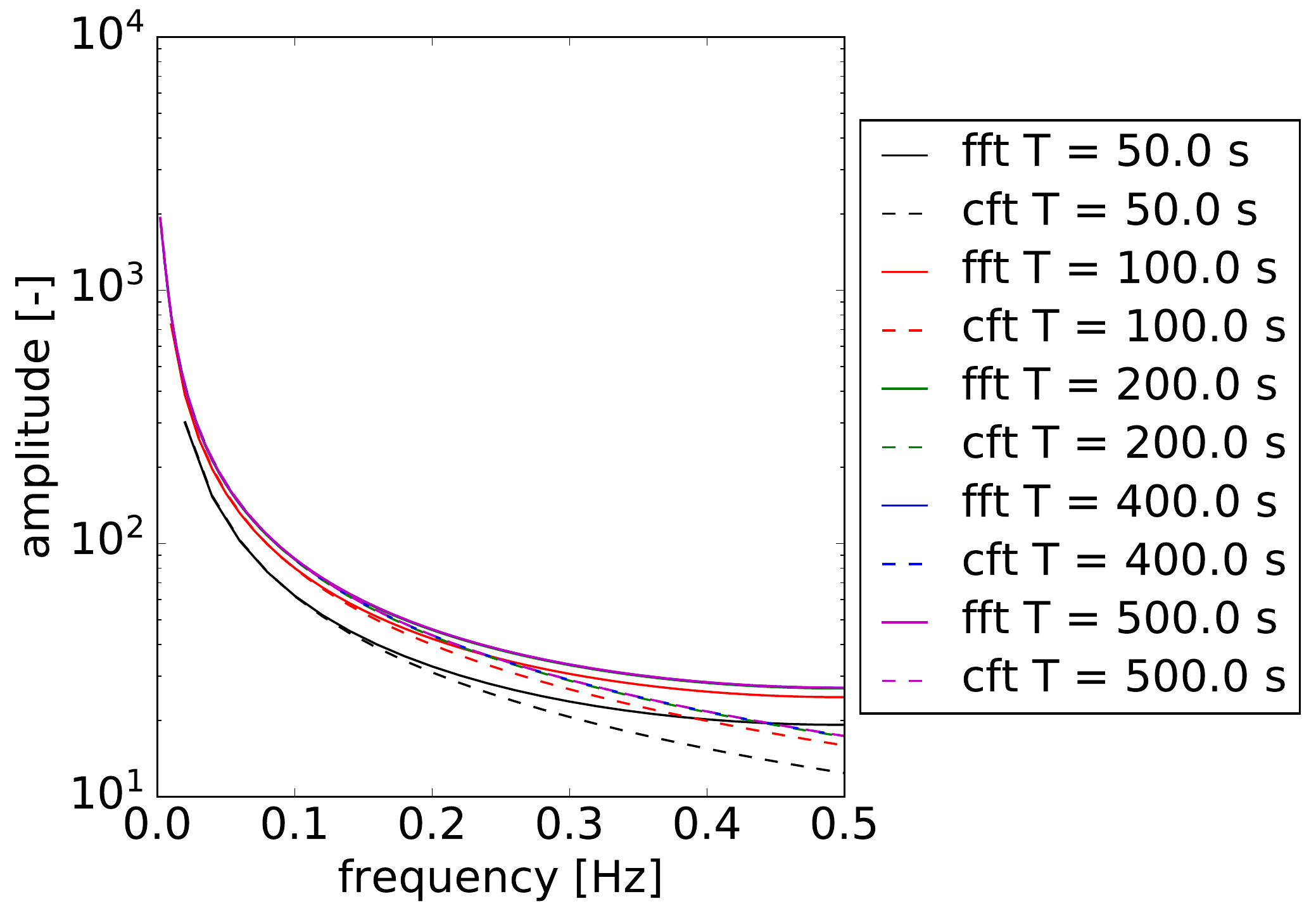}
\includegraphics[width=\figwidthhalf mm]{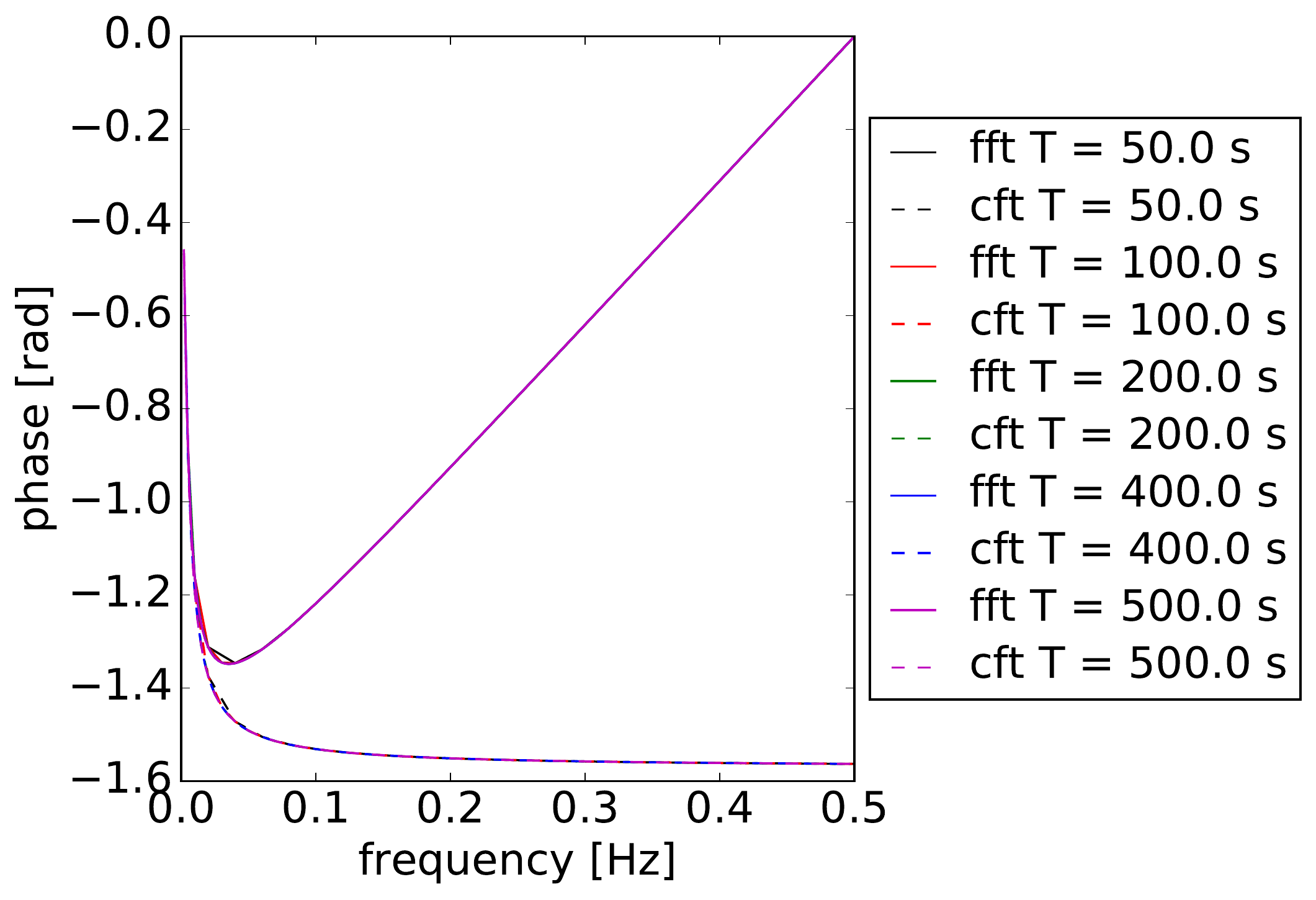}
\end{center}
\caption{Effect of sampling time on amplitude and phase values derived using CFT and FFT, at sampling rate of 1~Hz.
\label{fig:fft_vs_cftexp_tsamp}}
\end{figure}

What this means is that the sampling time must increase in order to have a higher resolution in frequency space. It also shows that the sampling time is the limiting factor for the lowest frequency that can be reasonably resolved from a signal. In PPT, the lower frequencies enable deeper defects and structures to be observed, so there is value in measuring over longer periods of time. This needs to be balanced with practical issues such as disk space, data size, and measurement turnaround time. For exponential functions it is relatively trivial to extrapolate and enable the visualisation of such defects. However, temperature-time decay curves may not be represented well by exponential functions, as shall be shown later. Based on the given exponential function for temperature, it is also interesting to investigate the dependence of CFT and FFT on sampling time. These results for sampling frequency of 1~Hz are shown in Fig.~\ref{fig:fft_vs_cftexp_tsamp}. It can be observed that there is a minimum sampling time of around 200 seconds required before the amplitude in the frequency space converge, for both CFT and FFT. Note that this value of 200 seconds is not universal, as it depends on the rate of decay. The relative impact of sampling time on phase values is negligible.

\begin{figure} 
\begin{center}
\includegraphics[width=\figwidthhalf mm]{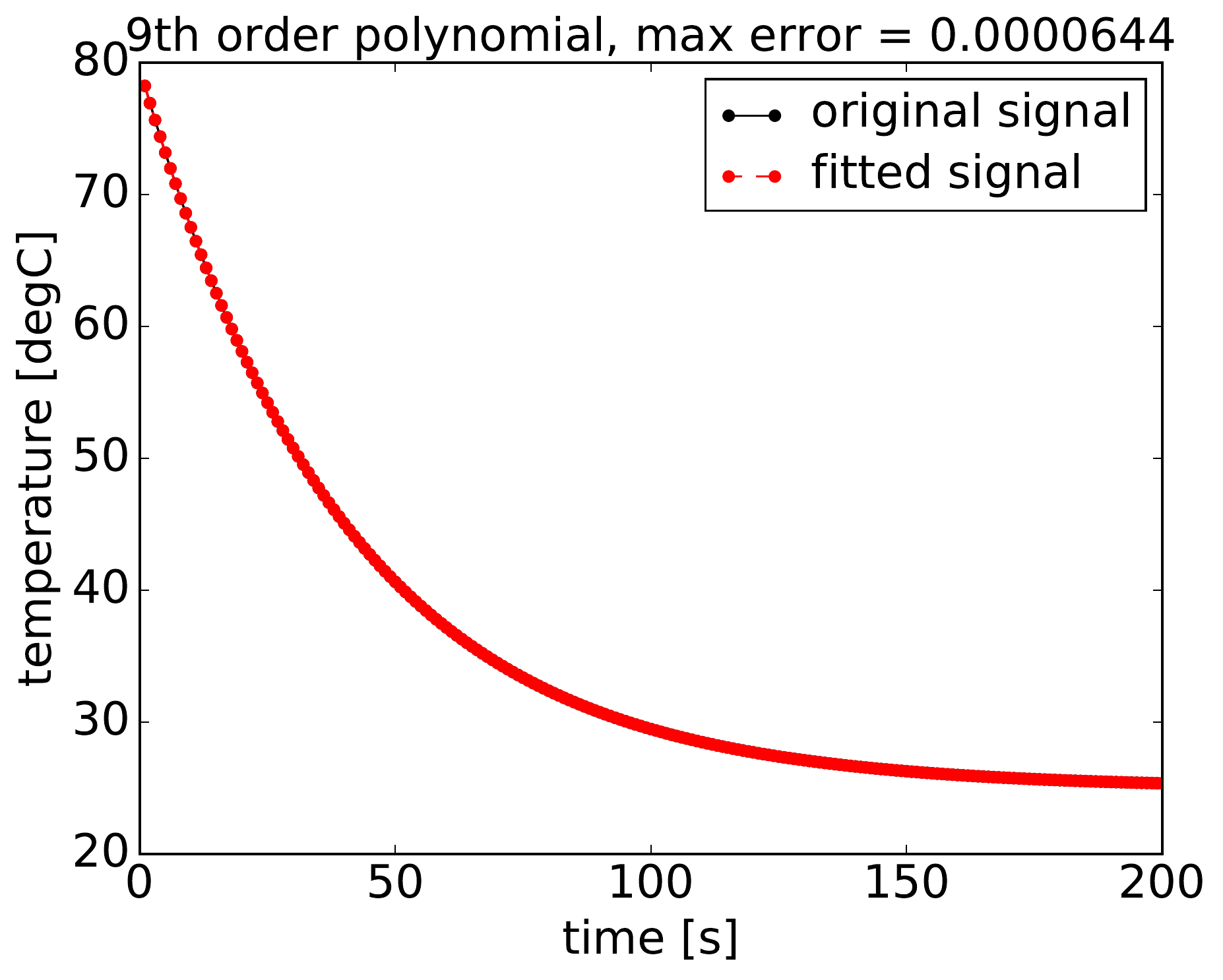}
\includegraphics[width=\figwidthhalf mm]{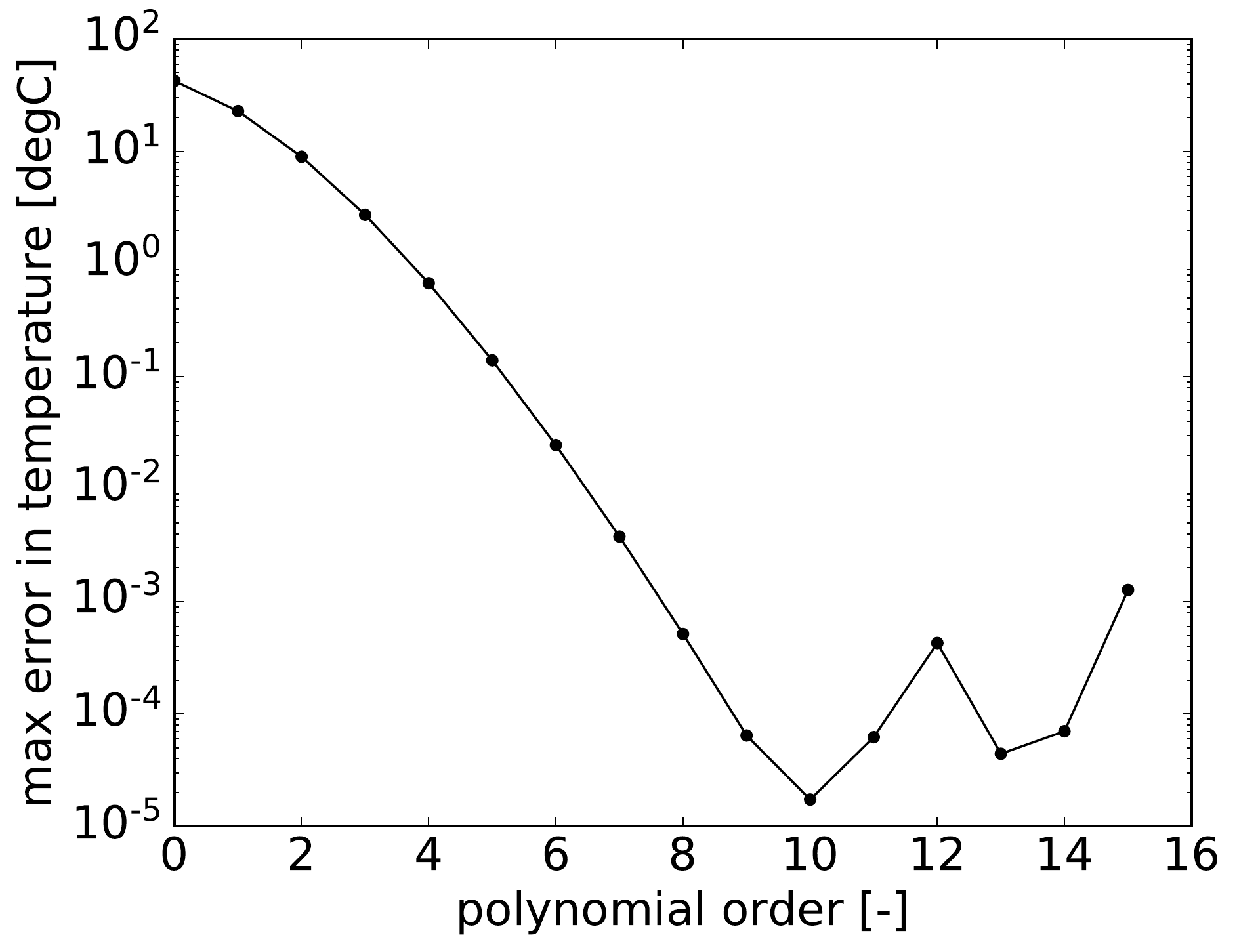}
\includegraphics[width=\figwidthhalf mm]{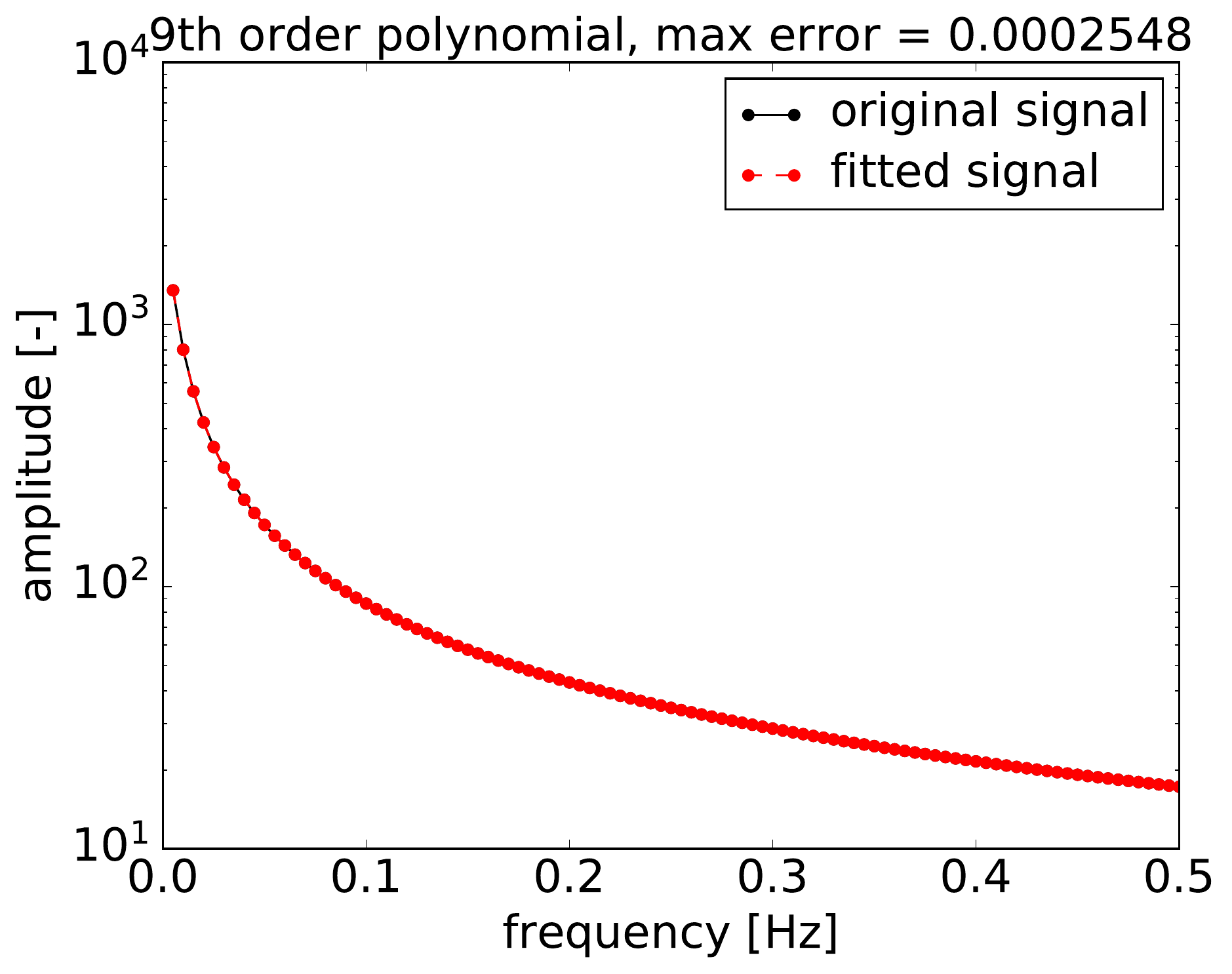}
\includegraphics[width=\figwidthhalf mm]{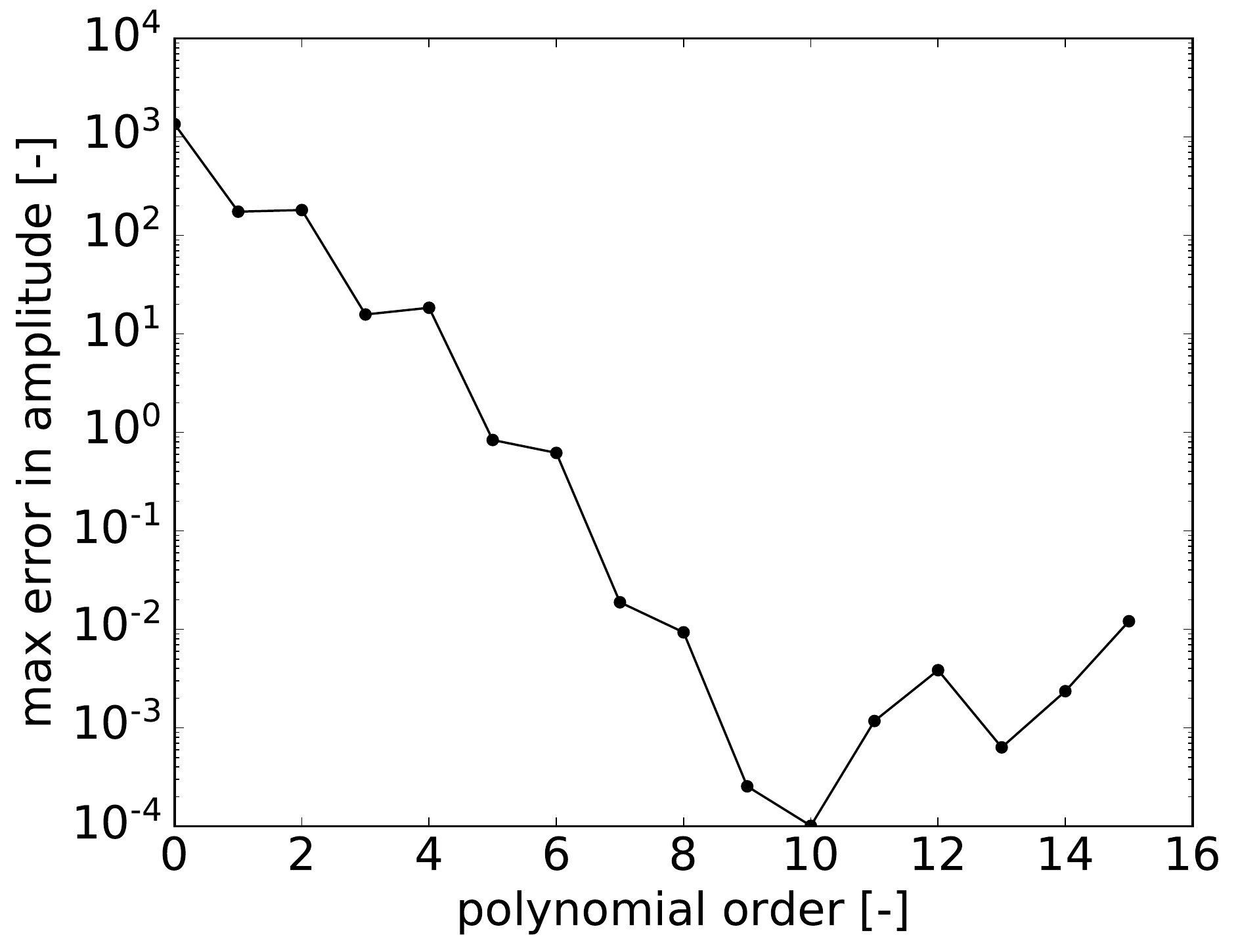}
\includegraphics[width=\figwidthhalf mm]{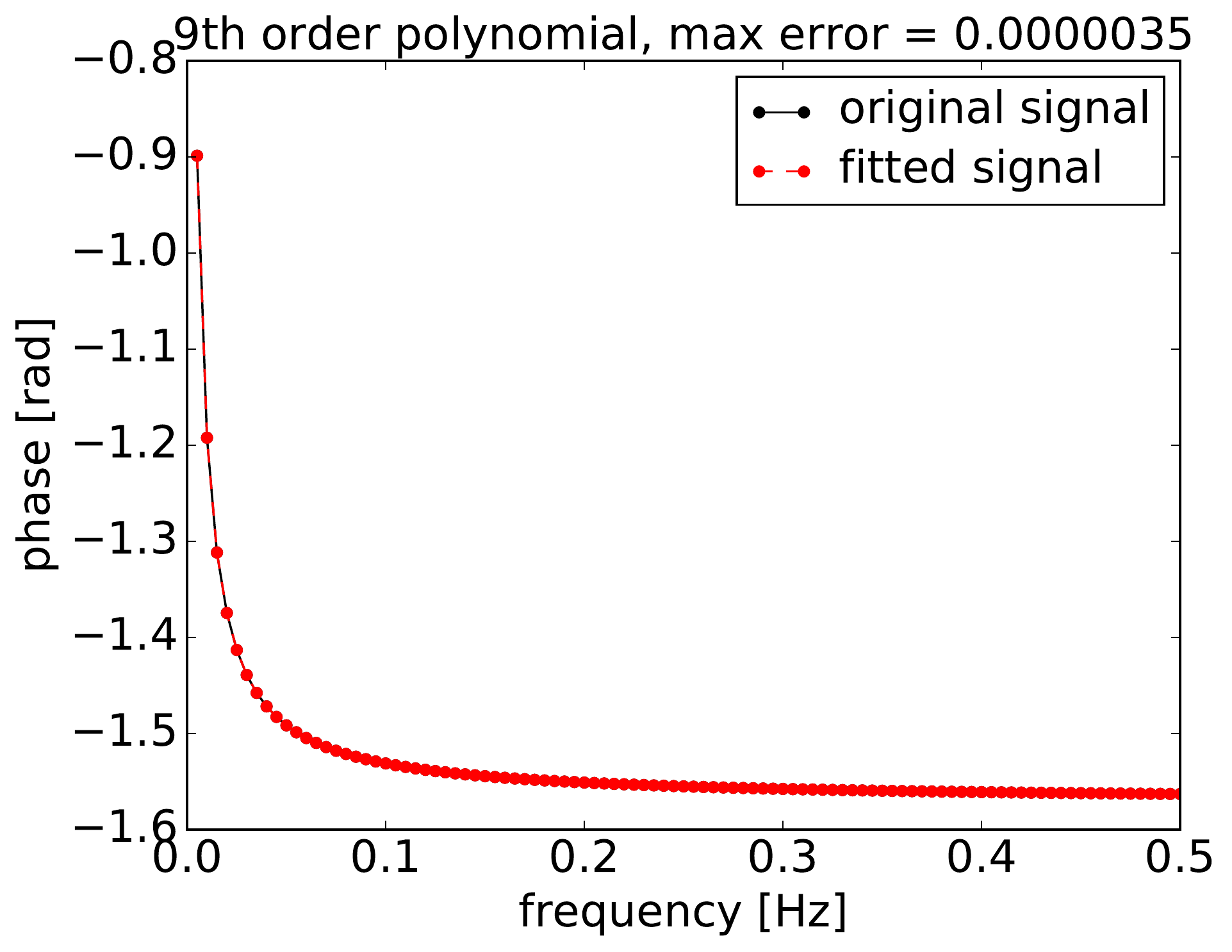}
\includegraphics[width=\figwidthhalf mm]{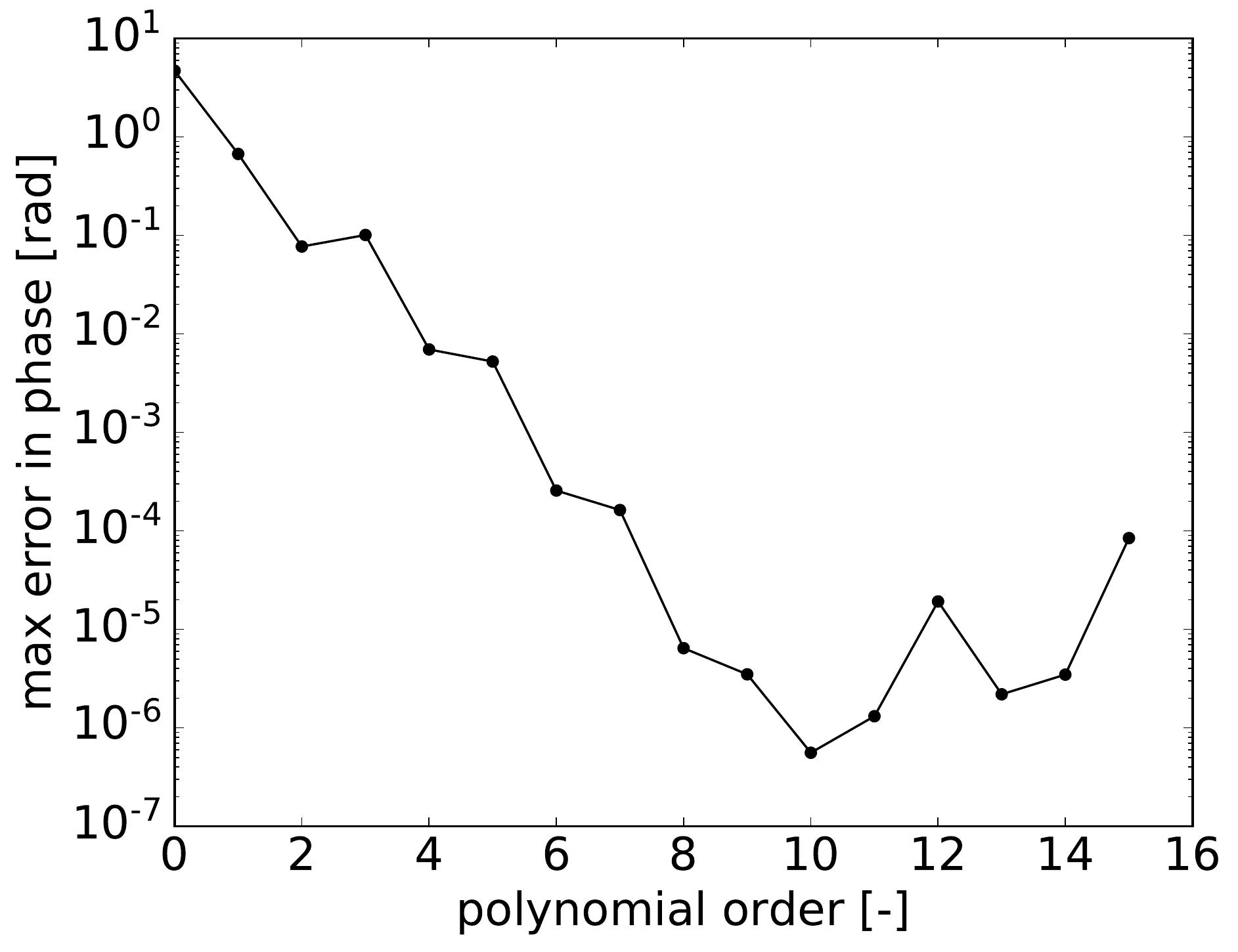}
\end{center}
\caption{Feasibility of polynomial fitting of temperature-time decay curves of exponential nature. 9th order polynomial fitting was used to show that errors are insignificant (left column). Corresponding error values for different orders are shown (right column). Data presented for temperature (top row), amplitude (middle row) and phase (bottom row). Sampling rate was kept constant at 1~Hz.
\label{fig:polyfitcftexp}}
\end{figure}

Before going on to real signals, it is known that researchers such as Ibarra-Castanedo and Maldague~\cite{,Ibarra2015TS} fit polynomials to temperature-time decay curves, to reconstruct the signals before applying the FFT. In order to establish polynomial fitting as fit for purpose in terms of analysing exponential signals in the frequency space, it is important to benchmark polynomial fitting algorithms. Results are shown in Fig.~\ref{fig:polyfitcftexp}, where a purpose written algorithm was used to find the optimal order of the polynomial for fitting the exponential function. It is interesting to note that the errors due to fitting decreased up to the 10th order. By setting the tolerance for temperature to 0.0001 degrees Celsius, it was found that optimum fits were obtained using 9th order polynomials. As shown in Fig.~\ref{fig:polyfitcftexp}, the errors in temperature, amplitude and phase were insignificant. Hence, it is possible to use polynomials to fit temperature-time decay curves that are similar to exponential curves with no loss in fidelity.

\begin{figure} 
\begin{center}
\includegraphics[width=\figwidthhalf mm]{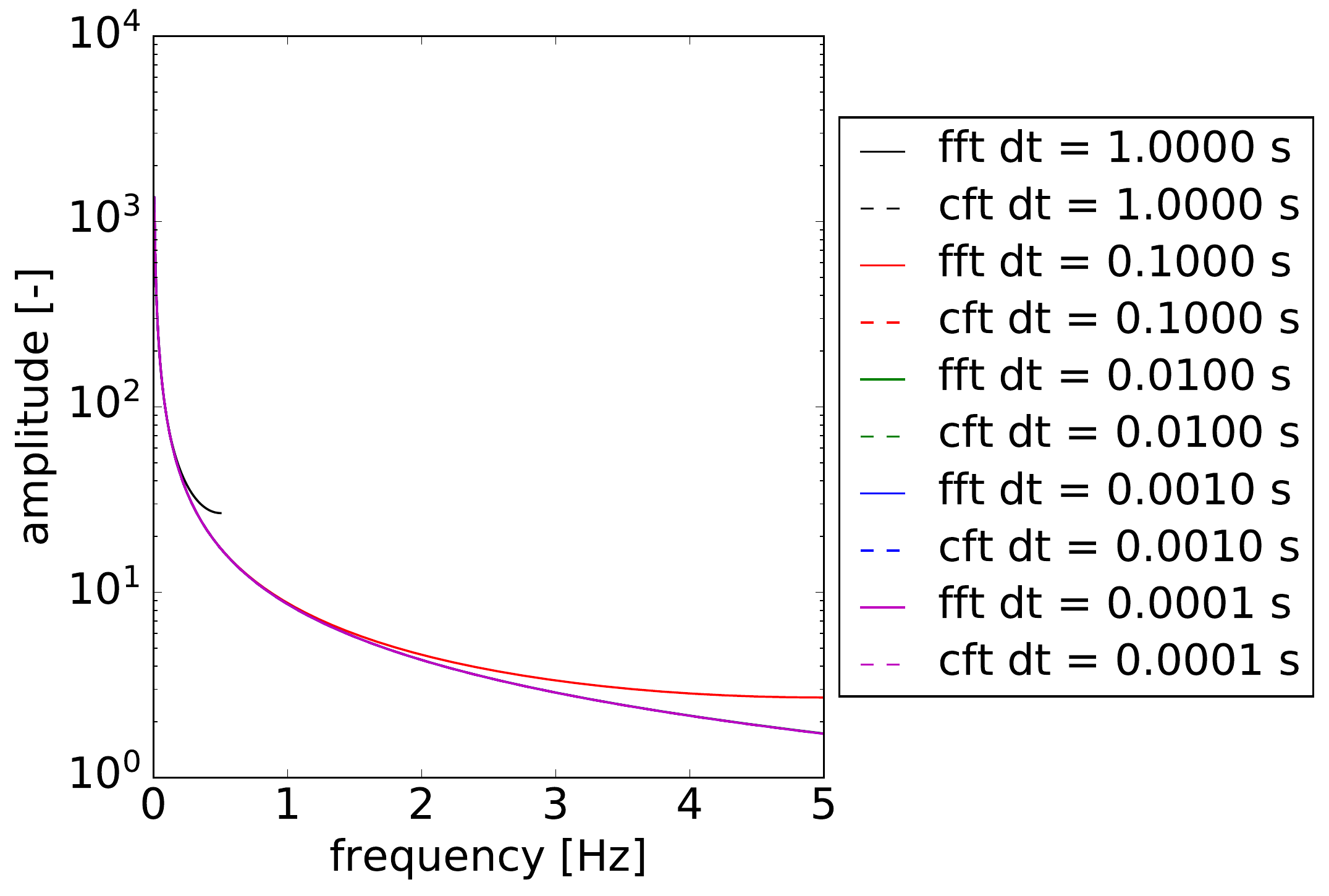}
\includegraphics[width=\figwidthhalf mm]{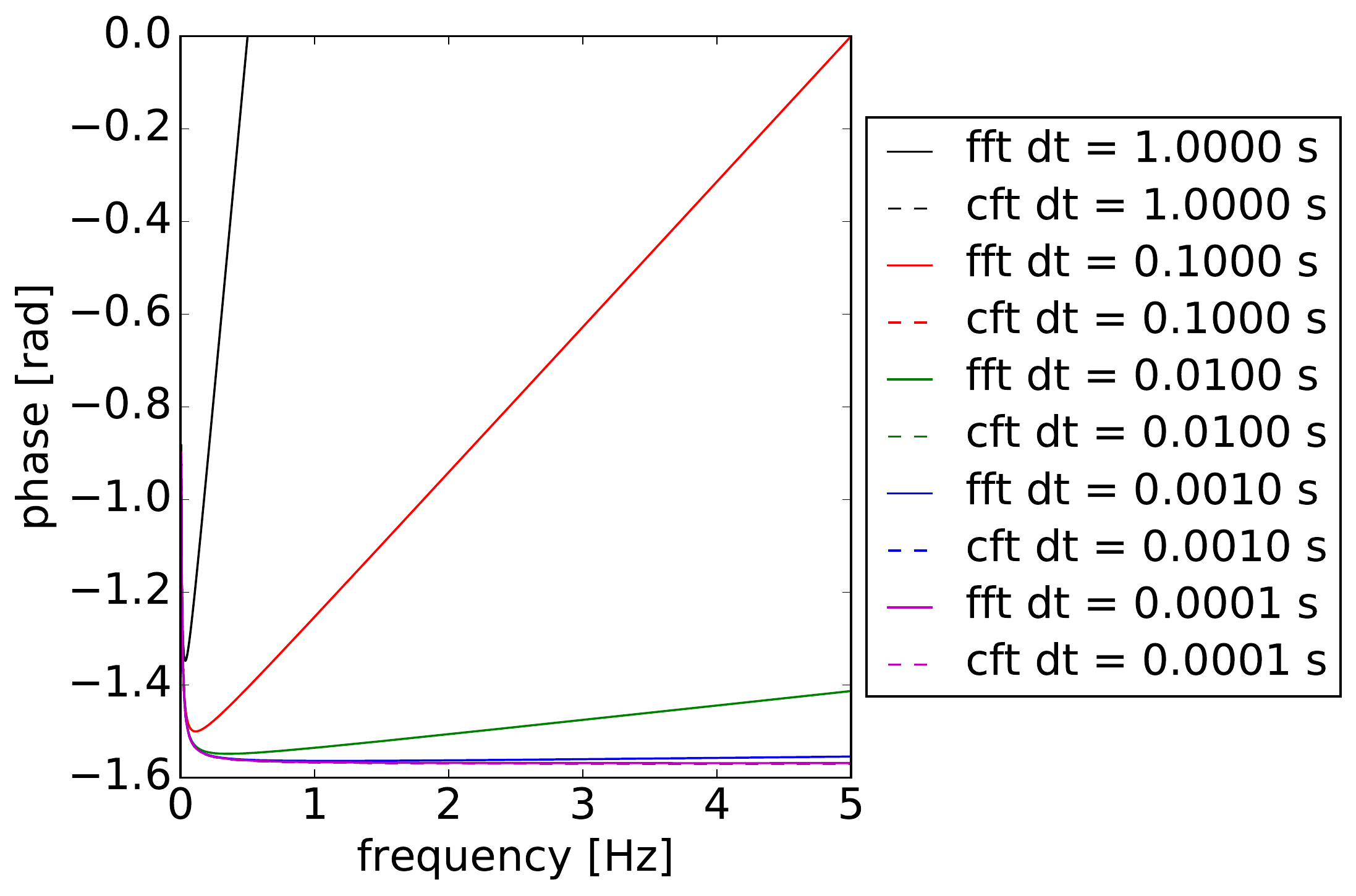}
\includegraphics[width=\figwidthhalf mm]{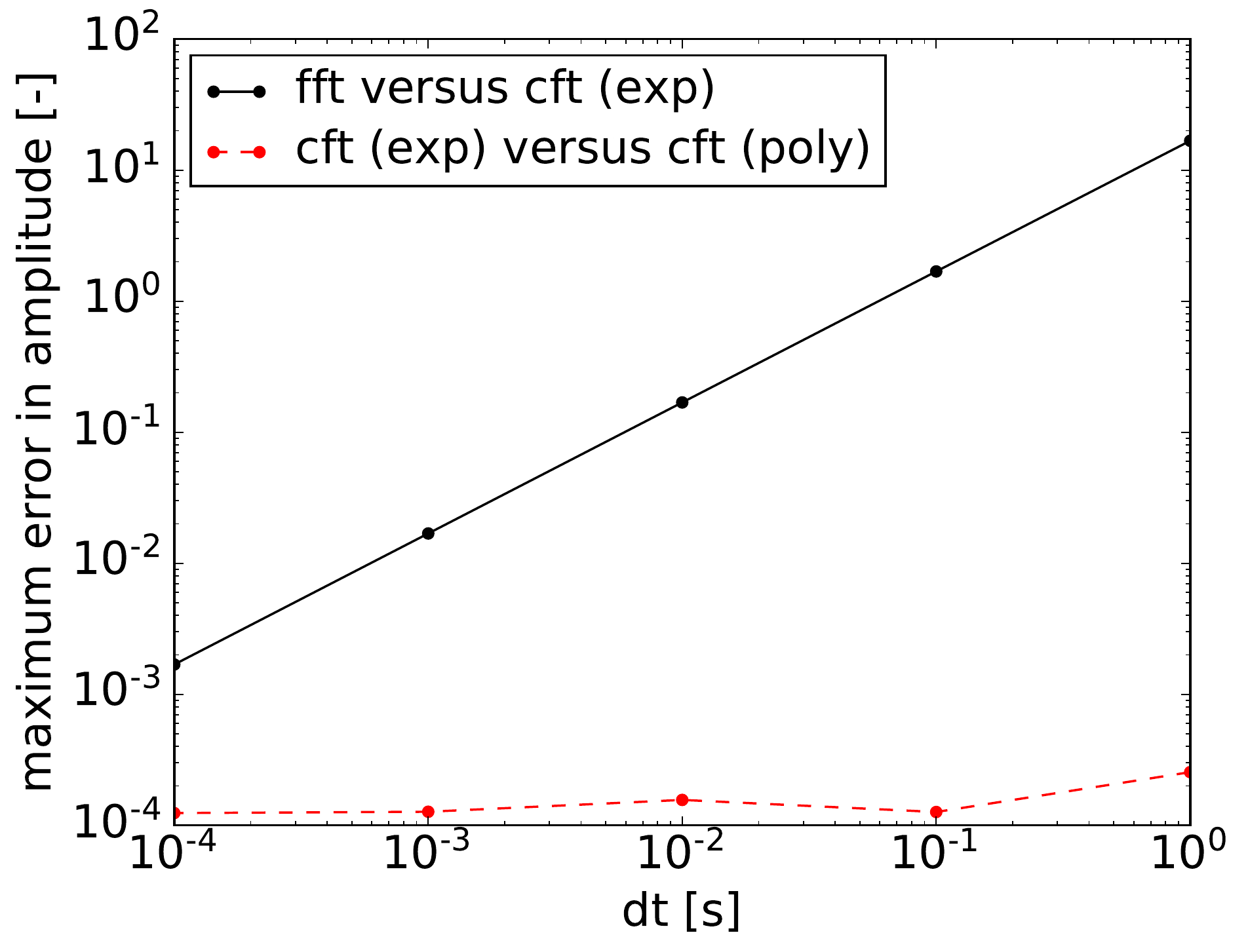}
\includegraphics[width=\figwidthhalf mm]{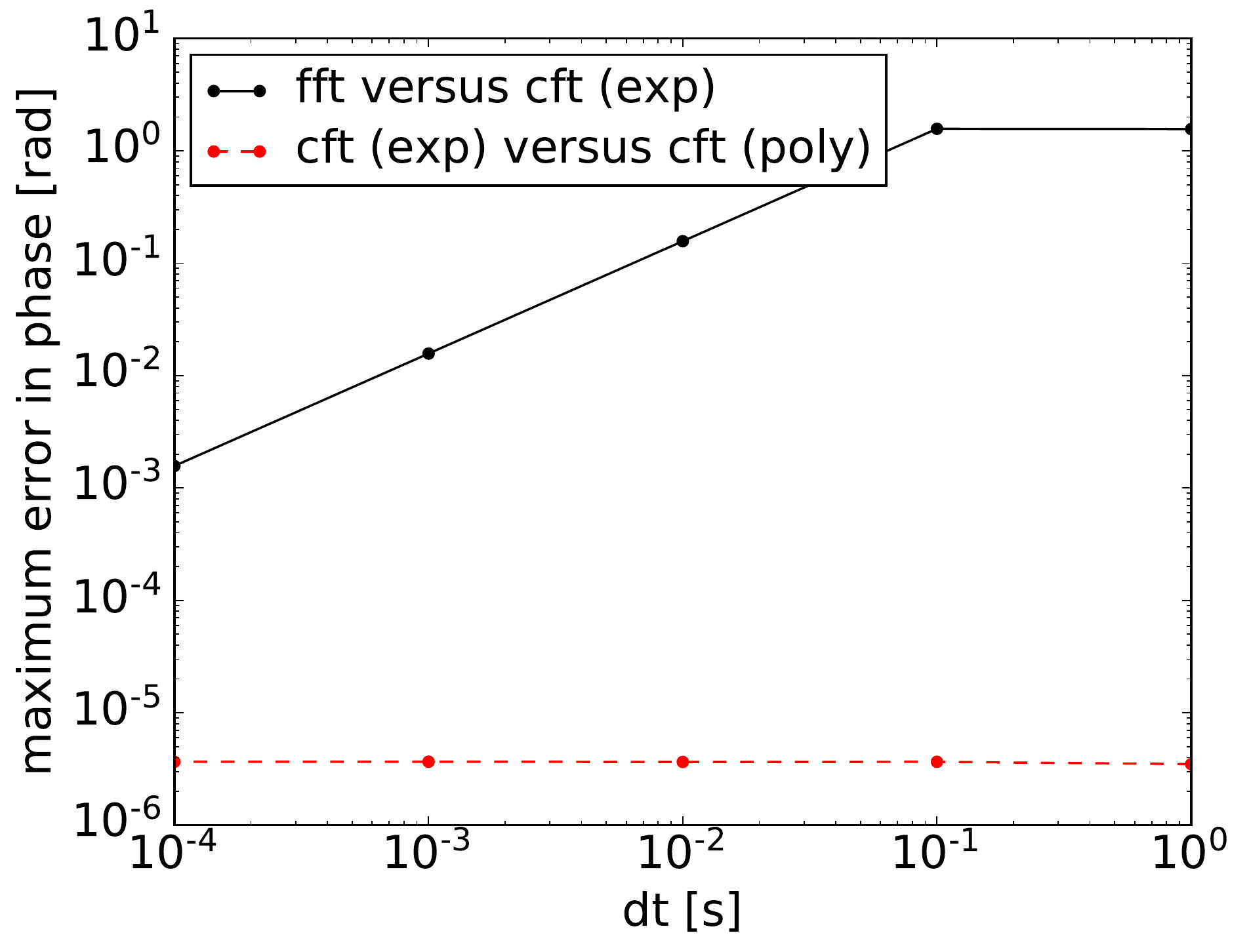}
\end{center}
\caption{Top row: Comparisons between FFT and CFT (without polynomial fitting) of exponential functions, for different sampling rates. Bottom row: Maximum errors in amplitude and phase for FFT and CFT (9th order polynomial fitting), for different sampling rates, compared with exact CFT of exponential function.
\label{fig:fft_vs_cftexppoly_dt}}
\end{figure}

Also, it is important to address the aliasing problem in PPT, to understand the motivation behind the current work. As shown before in Fig.~\ref{fig:fft_vs_cftexp_freq}, the amplitude and phase produced by FFT and CFT under the same conditions are radically different. For FFT, it is expected that the phase values will decrease with frequency initially, and increase up to the value of 0 at the point corresponding to half the sampling frequency. This is a feature of FFT rather than the signals, and cannot be physically removed even with changes to variables such as sampling rate. For the purpose of benchmarking the techniques for practical applications, assume a frequency range of interest up to 5~Hz. The relative effects of a vast range of sampling rates from 1~Hz to 10~kHz are shown in Fig.~\ref{fig:fft_vs_cftexppoly_dt}. It can be seen that the amplitude and phase of the exponential signal (using CFT) are not affected by the sampling rate, while equivalent values from FFT of the same signal vary significantly. It appears that if a range of frequency values of interest are defined, the results from FFT converge towards that of the analytical solutions. However, even at sampling rates of 10~kHz, the maximum errors in amplitude and phase are still much higher than equivalent errors if polynomial functions were used with CFT. While some infrared cameras may be capable of capturing images at 10~kHz, the equivalent storage space required for long duration measurements do not yet exist. As such, there is no practical experimental setup that enables the FFT method to give the precise solutions required for understanding the spectral response of temperature-time decay curves. In contrast, by applying CFT directly to polynomial functions used to fit said decay curves, the errors were almost constant for the range of sampling rates investigated. Hence, it is possible to understand the physical phenomena accurately, if CFT with polynomial fitting is used, even if cheaper infrared cameras are used at low frame rates. It can also mitigate issues such as measurement noise. However, the frame rates have to be sufficiently high to capture the phenomena, especially at the start of the decay curves. It is possible to do so with cheap hardware over a network as well.

Having established the precision with polynomial fitting with CFT on temperature-time decay curves, in terms of amplitude and phase in the frequency space, it is important to reiterate the purpose of this publication. In recent times, there are still new publications related to PPT and with FFT~\cite{Maldague1996JAP,Ibarra2005THESIS,Ibarra2004QIRT,Ibarra2015TS,Wang2018NDTE,DAccardi2019JNE,Saeed2018JNE,Ishikawa2019CS,Moradi2019CST}, which may not be sufficient representations of the true phenomena of interest. A concerted effort by the community will most certainly move the research in the right direction, as databases can be updated by applying the recommended polynomial fitting with CFT onto existing data, and sharing such information with others. A global effort in this area will accelerate the development of advanced technologies in Thermography, which will benefit both academia and industry, as the existing data is probably of the scale sufficient for a Big Data approach, similar to that of Tian~\etal~\cite{Tian2019AX}. It would certainly allow Machine Learning algorithms to identify geometries and defects automatically in complex structures, where it would have been deemed impossible. However, that cannot be done until a unified and precise approach is adopted for data curation. The first step to achieve this is to use polynomial fitting with CFT on PPT measurements, as well as any thermal measurements with temperature-time decay curves.

\section{Materials and Methods}
\label{sect:methods}

The experimental setup is the same as that of Goh~\etal~\cite{Goh2018AX}. The camera used in the current study is the FLIR SC7500, with synchronised images obtained in the same way as Goh~\etal~\cite{Goh2018AX}. A 3D printed sample of dimensions 70x56x8~mm,  made from ABS, was used. 9 defects of dimensions 4x4x0.5~mm were embedded at depths from 0.5~mm to 4.5~mm, in steps of 0.5~mm. These defects were spaced 10~mm apart, and placed in the centre of the sample to minimise edge effects. A drawing of the dimensions can be found in Goh~\etal~\cite{Goh2018AX}.

\begin{figure} 
\begin{center}
\includegraphics[width=\figwidthhalf mm]{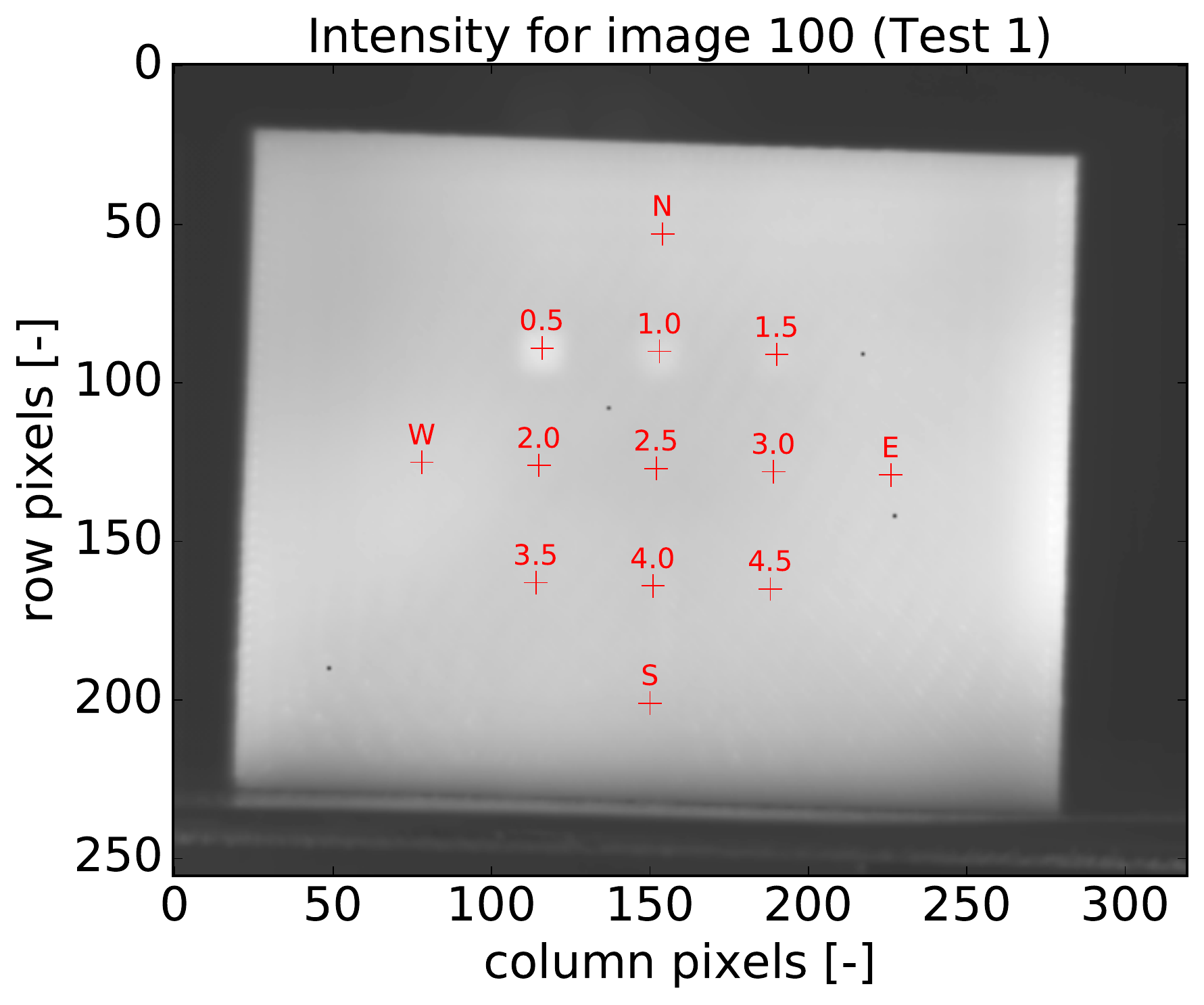}
\includegraphics[width=\figwidthhalf mm]{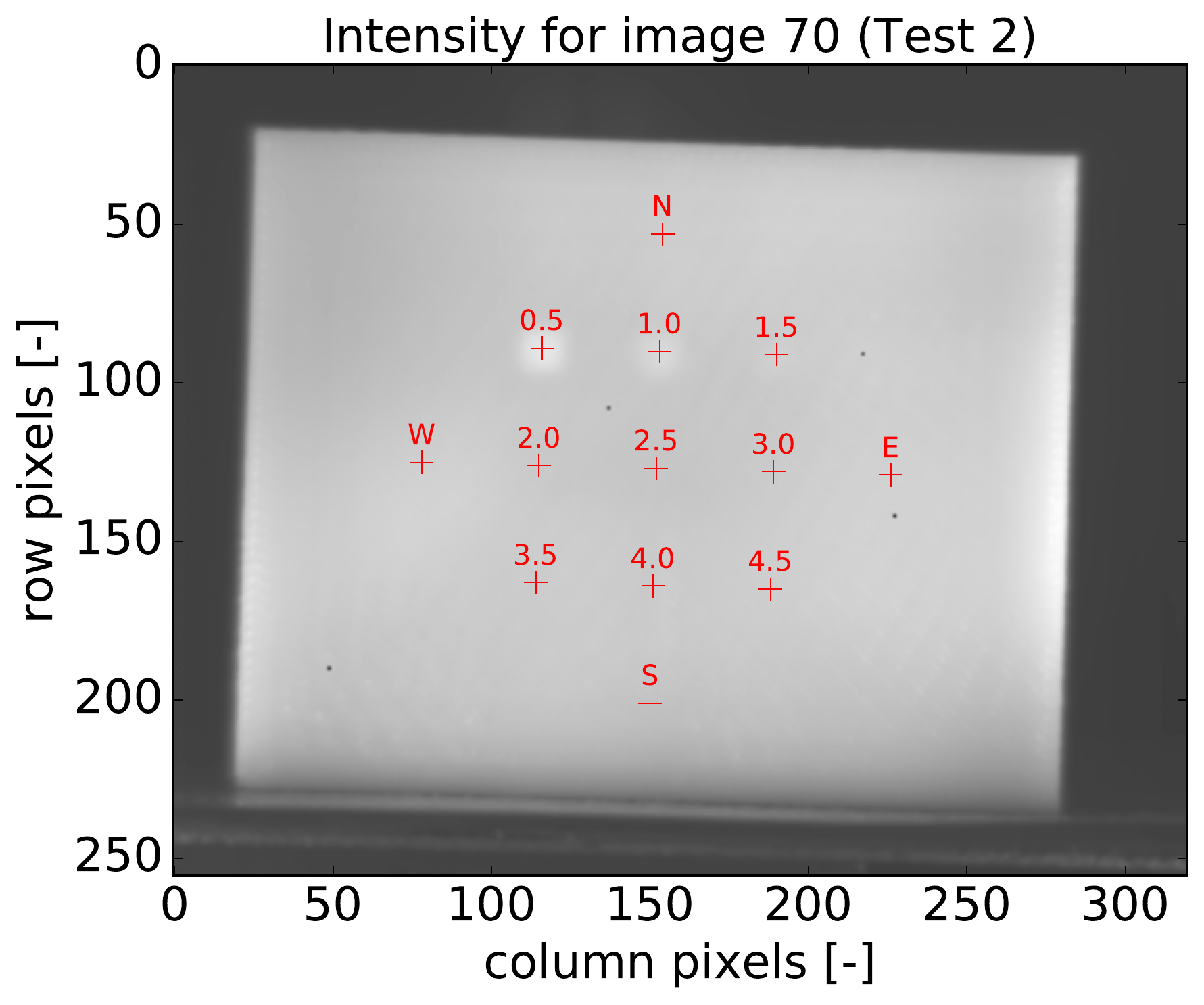}
\includegraphics[width=\figwidthhalf mm]{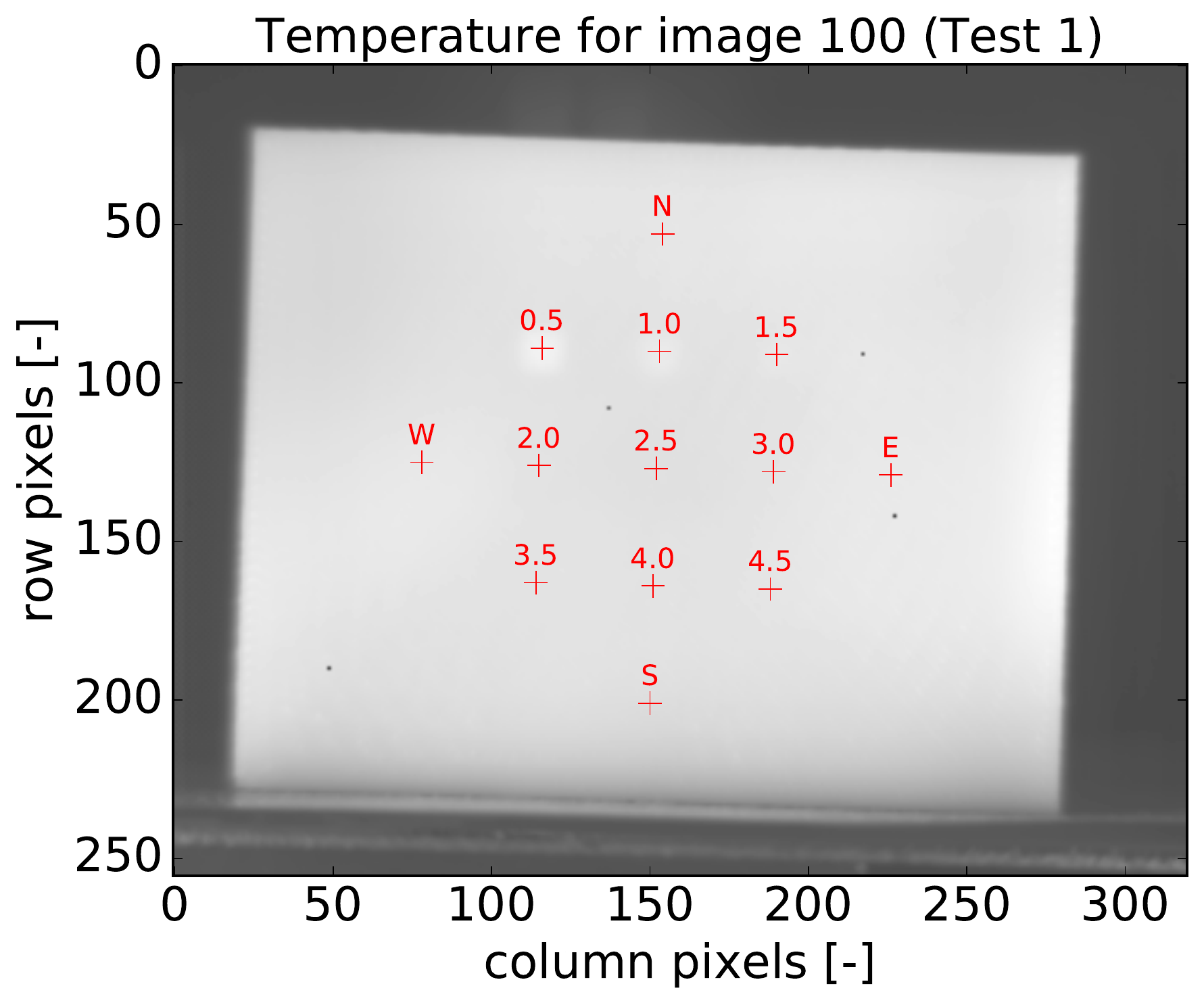}
\includegraphics[width=\figwidthhalf mm]{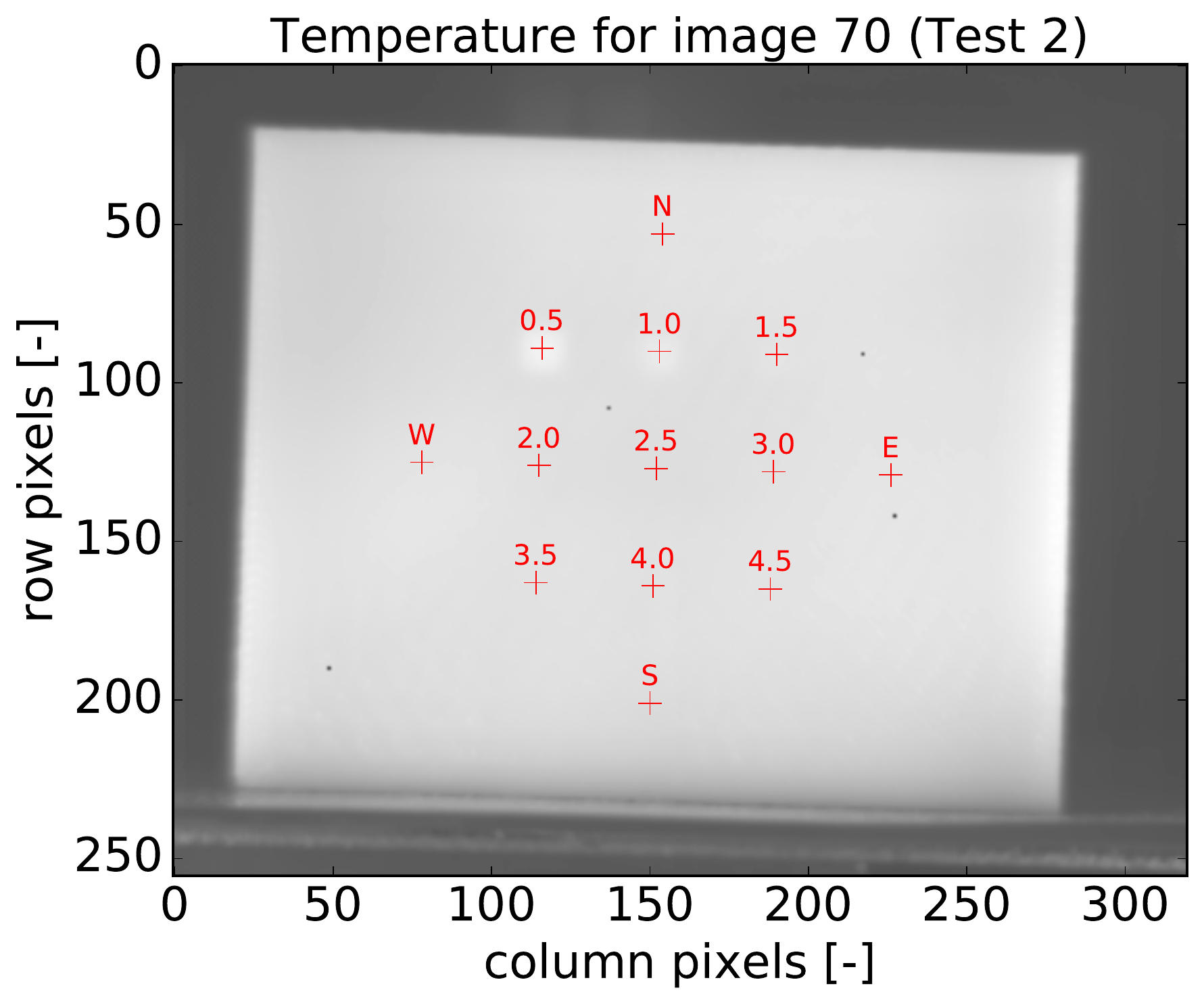}
\end{center}
\caption{Sample images for Tests 1 (left) and 2 (right), shown at the maximum temperature (end of heating), for intensity (top row) and calibrated temperatures (bottom row). 13 sampled points are also shown, with numbers indicating depth of defects 4x4x0.5~mm in size. N, S, E, and W are points taken from sound areas for benchmarking.
\label{fig:ppt_sampleimg}}
\end{figure}

Two sets of measurements were obtained for the current study, using a 200 W heating system made from halogen lamps. Images were captured for 200 seconds at 1 frame per second, for heating of 100 (Test 1) and 70 (Test 2) seconds respectively. Image intensity and temperature values were obtained for comparisons. For the purpose of brevity, only 13 points were used to obtain temperature-time decay curves for comparisons. 9 points were selected from the centroids of embedded defects, and 4 reference points from sound regions (N, S, E, W) were selected to benchmark the proposed technique against existing ones. Sample images are shown in Fig.~\ref{fig:ppt_sampleimg}. Cooling curves of 100 seconds in duration were extracted for the purpose of this study. From the images, it can be seen that there appears to be better contrast for intensity than calibrated temperatures. Each image shows consistent dark pixels that do not change with time, which are due to dead pixels in the camera, probably due to damage from lasers. Purpose written algorithms were written by Goh~\etal~\cite{Goh2018AX} to overcome such issues when extracting phase profiles and images for comparisons.

\section{Results and Discussion}
\label{sect:results}

For a start, it is important to establish the optimum order for polynomial fitting of temperature-time decay curves obtained from experiments. This is an important step as the curves may not be exponential in nature. For the 13 points of interest, a purpose written algorithm was written to find the lowest possible order such that the  maximum and mean errors of the fitted curves were below set tolerance values. The corresponding tolerance values for selected orders are shown in Table~\ref{tab:polyfit}.  It is interesting to note that as the polynomial order increased from 7 to 9, the maximum and mean errors of the fitted curves were reduced significantly, with much smaller improvements between the 9th and 11th orders. For consistency in comparing the experimental results, the order of polynomial fit was set to 9.

\begin{figure} 
\begin{center}
\includegraphics[width=\figwidthhalf mm]{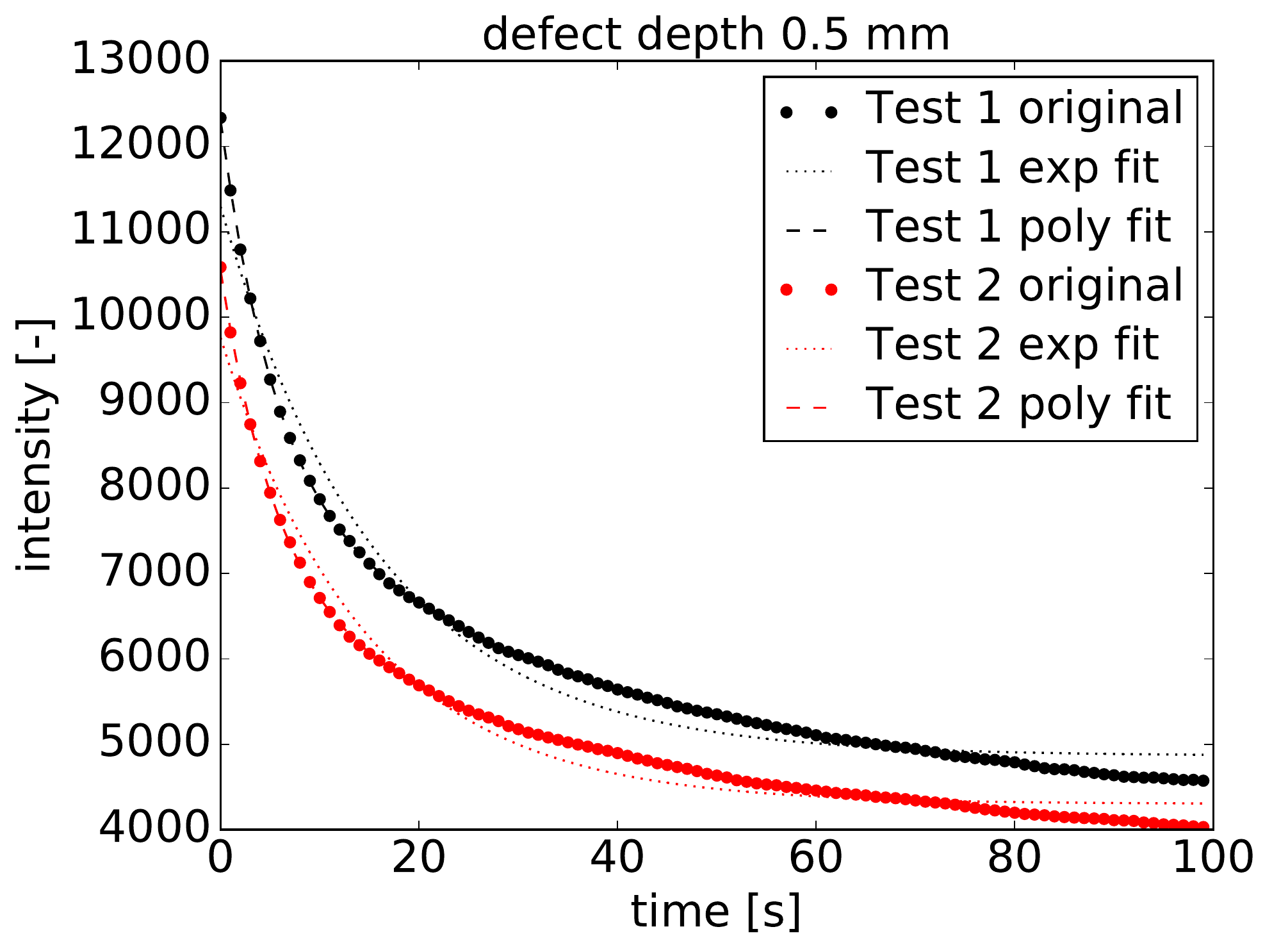}
\includegraphics[width=\figwidthhalf mm]{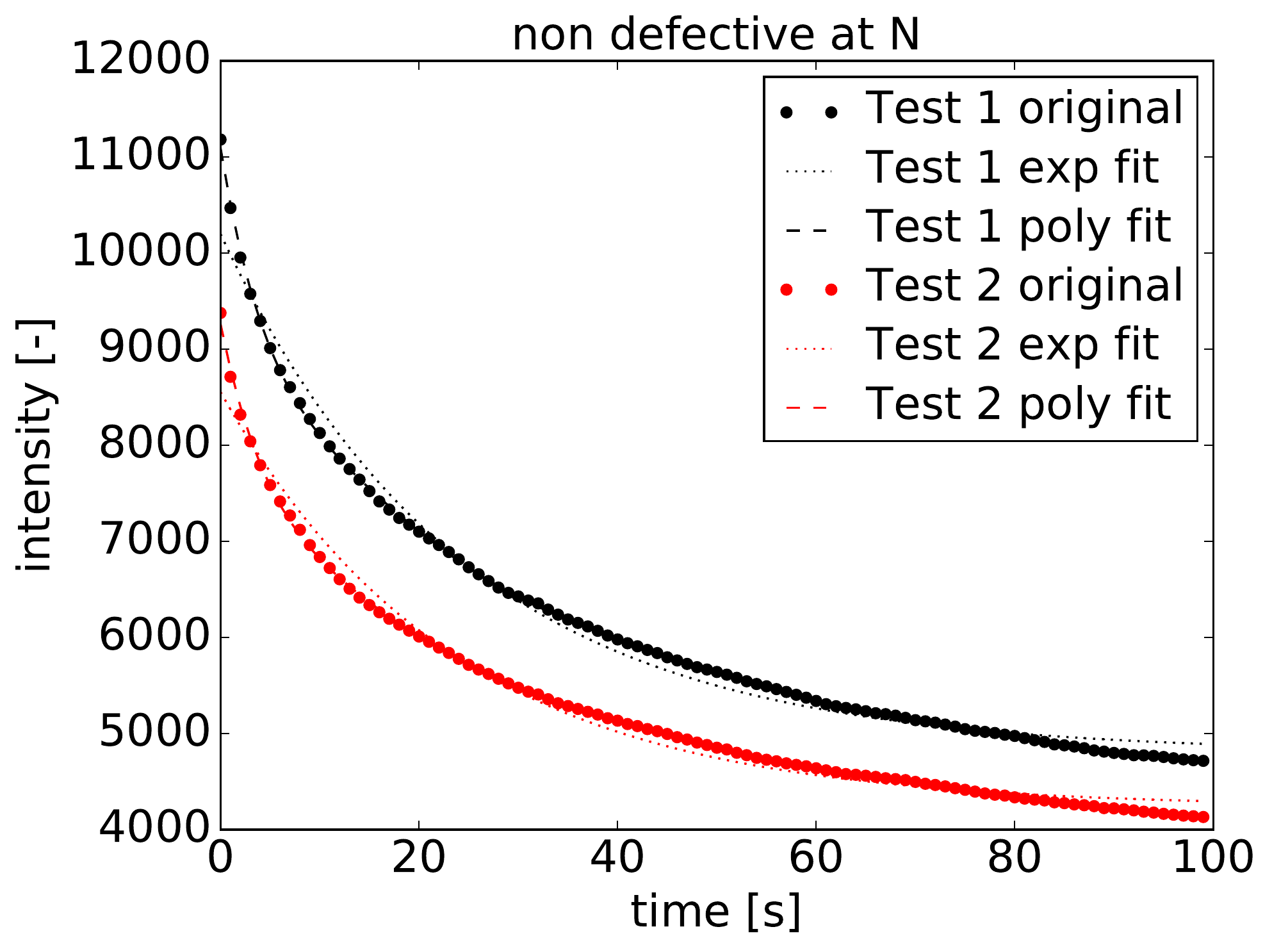}
\includegraphics[width=\figwidthhalf mm]{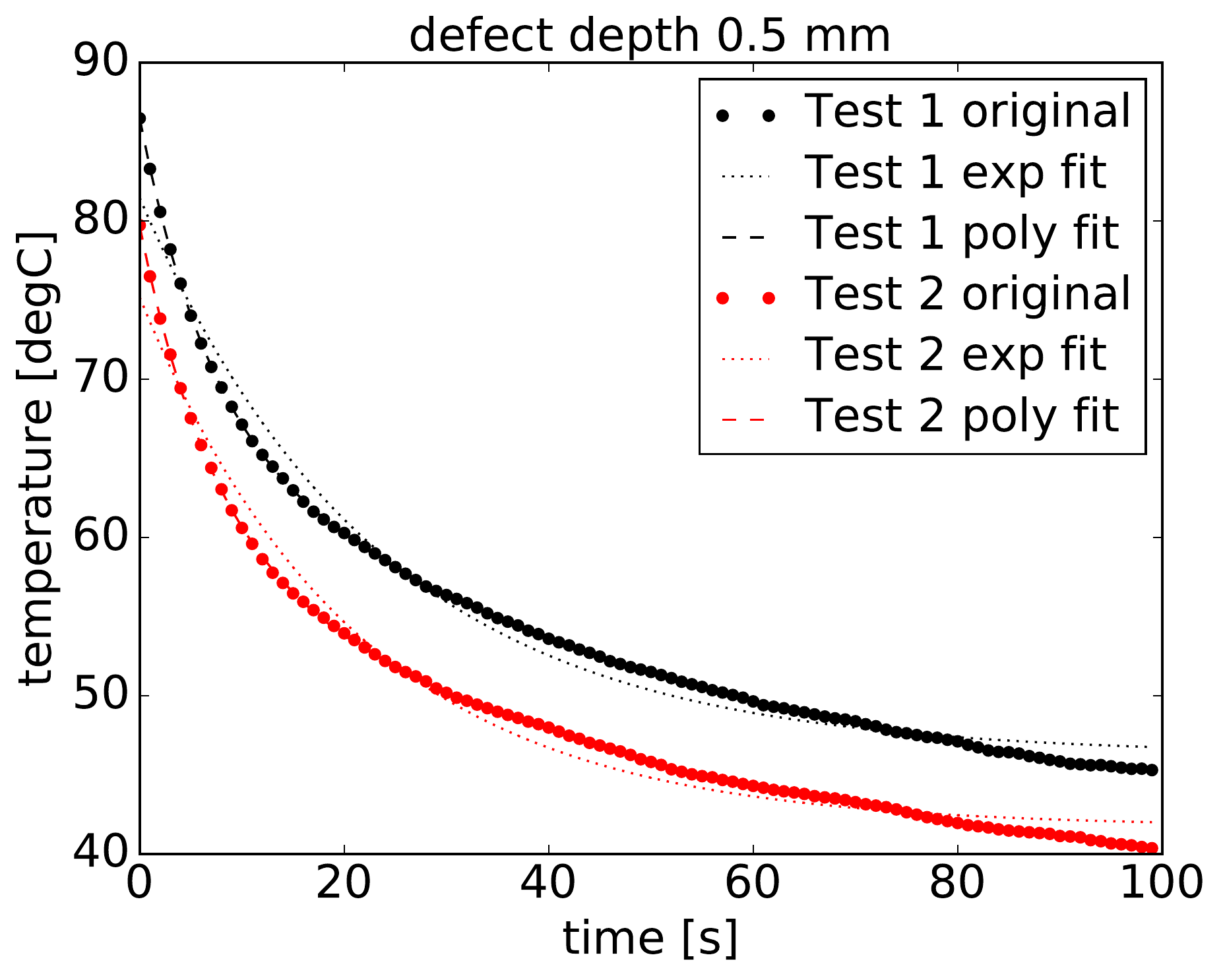}
\includegraphics[width=\figwidthhalf mm]{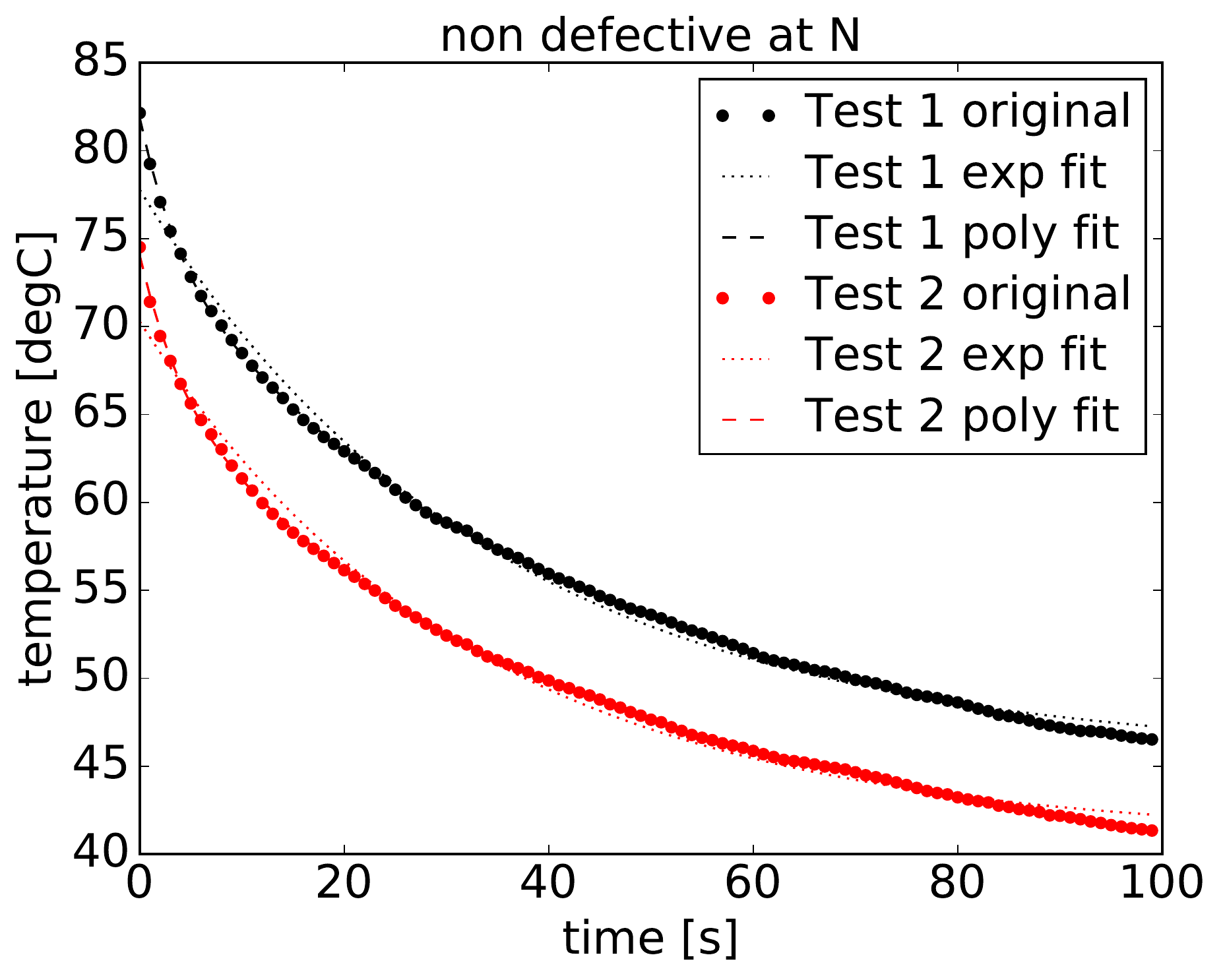}
\end{center}
\caption{Sample temperature-time decay curves for Tests 1 and 2, for both intensity and temperature. Data taken from point with defect at 0.5~mm depth, as well as sound area north of the defects. Exponential and polynomial fitted curves are shown for comparision.
\label{fig:ppt_samplesig}}
\end{figure}

Temperature-time decay curves were obtained for the point where defect was 0.5~mm below surface, and the point N (north of the array of defects). Fig.~\ref{fig:ppt_samplesig} shows the curves at these points, for both intensity and temperature. Thes selected points represent the shallowest defect and non-defective areas. It is quite obvious that as heating time increased, the peak temperature increased. The temperature and intensity values did not decrease at equal rates for Tests 1 and 2. For the shallow defect, the cooling rate was much higher. However, the temperatures were similar at these two points after cooling for 100 seconds. Due to the non-linear calibration curves in the infrared camera, the relative scales in intensity and temperature are very different. Hence, it is important to present data for both intensity and temperature, to ascertain the differences between the two. From the fitted curves, it is clear that polynomial functions fit the cooling curves better, across all the sample curves. Having established that polynomial functions fit the cooling curves better, different parameters were explored to understand their relative effects on the results.

\begin{figure} 
\begin{center}
\includegraphics[width=\figwidthhalf mm]{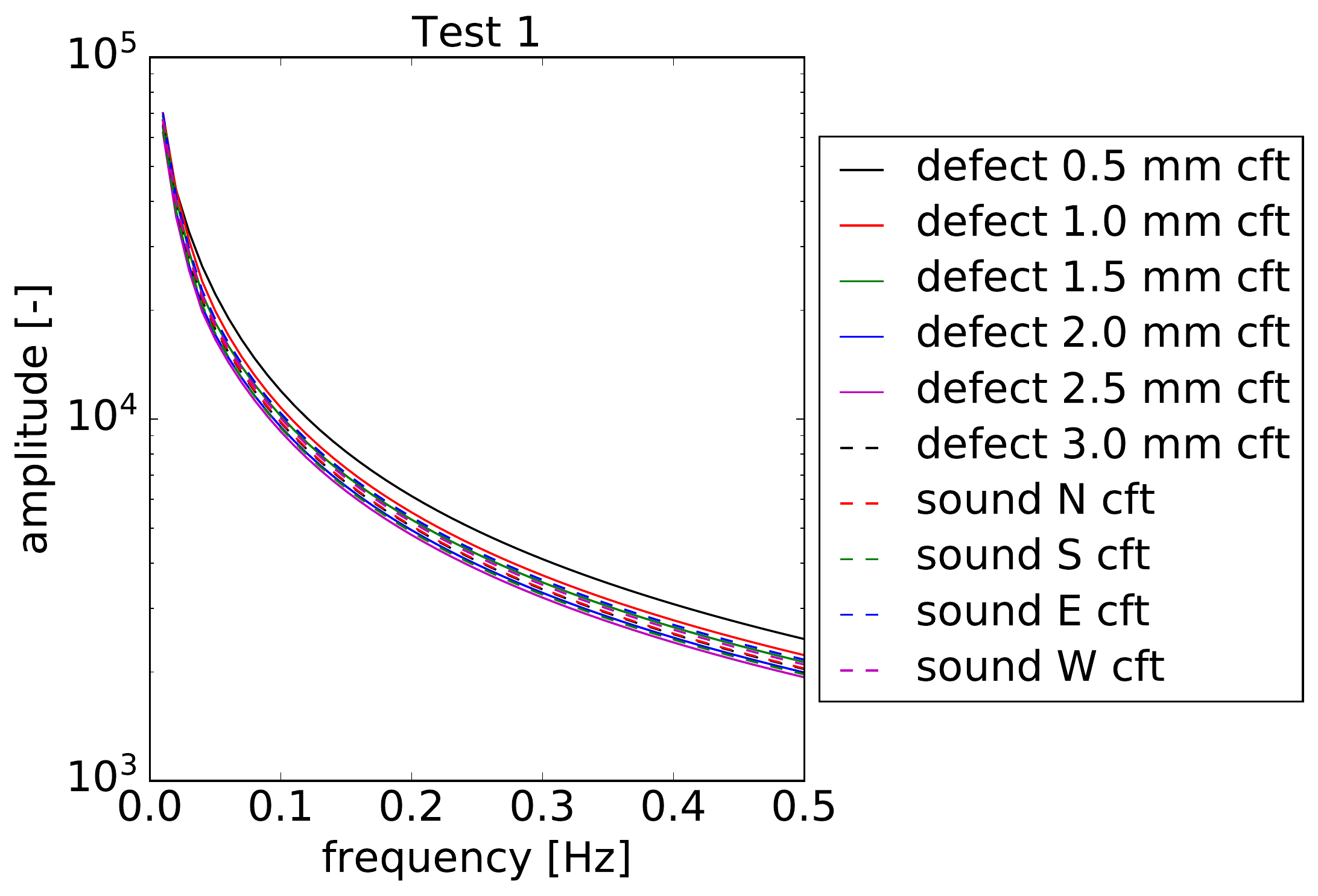}
\includegraphics[width=\figwidthhalf mm]{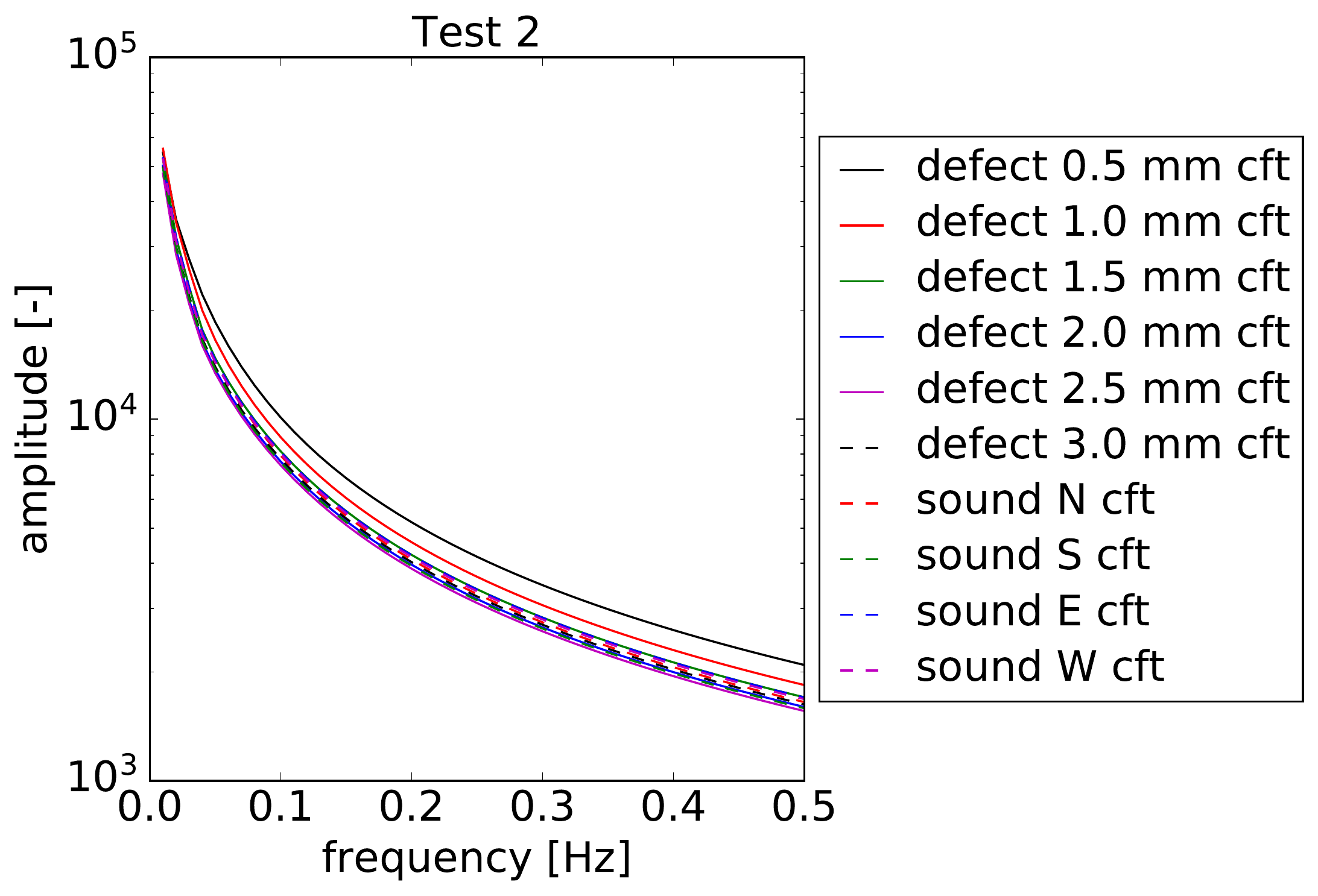}
\includegraphics[width=\figwidthhalf mm]{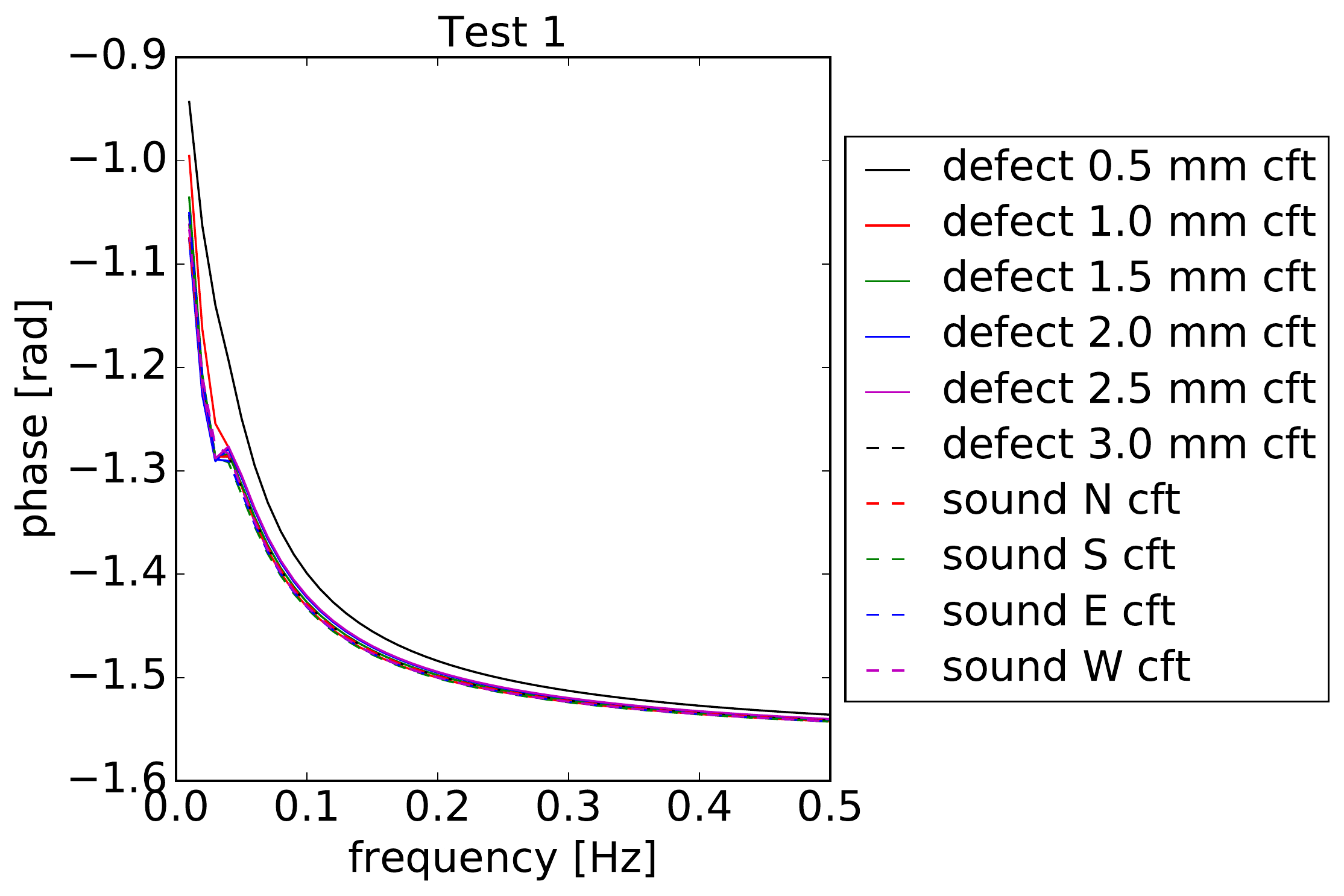}
\includegraphics[width=\figwidthhalf mm]{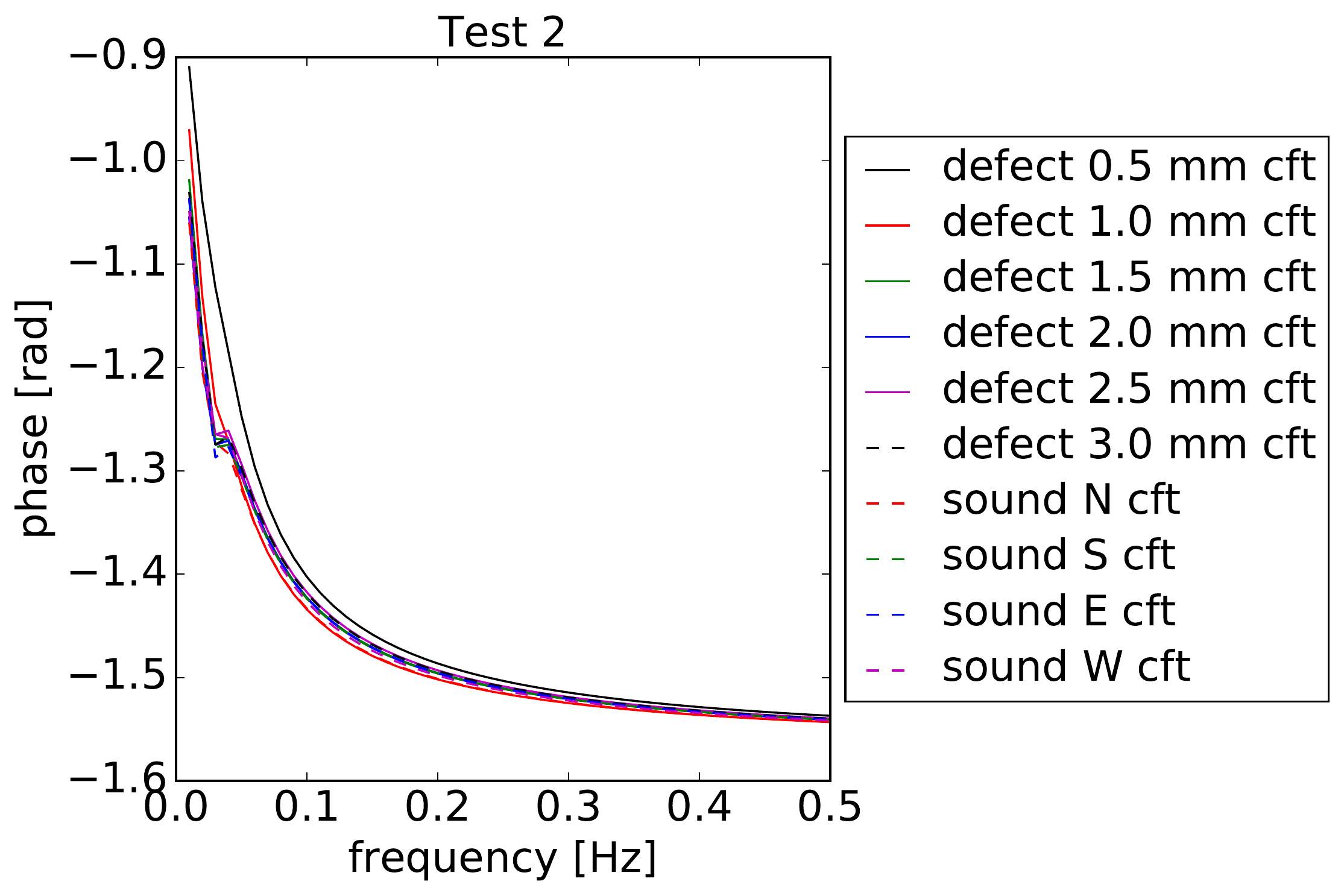}
\end{center}
\caption{Amplitude and phase values produced using CFT, at different defect depths, as well as sound areas (intensity).
\label{fig:ppt_diffdepthI}}
\end{figure}

\begin{figure} 
\begin{center}
\includegraphics[width=\figwidthhalf mm]{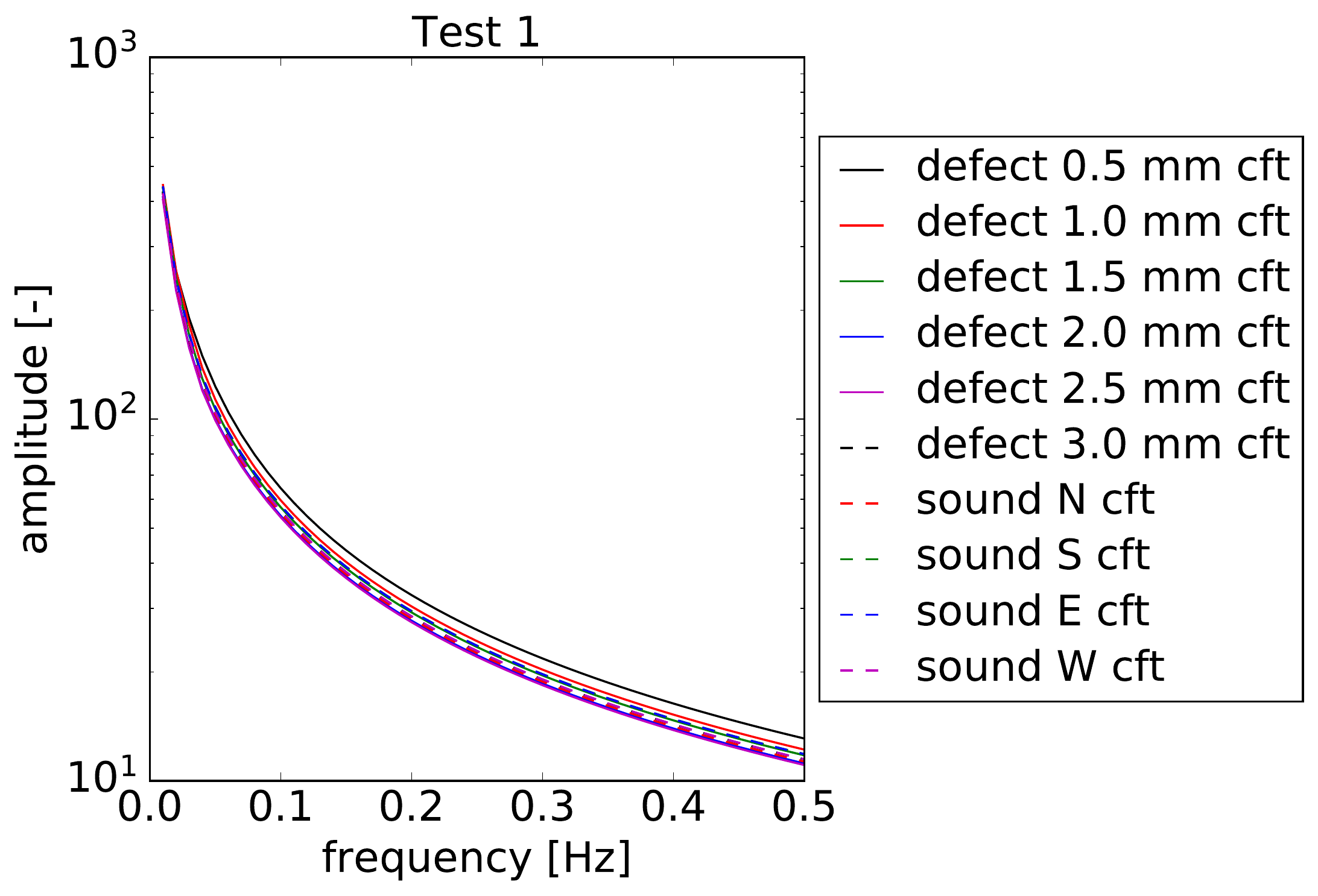}
\includegraphics[width=\figwidthhalf mm]{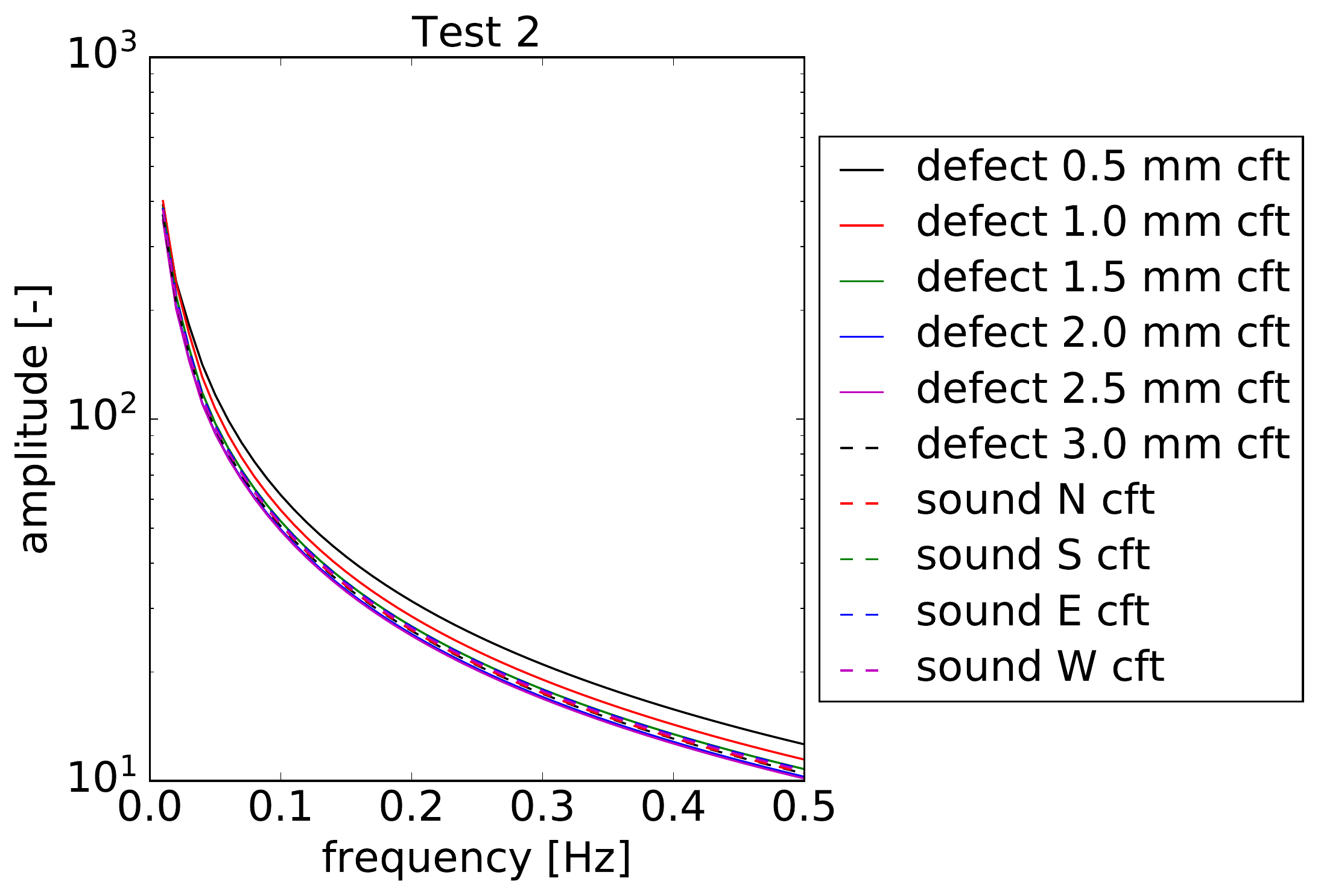}
\includegraphics[width=\figwidthhalf mm]{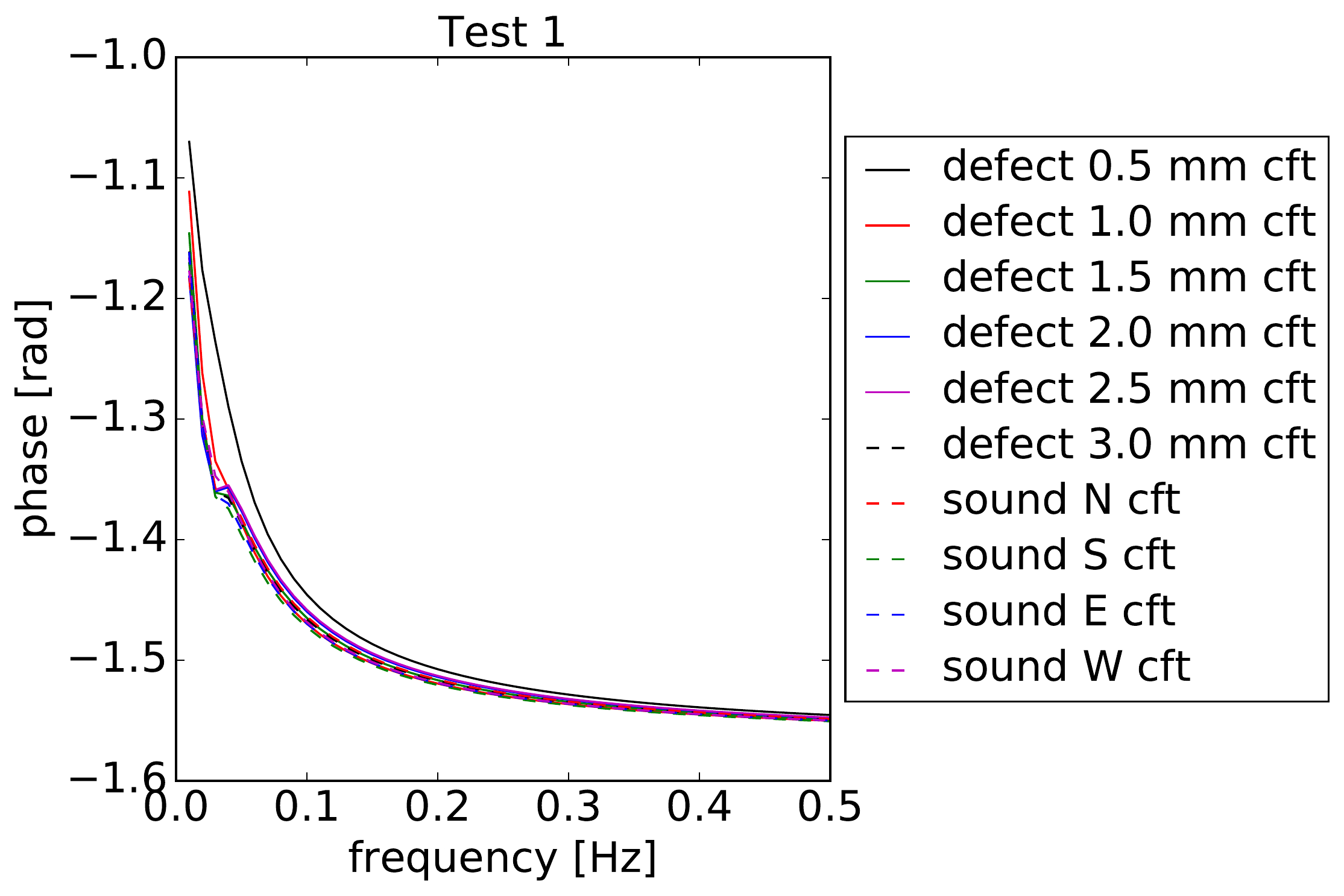}
\includegraphics[width=\figwidthhalf mm]{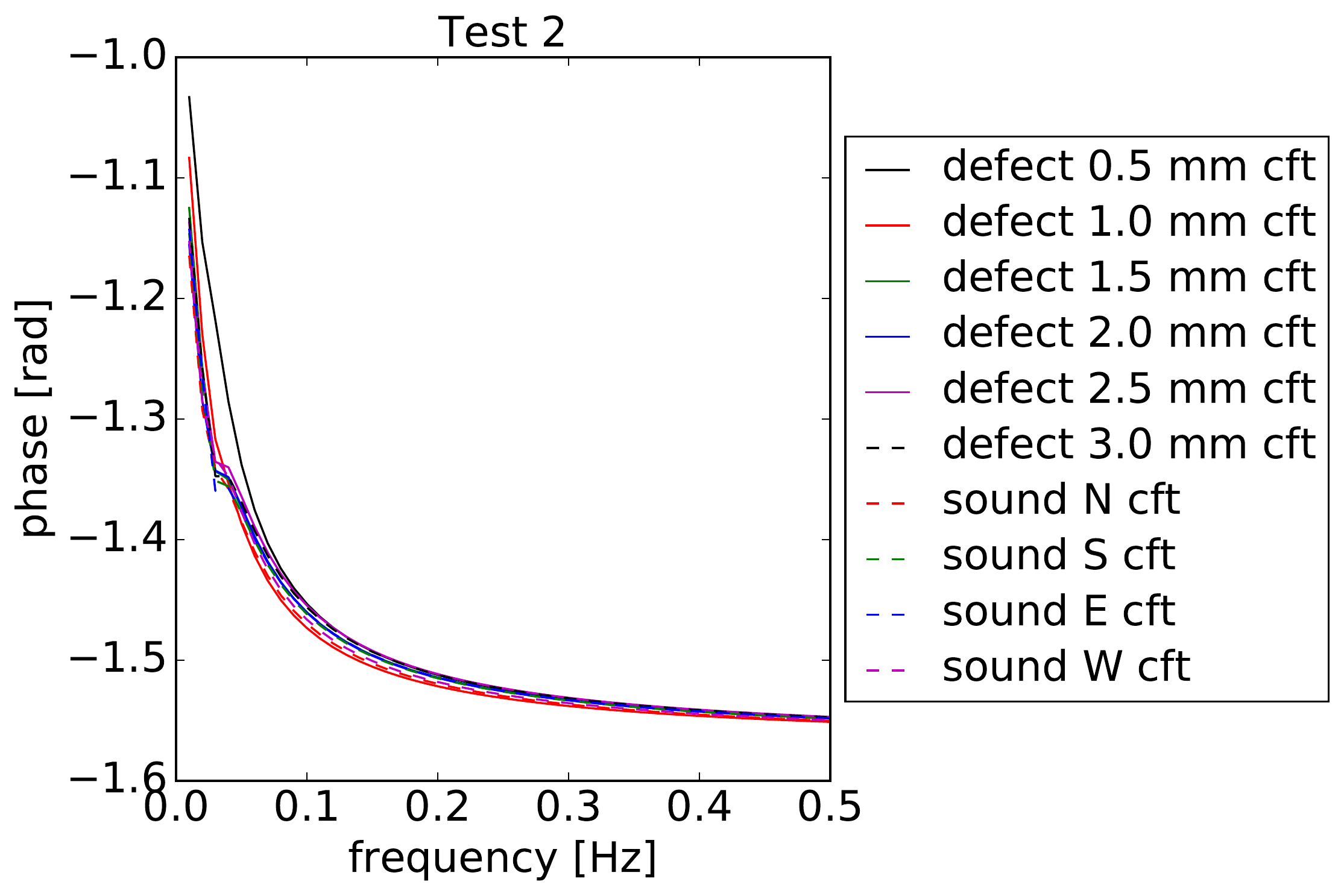}
\end{center}
\caption{Amplitude and phase values produced using CFT, at different defect depths, as well as sound areas (temperature).
\label{fig:ppt_diffdepthT}}
\end{figure}

By applying the proposed CFT technique on the polynomial fitted functions, it was possible to calculate the amplitude and phase in frequency space. Figs.~\ref{fig:ppt_diffdepthI} and \ref{fig:ppt_diffdepthT} show these results for defects up to 3.0~mm deep, as well as the 4 points representing sound areas, for intensity and temperature respectively. It is interesting to note that while the amplitude curves were smooth, there was a local discontinuity between 0.03 and 0.04 Hz, for detects deeper than 1.0 mm, in the phase results.  As the fitted curves produced minimal errors compared to measured signals, the integrals in CFT should be as close to actual as possible. Hence, these discontinuities may be representative of the phenomena measured.

\begin{figure} 
\begin{center}
\includegraphics[width=\figwidthhalf mm]{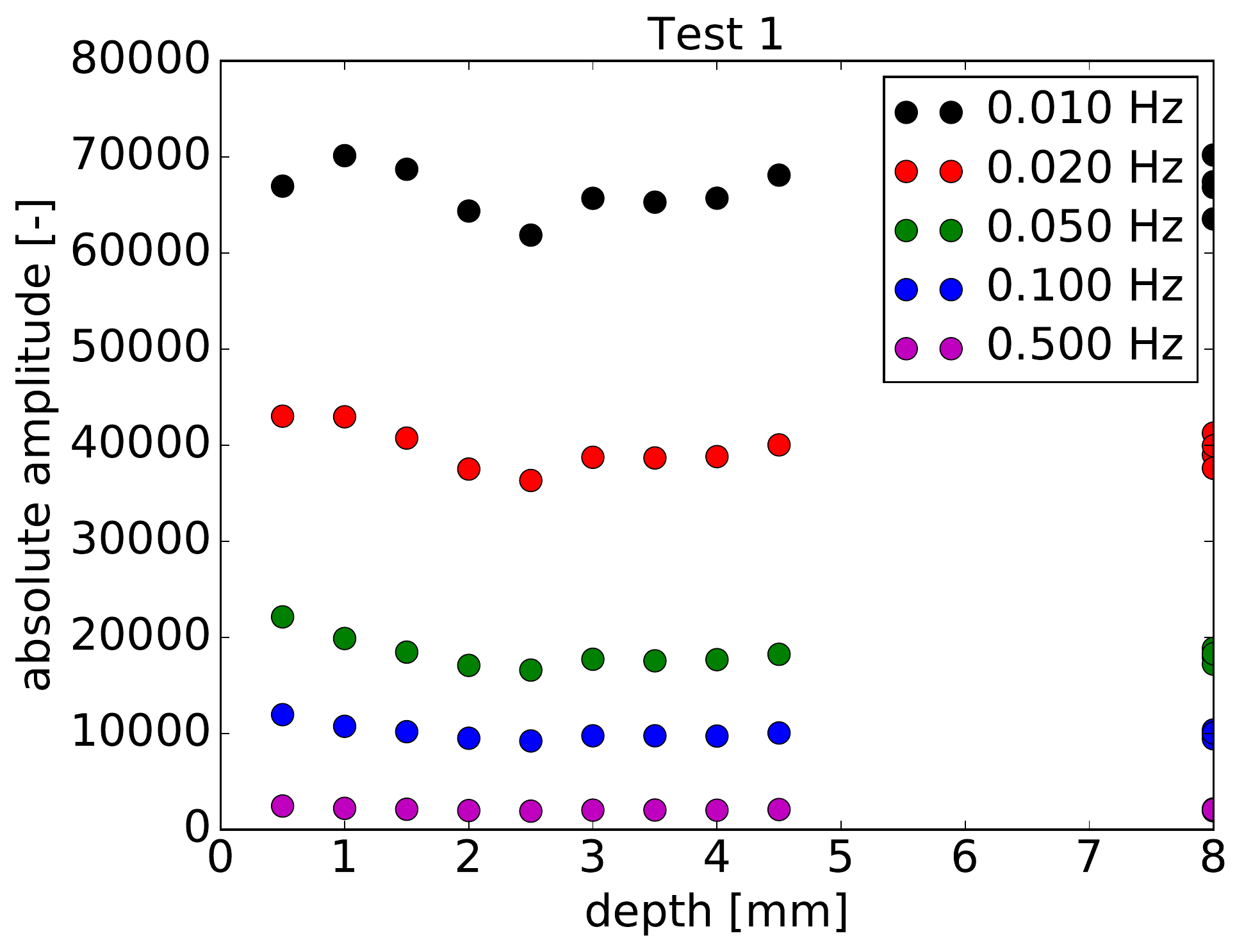}
\includegraphics[width=\figwidthhalf mm]{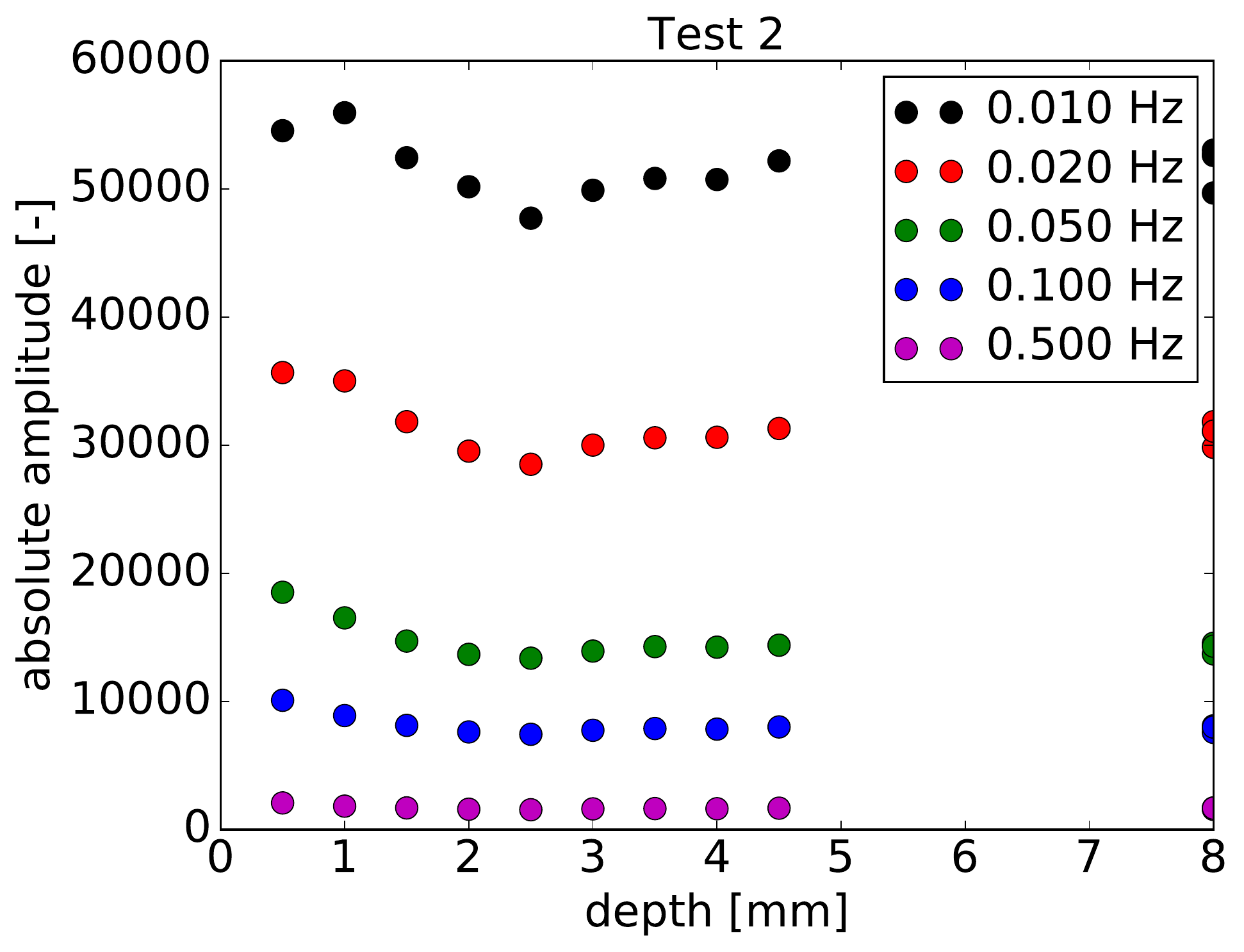}
\includegraphics[width=\figwidthhalf mm]{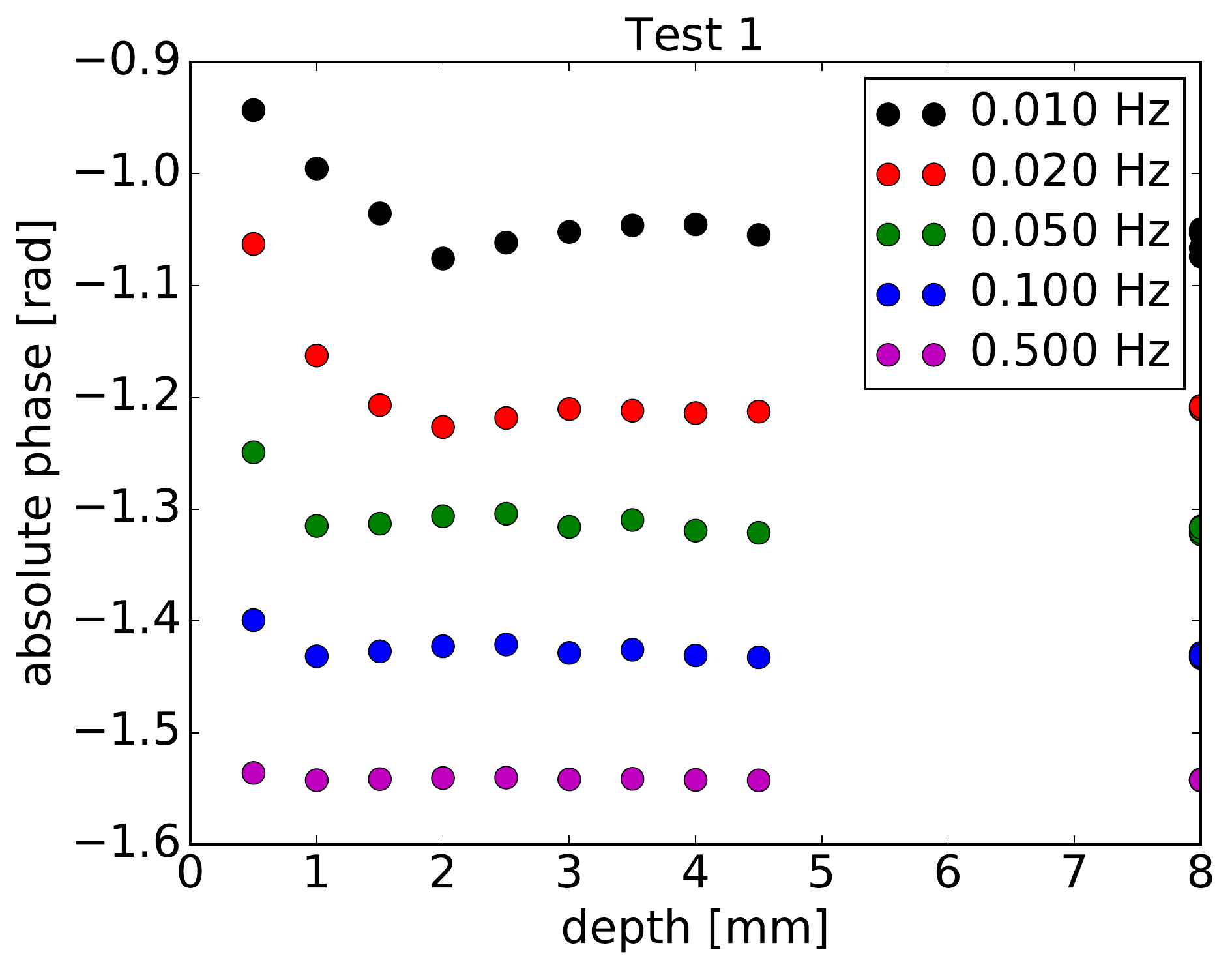}
\includegraphics[width=\figwidthhalf mm]{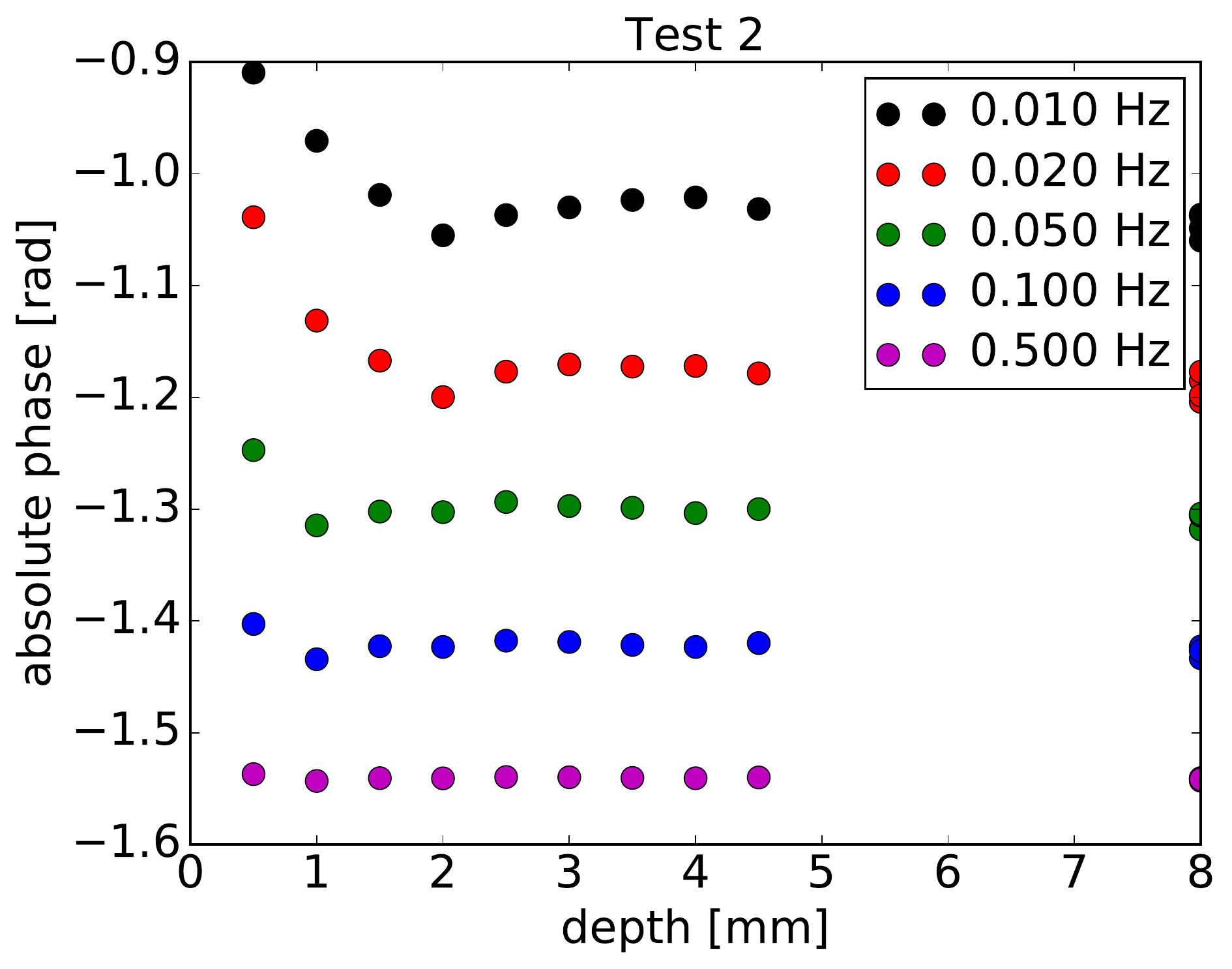}
\end{center}
\caption{Absolute contrast due to depth of defects, for values of amplitude and phase using CFT, at selected frequency values (intensity).
\label{fig:ppt_difffreqI}}
\end{figure}

\begin{figure} 
\begin{center}
\includegraphics[width=\figwidthhalf mm]{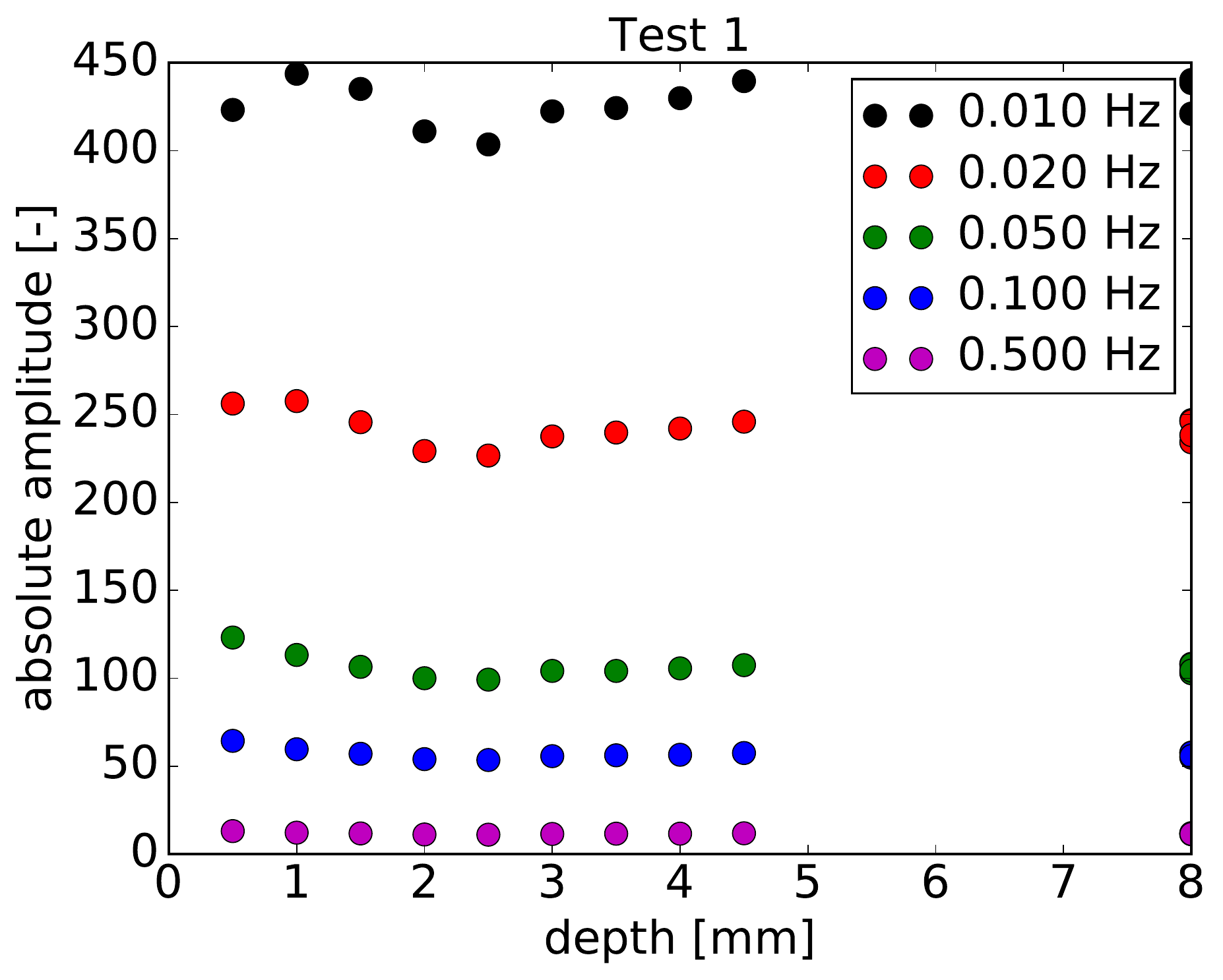}
\includegraphics[width=\figwidthhalf mm]{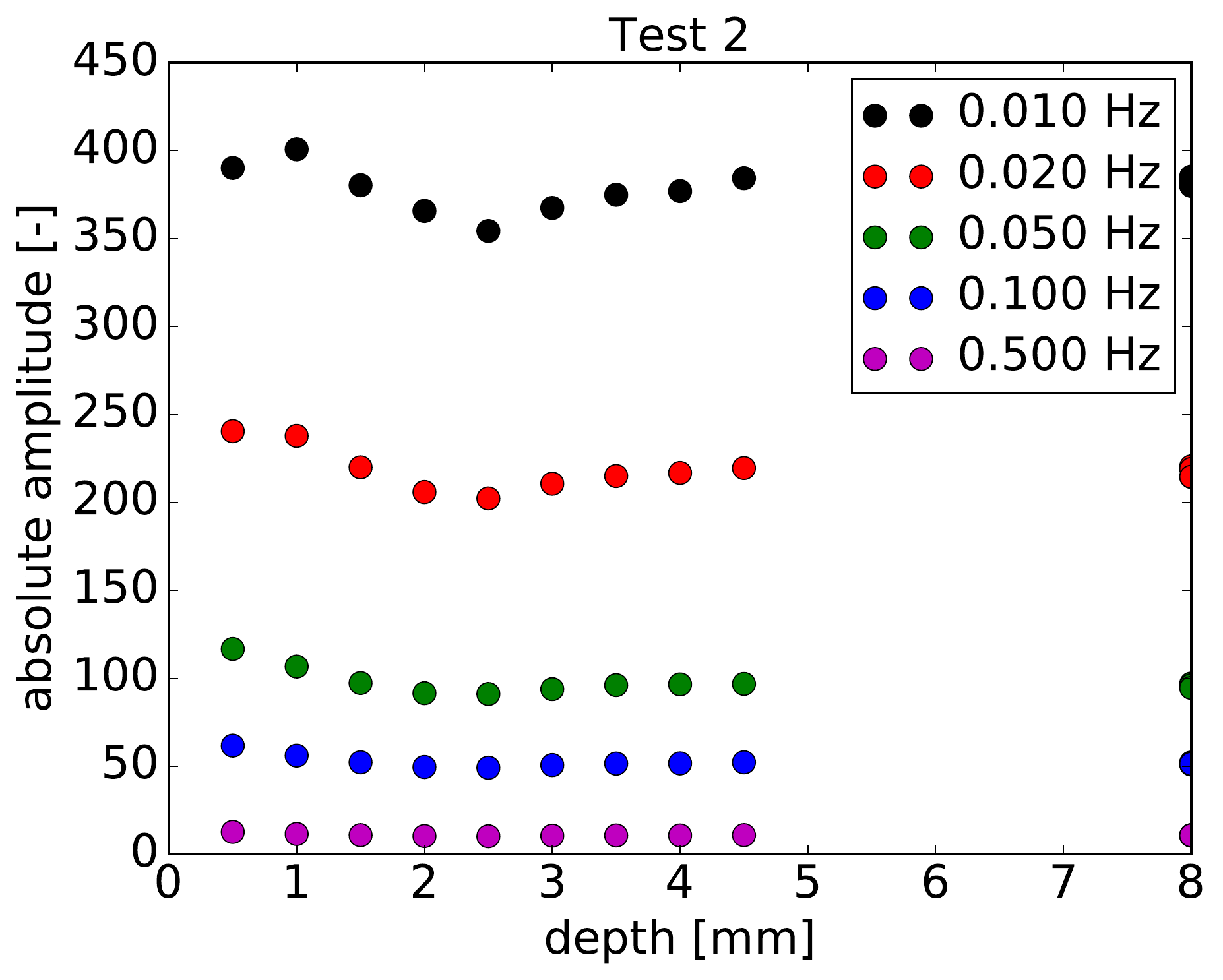}
\includegraphics[width=\figwidthhalf mm]{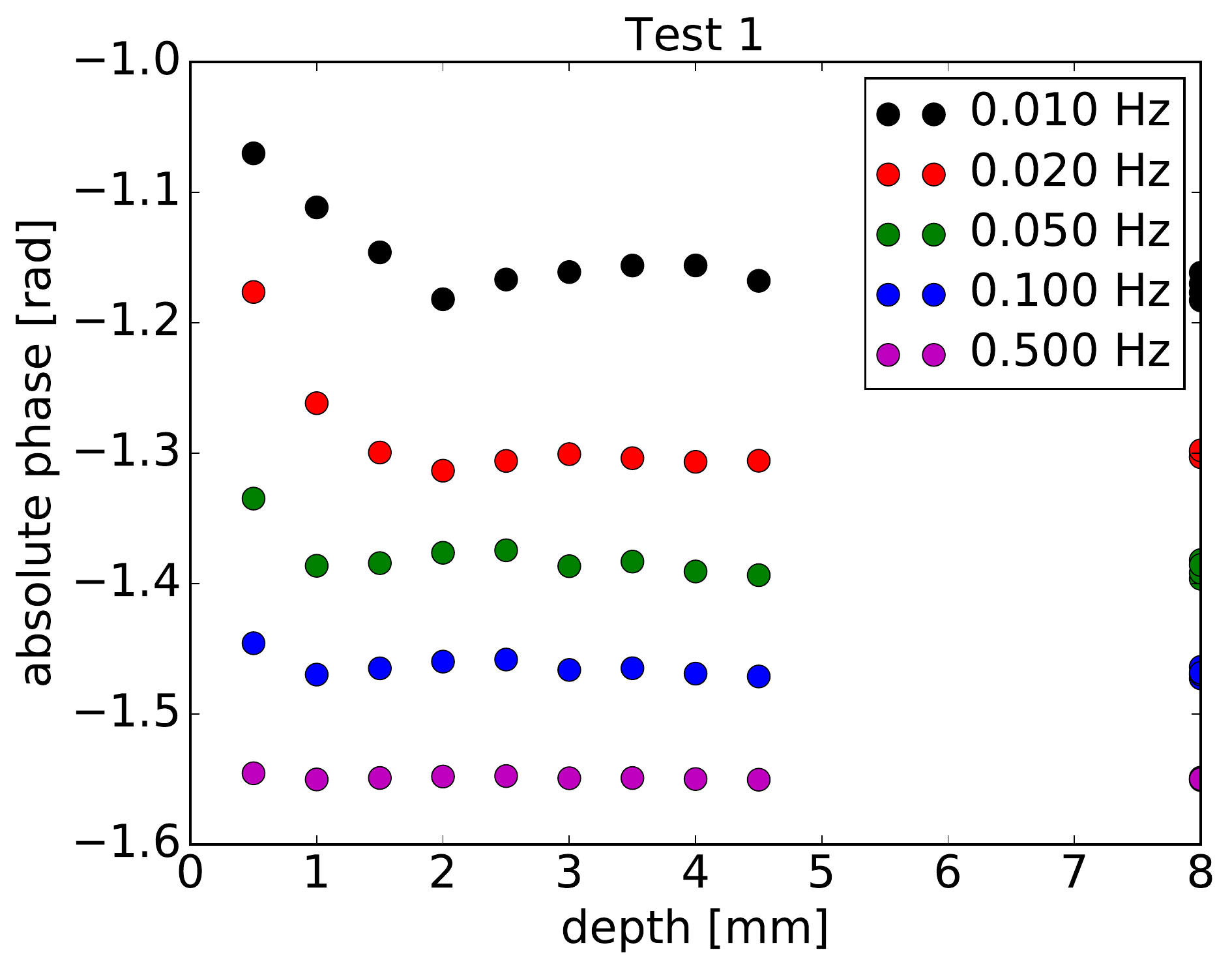}
\includegraphics[width=\figwidthhalf mm]{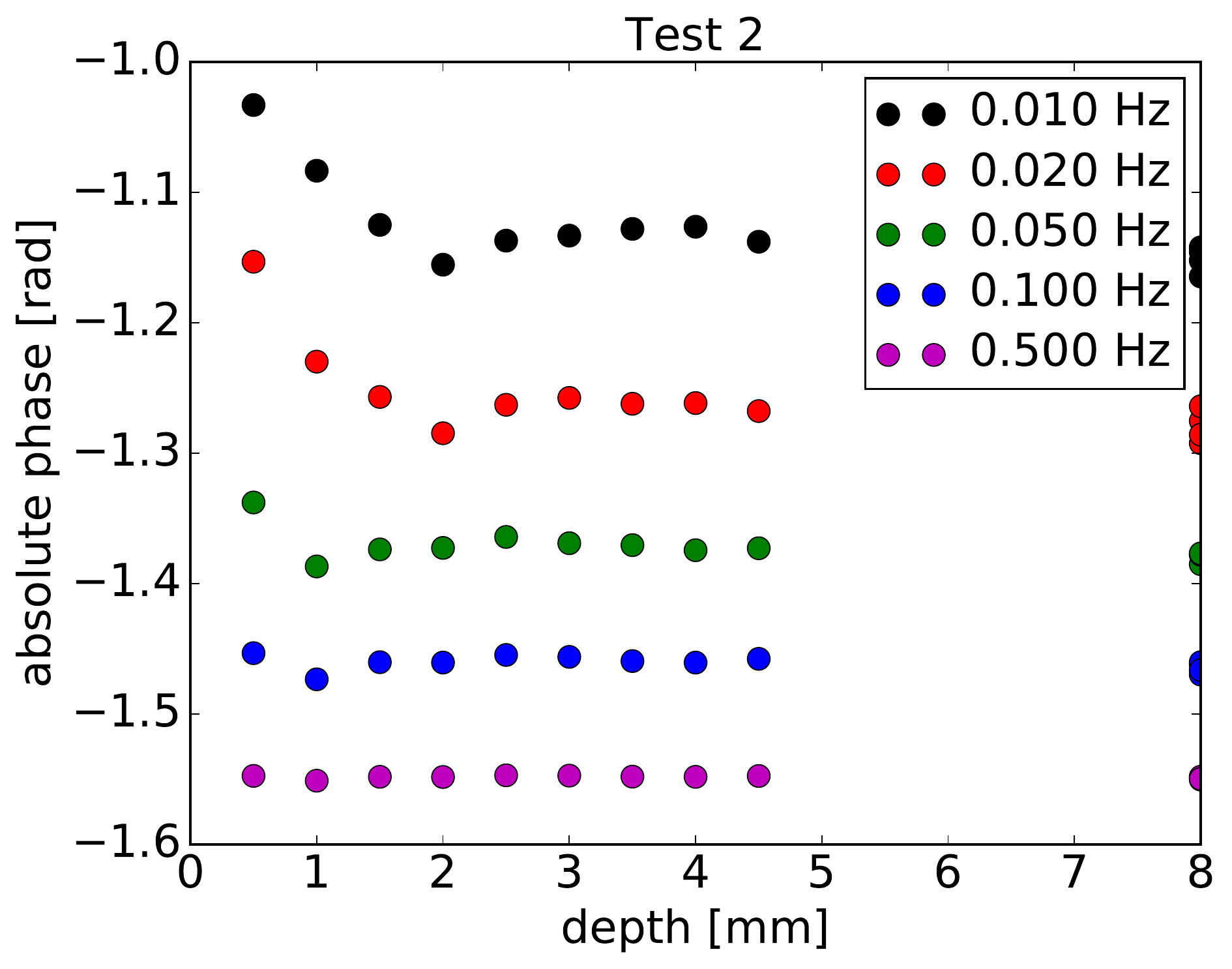}
\end{center}
\caption{Absolute contrast due to depth of defects, for values of amplitude and phase using CFT, at selected frequency values (temperature).
\label{fig:ppt_difffreqT}}
\end{figure}

In order to understand the relative contrast at different frequencies, selected frequencies were used, from 0.01 to 0.5~Hz. Results were obtained for the 13 points, and shown in Figs.~\ref{fig:ppt_difffreqI} and \ref{fig:ppt_difffreqT}, where amplitude and phase are shown against depth of defect (8~mm for sound areas). As expected, the phase contrast between defects at different depths were more prominent at lower frequencies, similar to the findings of Goh~\etal~\cite{Goh2018AX}. However, there is about 2 to 3 times better phase contrast obtained by using Lock-In Thermography, compared to PPT~\cite{Goh2018AX}, which is shown in these figures. The amplitude contrast is also more prominent at lower frequencies. It is also interesting to note that the absolute amplitude increased with decreasing frequency, which is an expected feature of Fourier Transform on decaying signals. The amplitude and phase results differ between CFT of intensity and temperature, as expected, because both values exist on different scales. Note that the values of amplitude and phase are also different between Tests 1 and 2. In order for reproducibility, the samples should reach thermal equilibrium before measurements commence. However, that may result in significant increase in measurement time.

\begin{figure} 
\begin{center}
\includegraphics[width=\figwidthhalf mm]{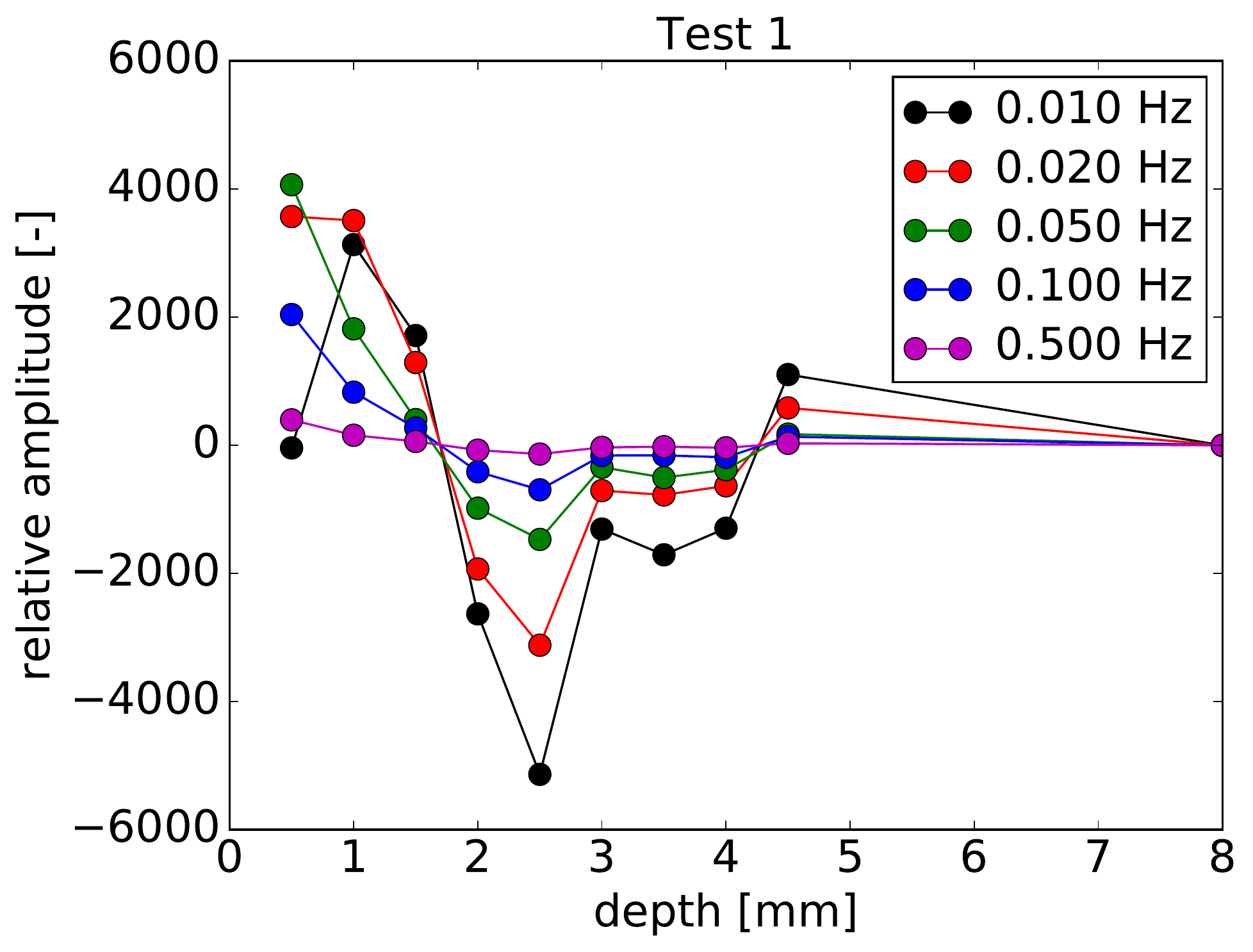}
\includegraphics[width=\figwidthhalf mm]{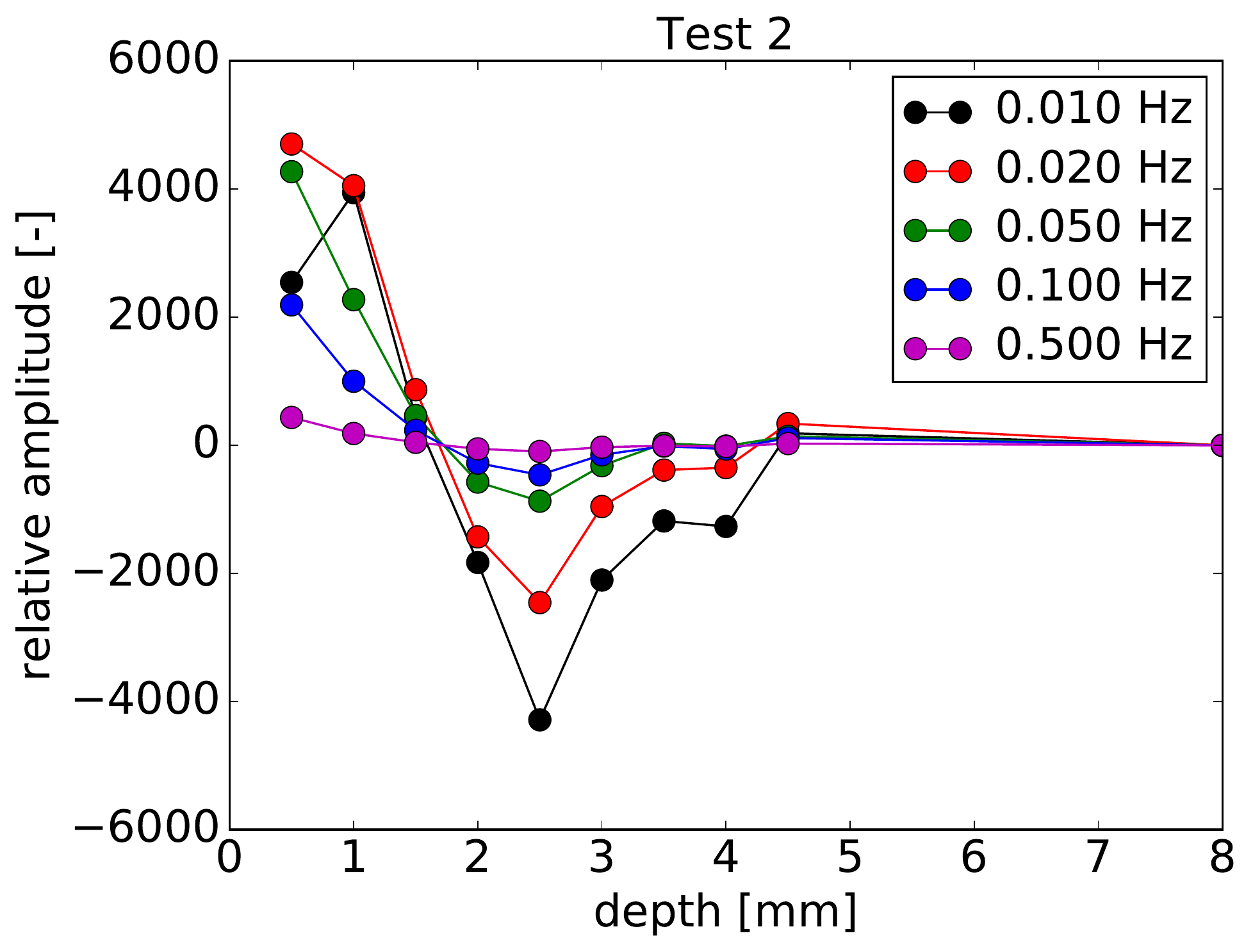}
\includegraphics[width=\figwidthhalf mm]{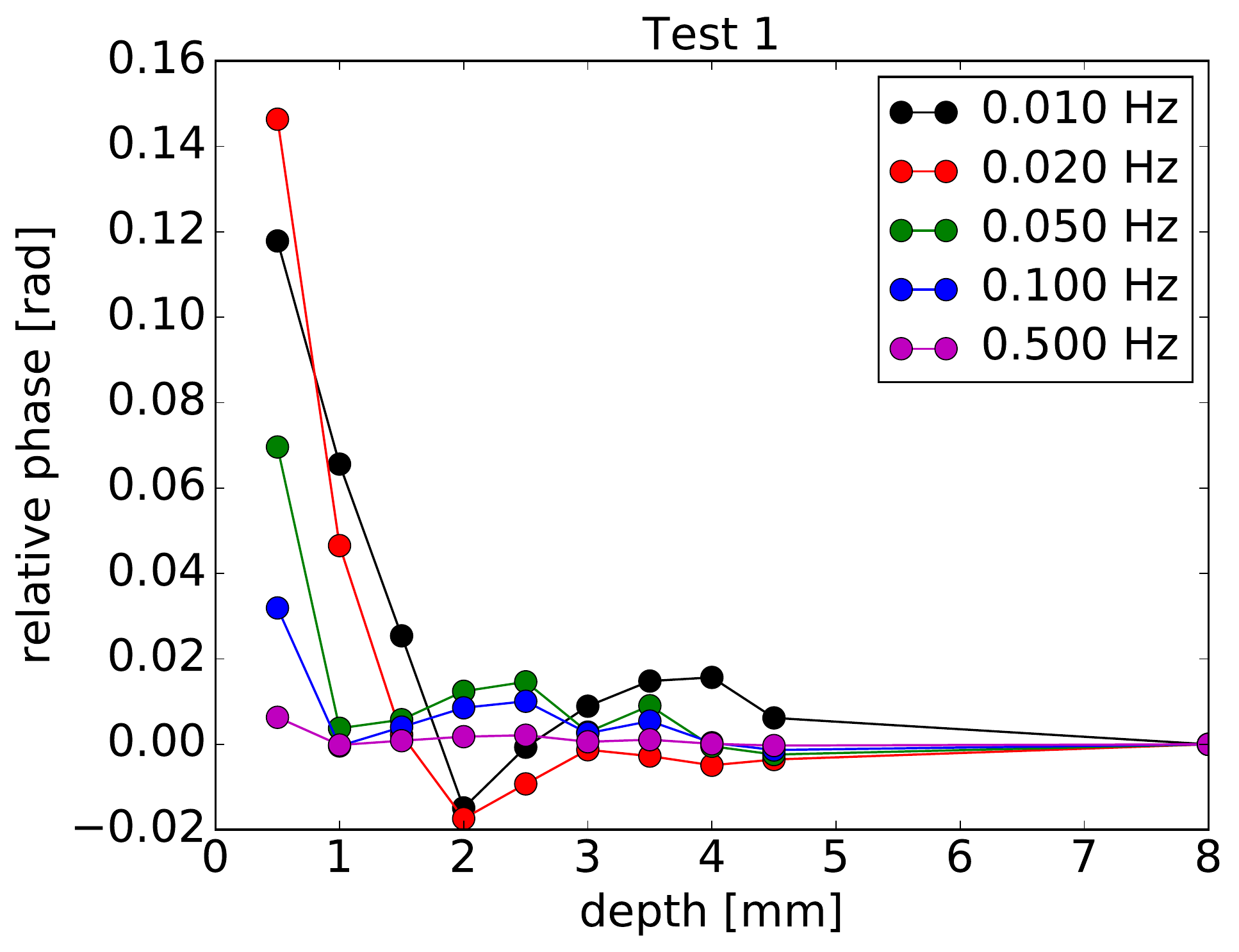}
\includegraphics[width=\figwidthhalf mm]{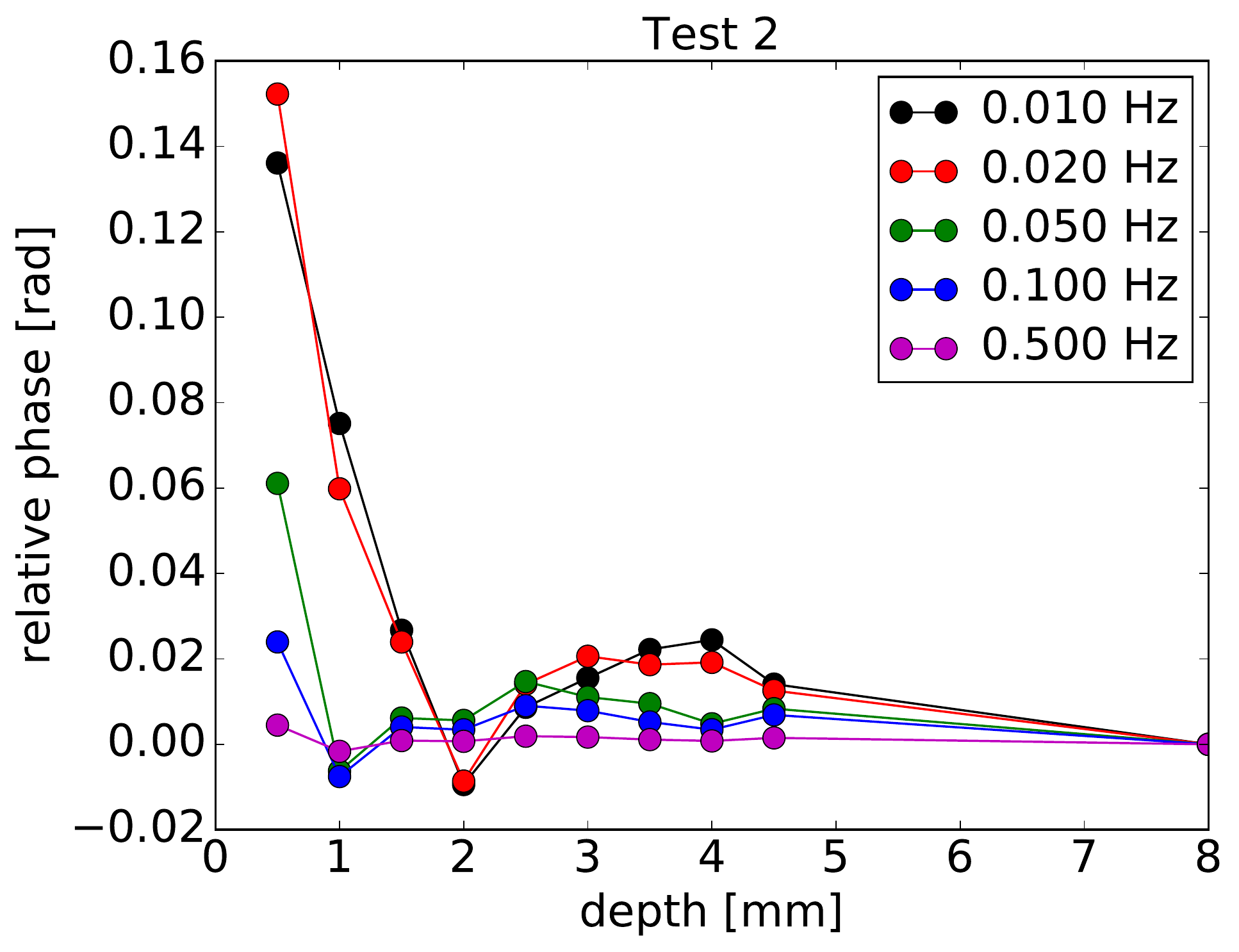}
\end{center}
\caption{Relative contrast due to depth of defects, for values of amplitude and phase using CFT, at selected frequency values (intensity).
\label{fig:ppt_difffreqIrel}}
\end{figure}

\begin{figure} 
\begin{center}
\includegraphics[width=\figwidthhalf mm]{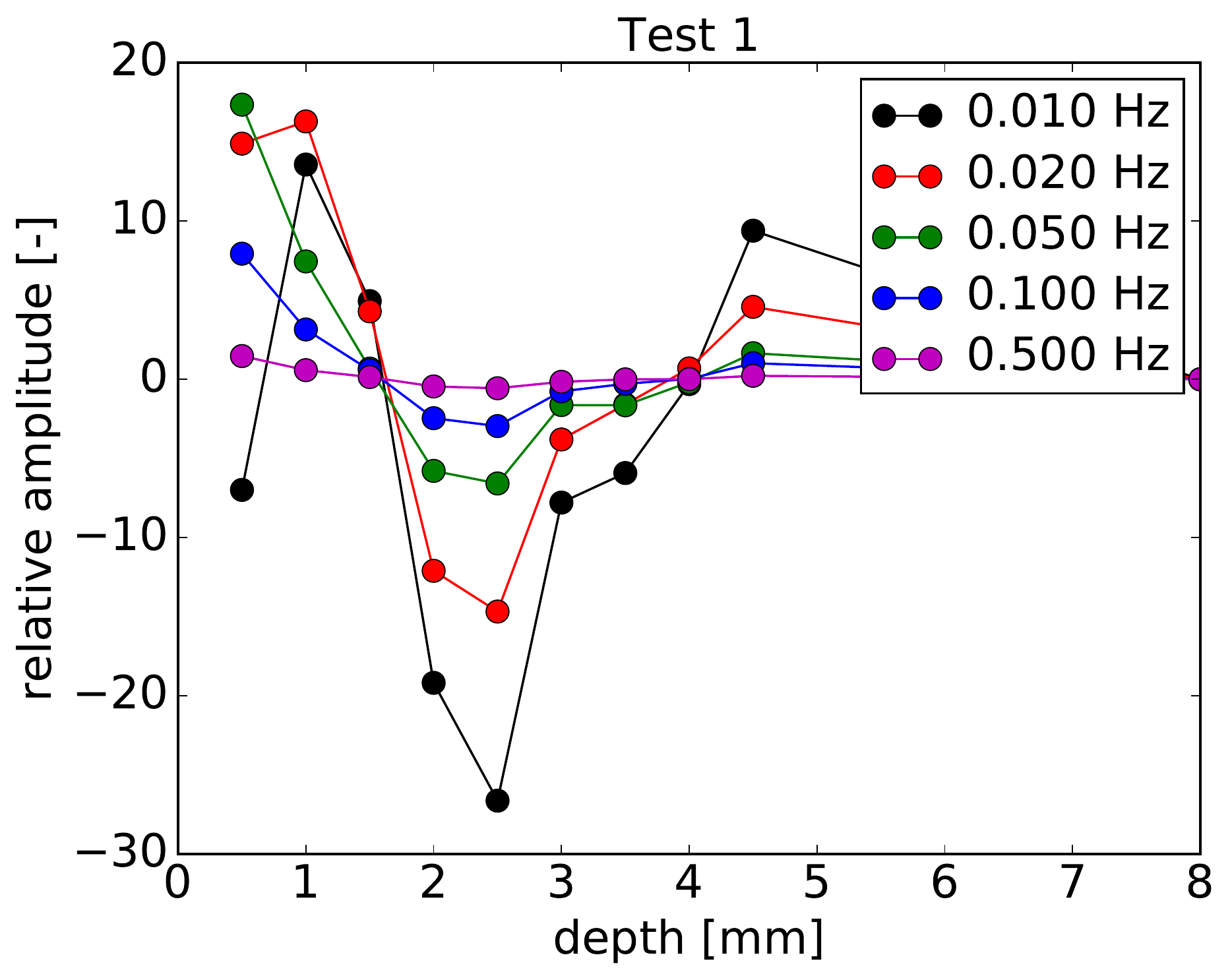}
\includegraphics[width=\figwidthhalf mm]{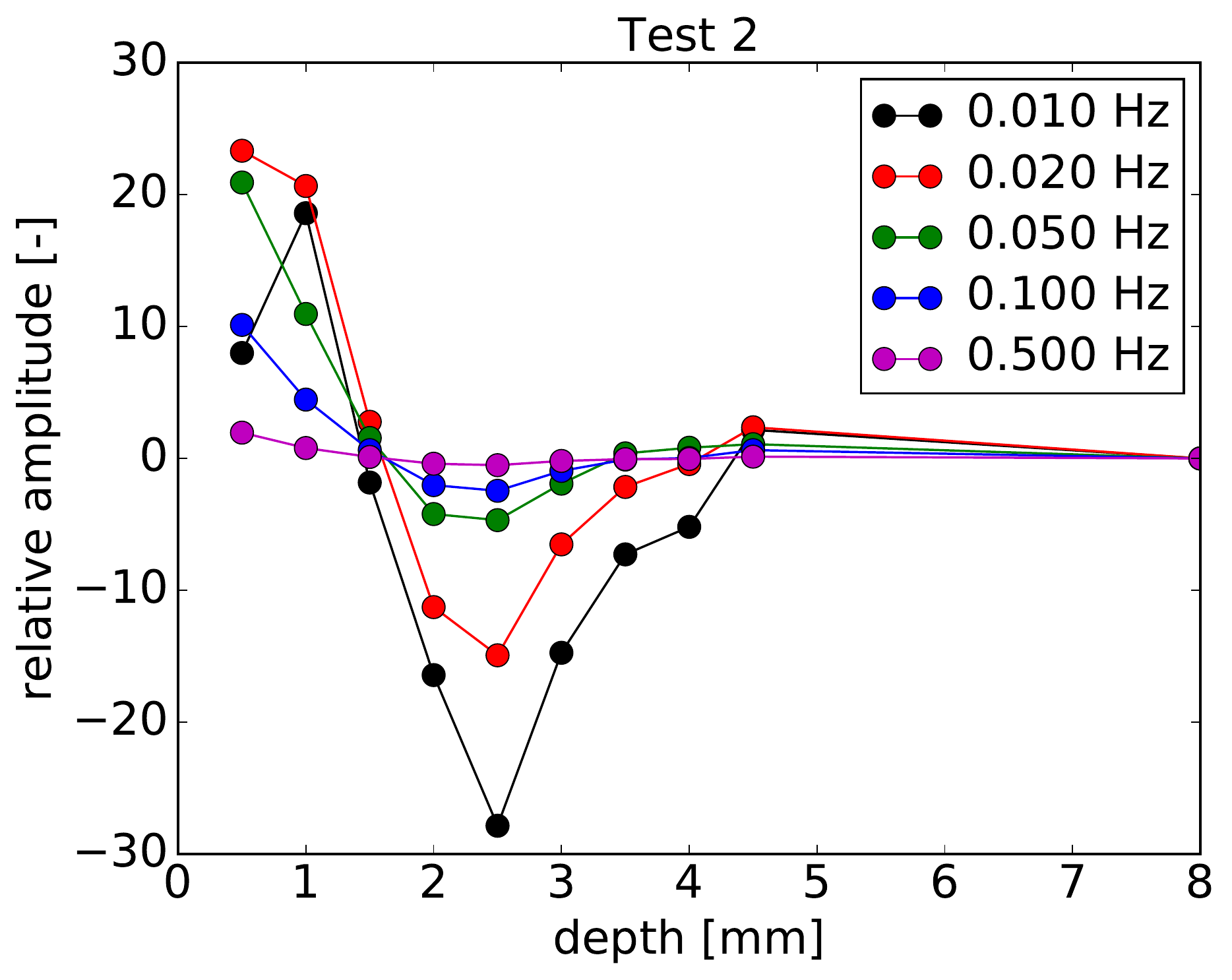}
\includegraphics[width=\figwidthhalf mm]{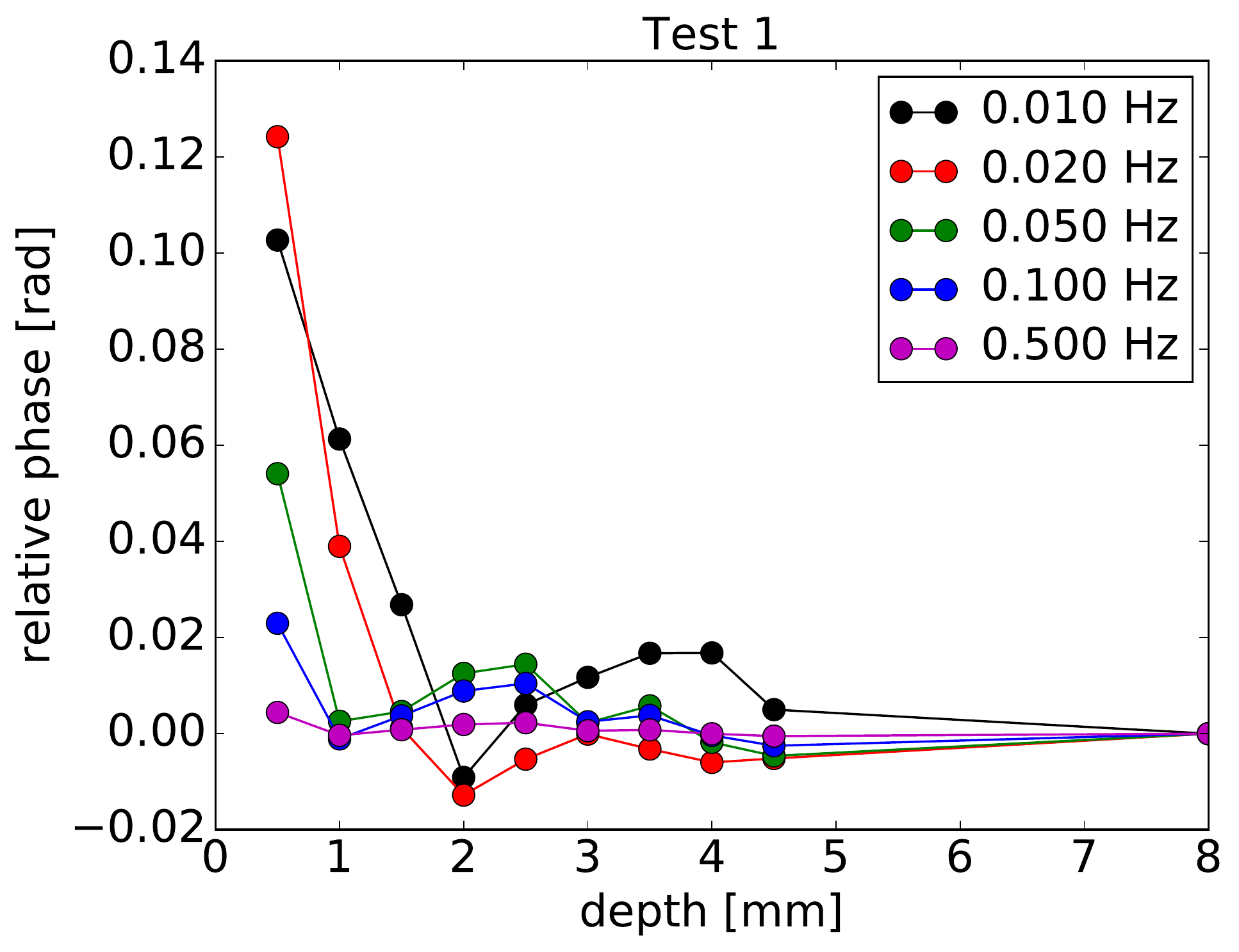}
\includegraphics[width=\figwidthhalf mm]{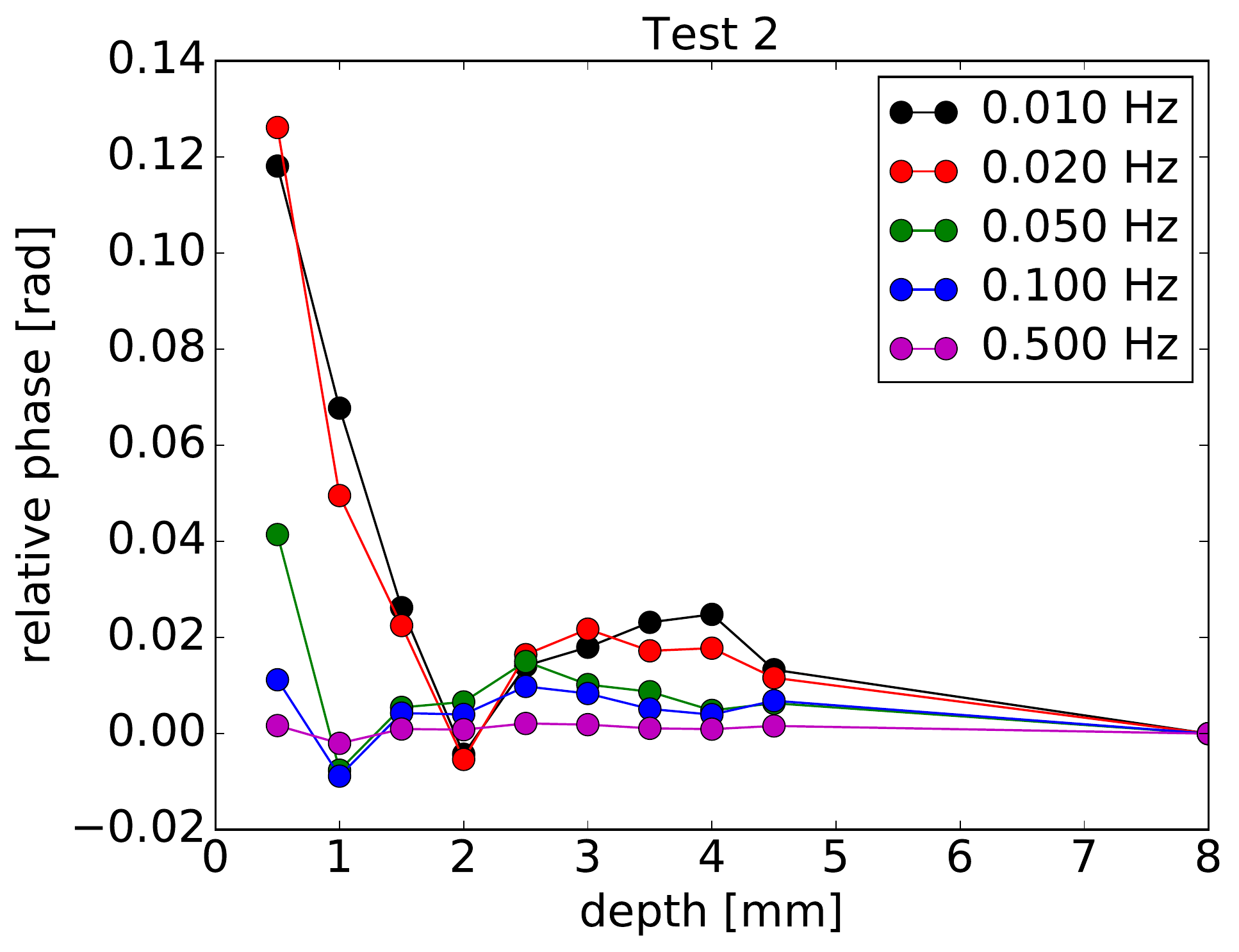}
\end{center}
\caption{Relative contrast due to depth of defects, for values of amplitude and phase using CFT, at selected frequency values (temperature).
\label{fig:ppt_difffreqTrel}}
\end{figure}

In order to visualise the relative effects of frequency on amplitude and phase contrast at different depths, the results were shifted such that the mean value of the 4 sound points was zero. These results are shown in Figs.~\ref{fig:ppt_difffreqIrel} and \ref{fig:ppt_difffreqTrel} for intensity and temperature respectively. It is clearly shown that lower frequencies result in better contrast, for both amplitude and phase. However, there are turning points shown in these results, similar to the findings of Goh~\etal~\cite{Goh2018AX} as well as Wong and Goh~\cite{Wong2018AX} for phase contrast. Hence, it is a challenging task to measure the thickness of samples using Thermography, as shown by Wong and Goh~\cite{Wong2018AX}, where the thickness could only be reasonably measured up to 1.8~mm. Recent breakthroughs by Tian~\etal~\cite{Tian2019AX} show that it is possible to measure up to 8.4~mm thickness in ABS, by using multidimensional data. The success rates with a tolerance of 0.4~mm were at least 90\% for samples 4.4~mm thick, even with the difference in cooling at the edges, as well as non-uniform heating of the samples. By comparing the relative contrast for amplitude and phase, for both intensity and temperature, it is obvious that Tests 1 and 2 produce different results, which is a clear indication of dependence on heating time. This is especially prominent at lower frequencies. While it may seem a monumental task to move PPT towards a common standard, there is probably sufficient information from different research groups to create a better understanding of the data. While it would require a concerted effort from researchers to do this, the standardisation of data into a large collective database will move the research forward in the right direction.

\begin{figure} 
\begin{center}
\includegraphics[width=\figwidthhalf mm]{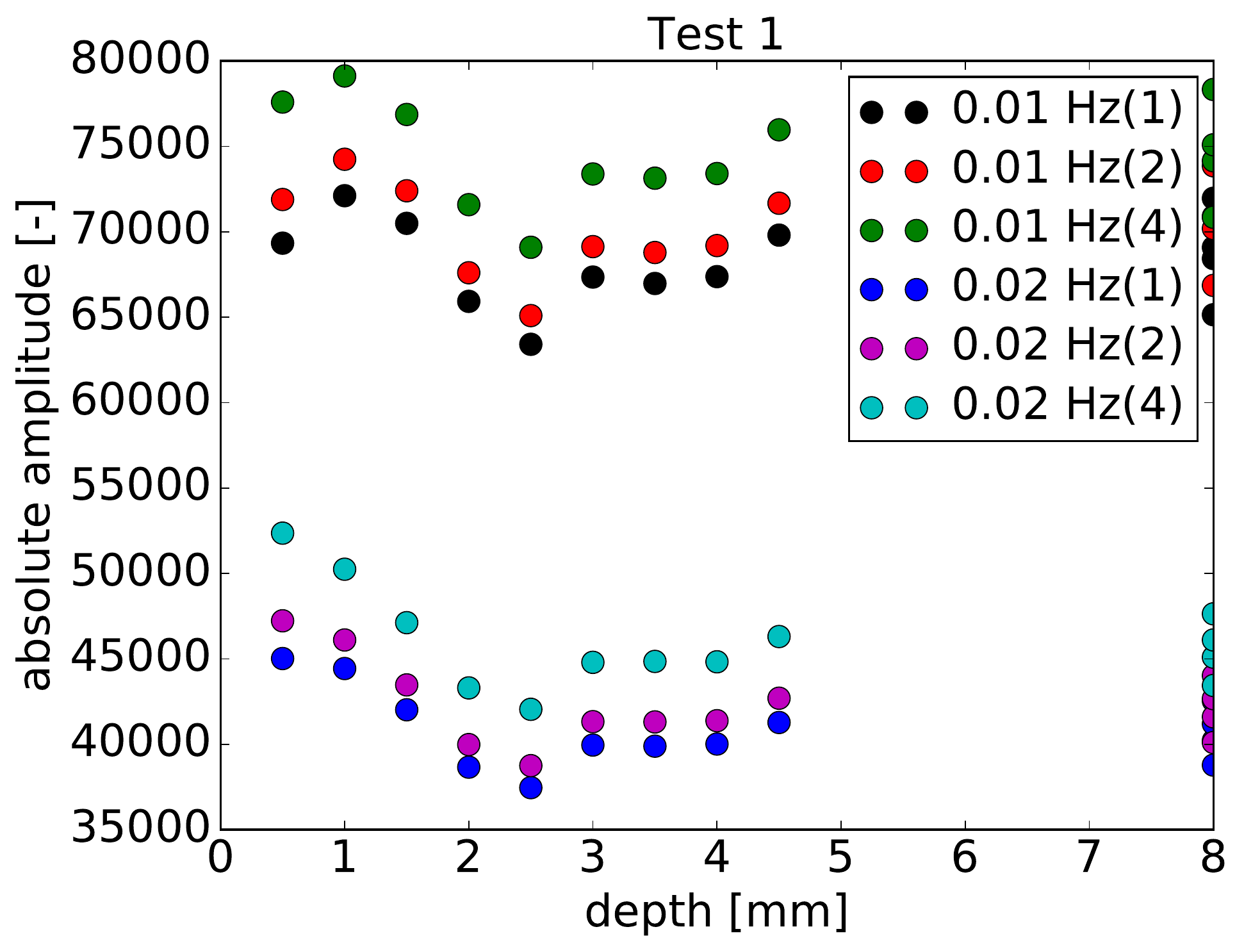}
\includegraphics[width=\figwidthhalf mm]{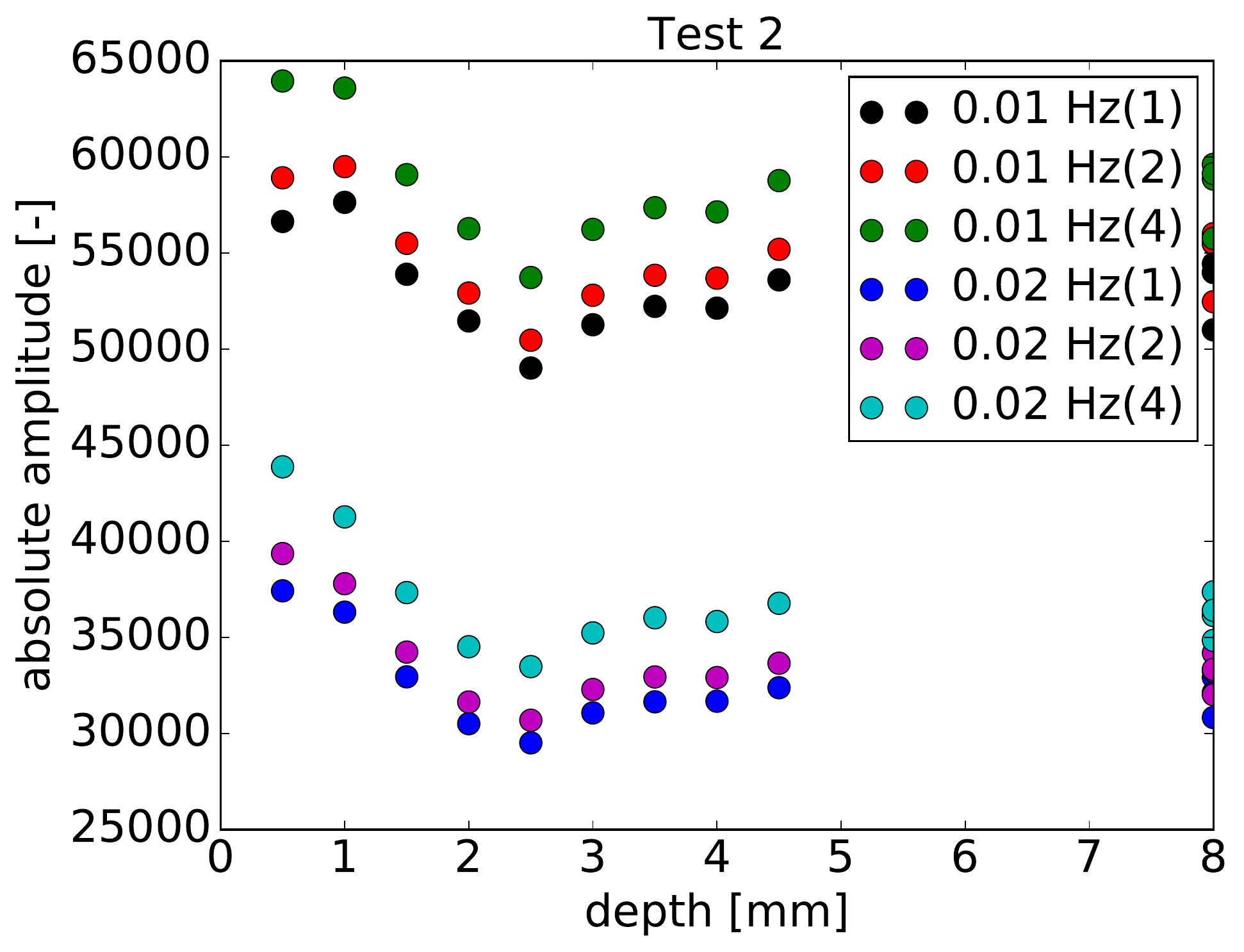}
\includegraphics[width=\figwidthhalf mm]{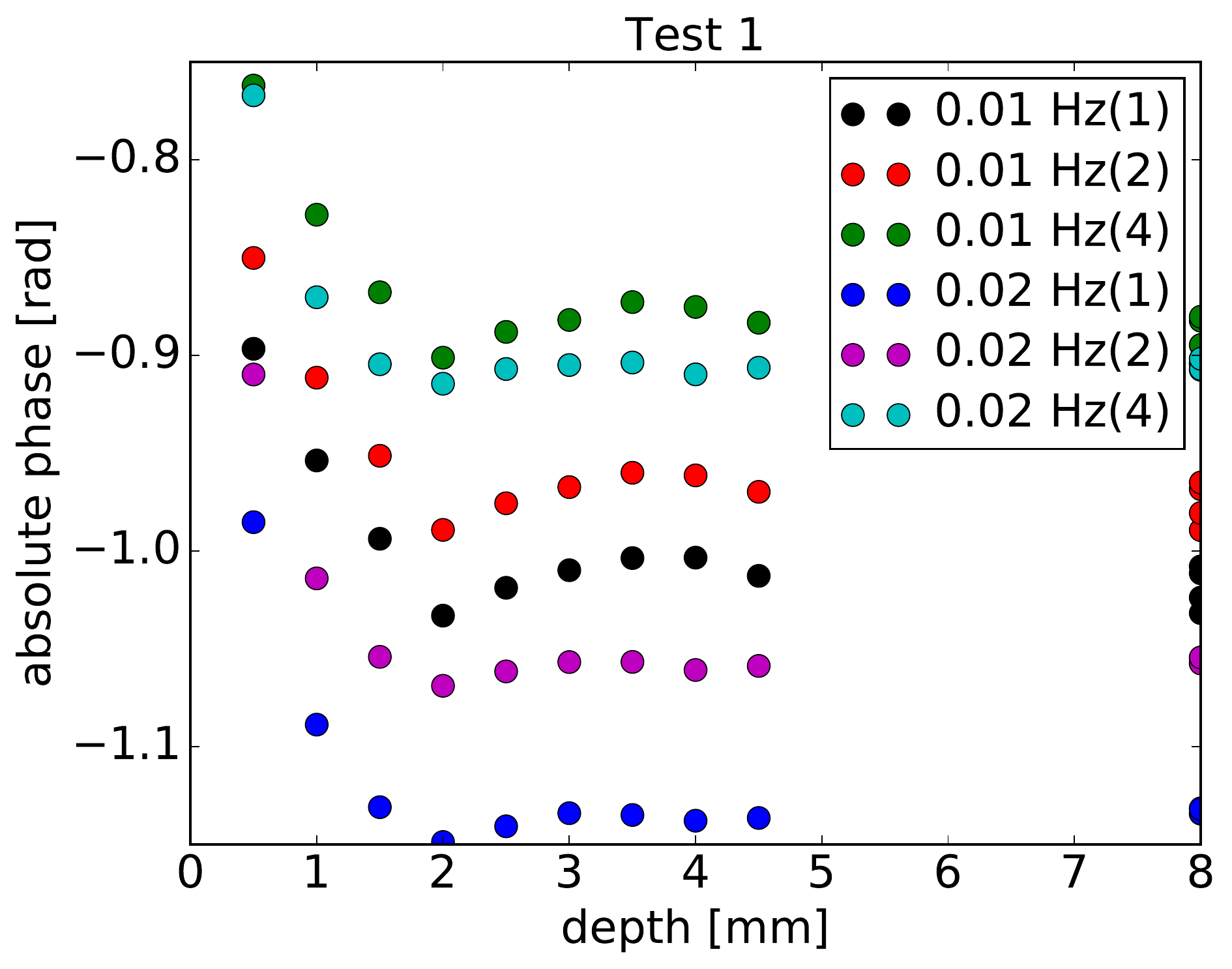}
\includegraphics[width=\figwidthhalf mm]{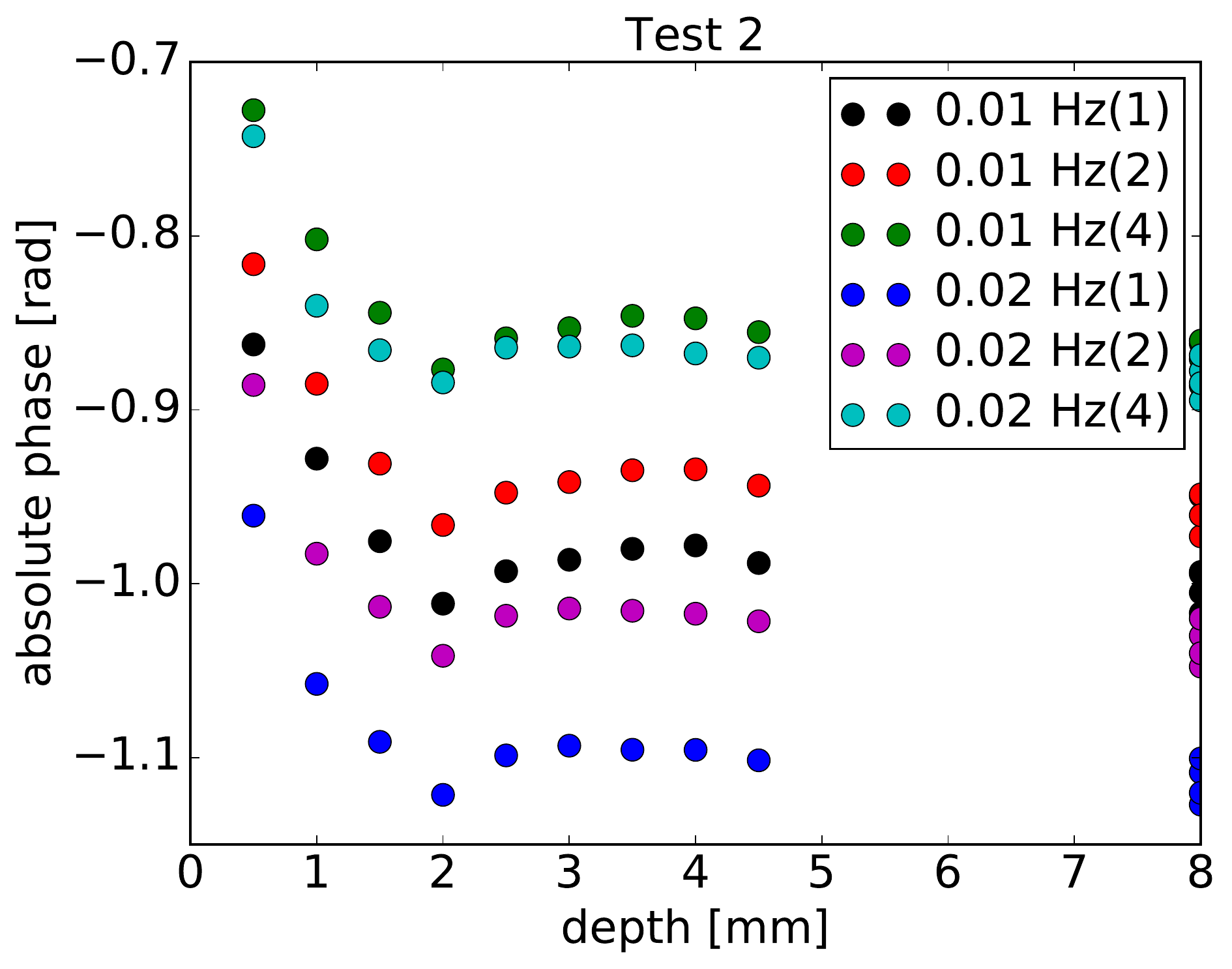}
\end{center}
\caption{Absolute contrast due to depth of defects, for values of amplitude and phase using FFT, at selected frequency values with subsampling of every 2 and 4 frames (intensity).
\label{fig:ppt_difffreqsubsI}}
\end{figure}

\begin{figure} 
\begin{center}
\includegraphics[width=\figwidthhalf mm]{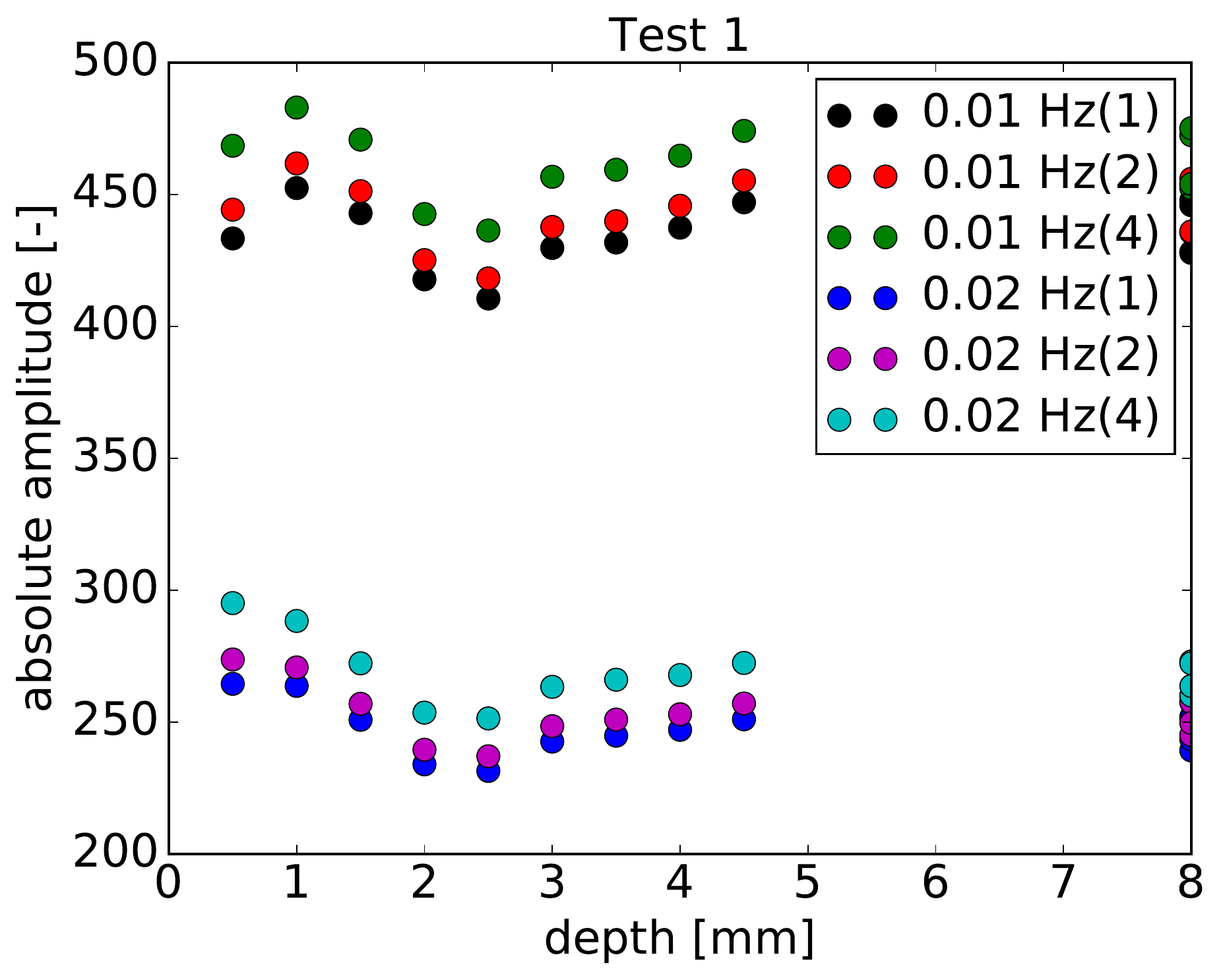}
\includegraphics[width=\figwidthhalf mm]{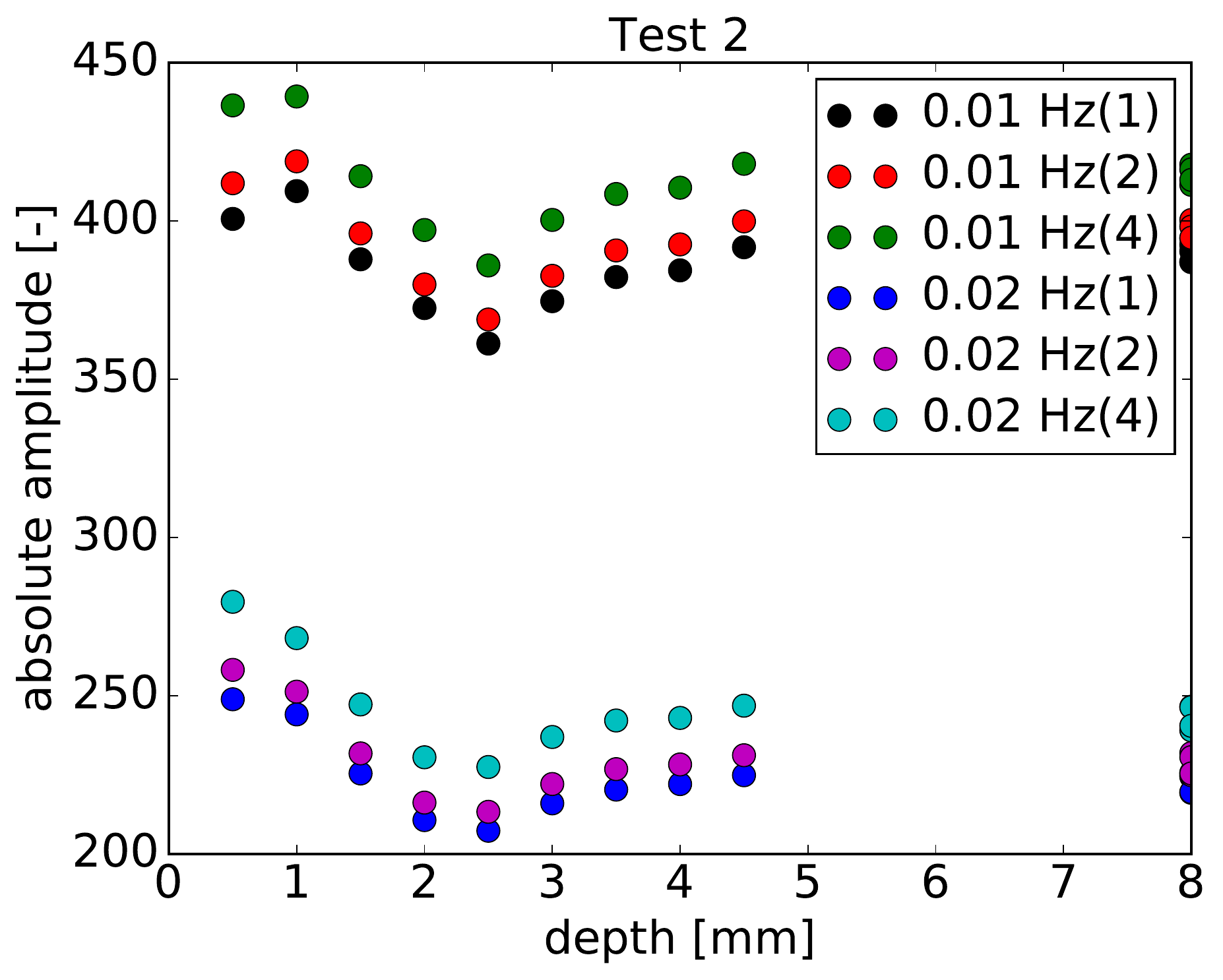}
\includegraphics[width=\figwidthhalf mm]{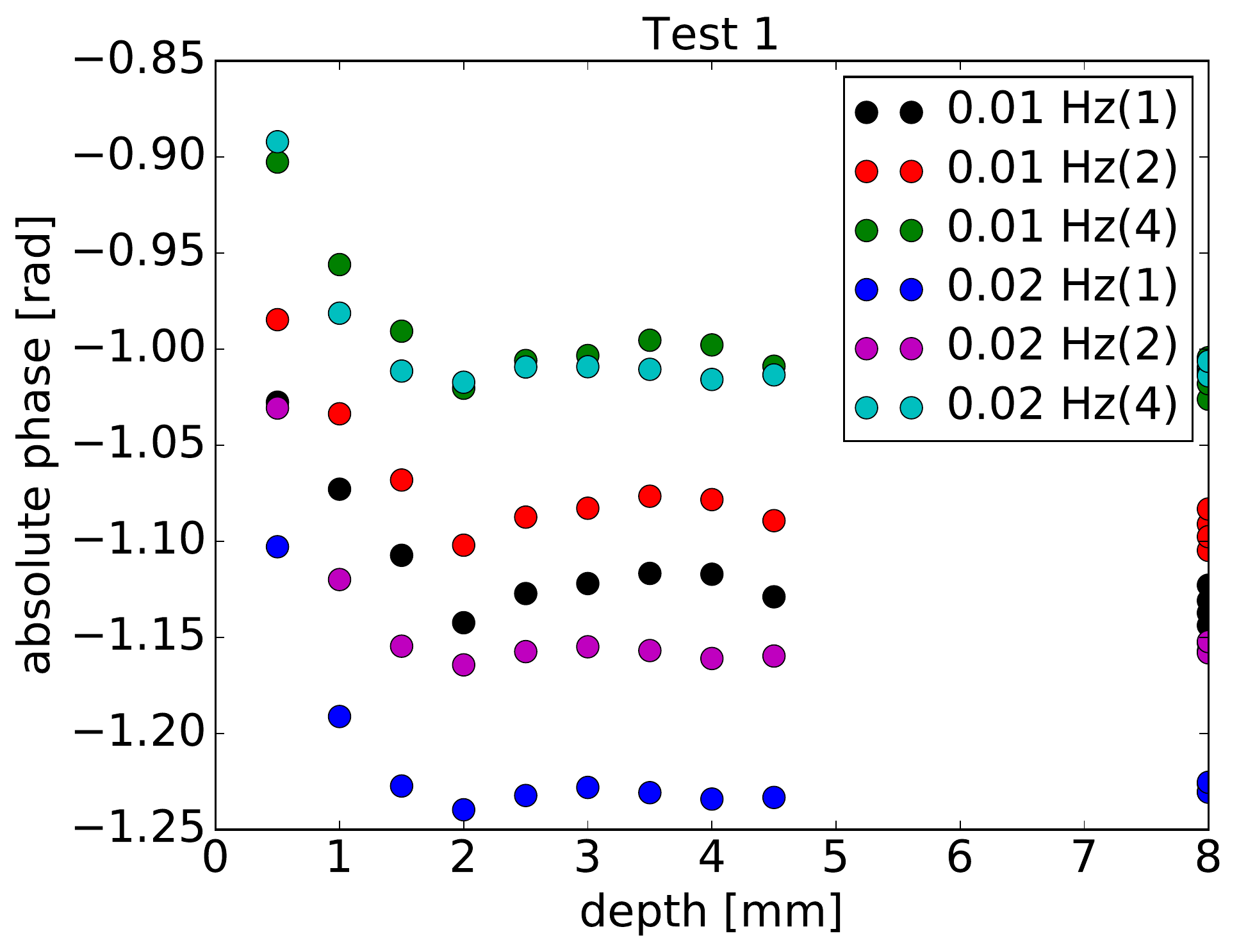}
\includegraphics[width=\figwidthhalf mm]{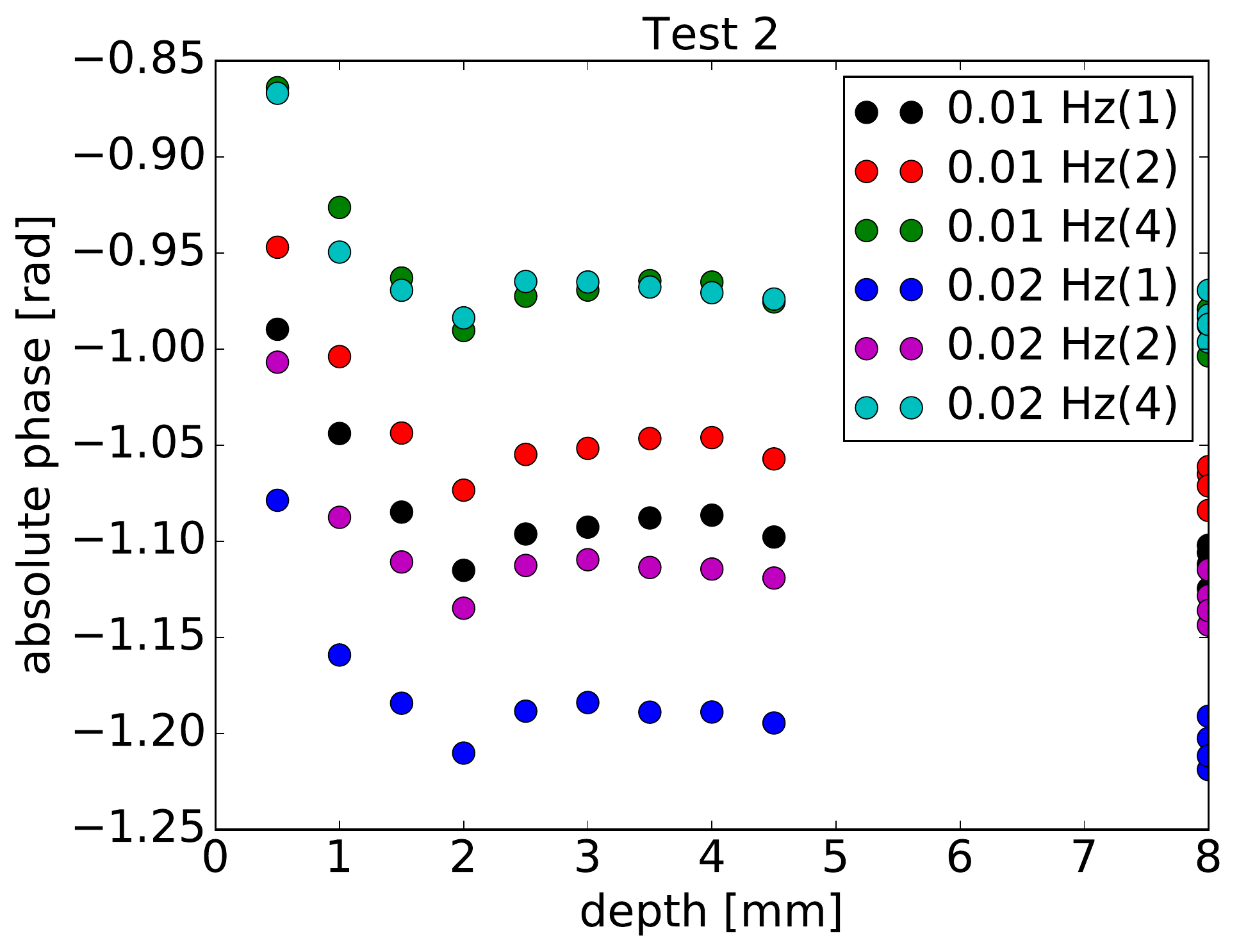}
\end{center}
\caption{Absolute contrast due to depth of defects, for values of amplitude and phase using FFT, at selected frequency values with subsampling of every 2 and 4 frames (temperature).
\label{fig:ppt_difffreqsubsT}}
\end{figure}

For PPT using FFT, it was established in the Introduction section that the effects of subsampling were prominent. Hence, it is of interest to understand the relative effects of subsampling on the amplitude and phase contrast, using FFT on selected frequencies of 0.01 and 0.02~Hz. Sample points were taken for the full 100 seconds for Tests 1 and 2, with full 100 points, as well as sample points taken every 2 and 4 points. These results are shown in Figs.~\ref{fig:ppt_difffreqsubsI} and \ref{fig:ppt_difffreqsubsT}. Both the amplitude and phase values shifted upwards (increased) as fewer points were taken. This is similar to the findings of Goh~\etal~\cite{Goh2018AX}.

\begin{figure} 
\begin{center}
\includegraphics[width=\figwidthhalf mm]{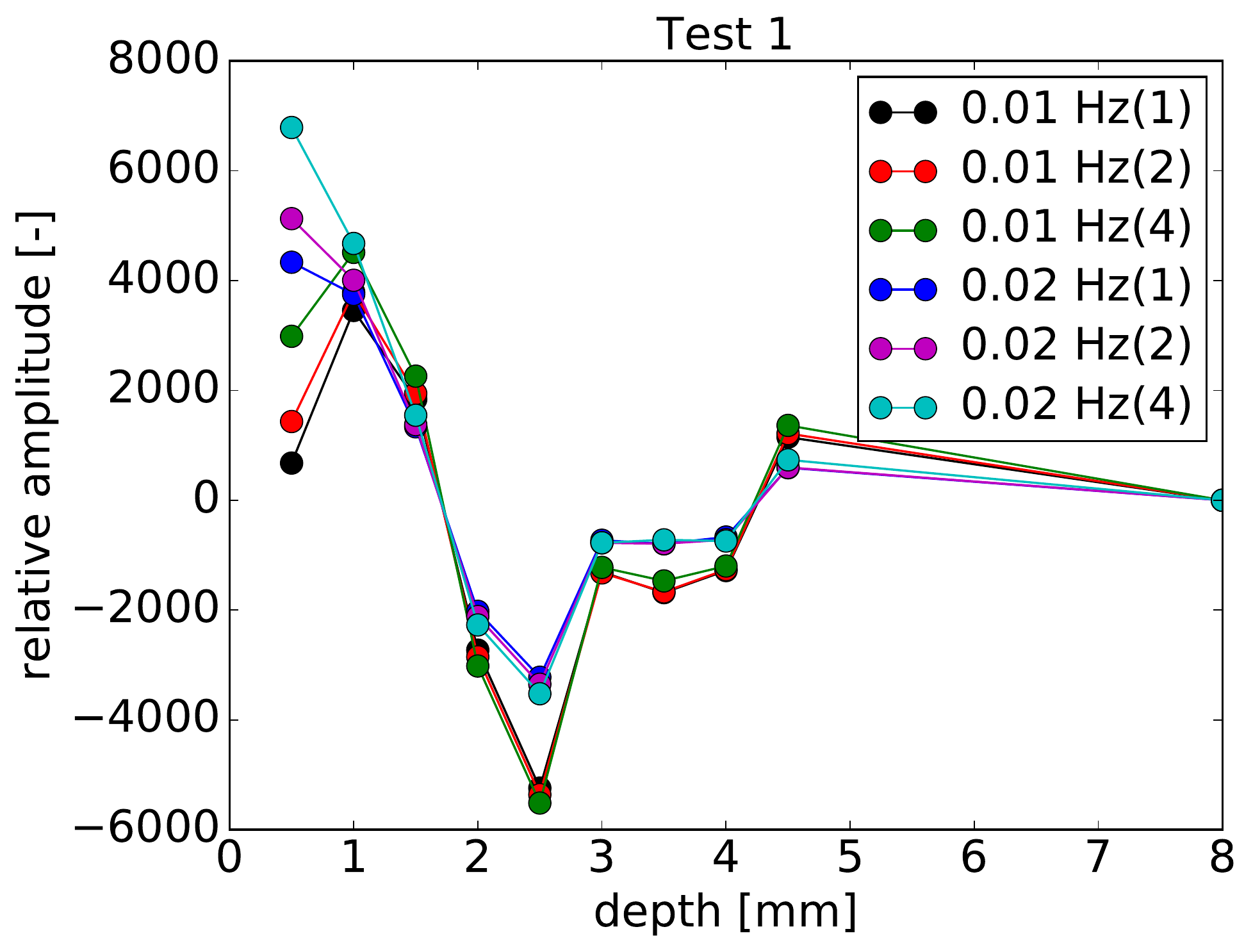}
\includegraphics[width=\figwidthhalf mm]{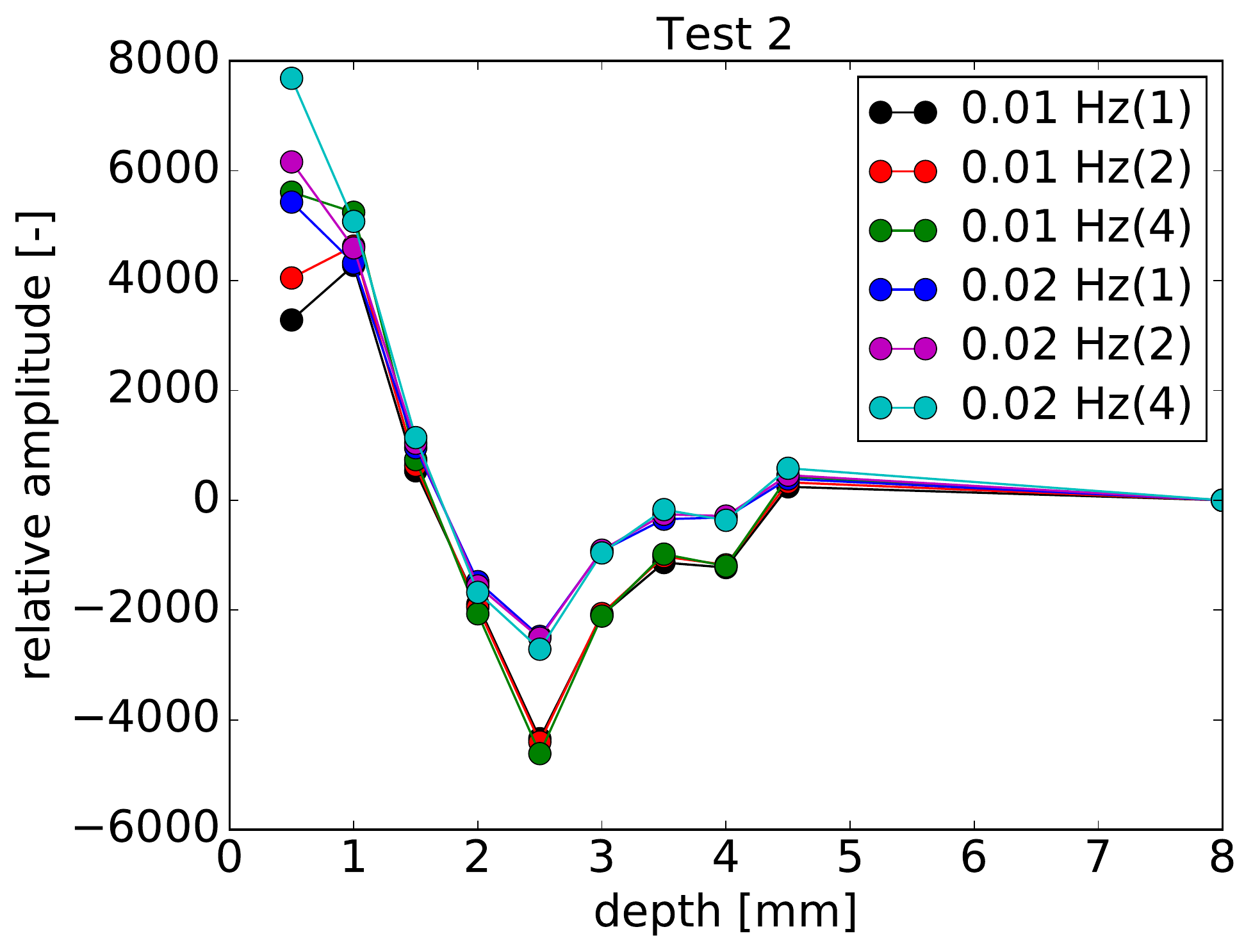}
\includegraphics[width=\figwidthhalf mm]{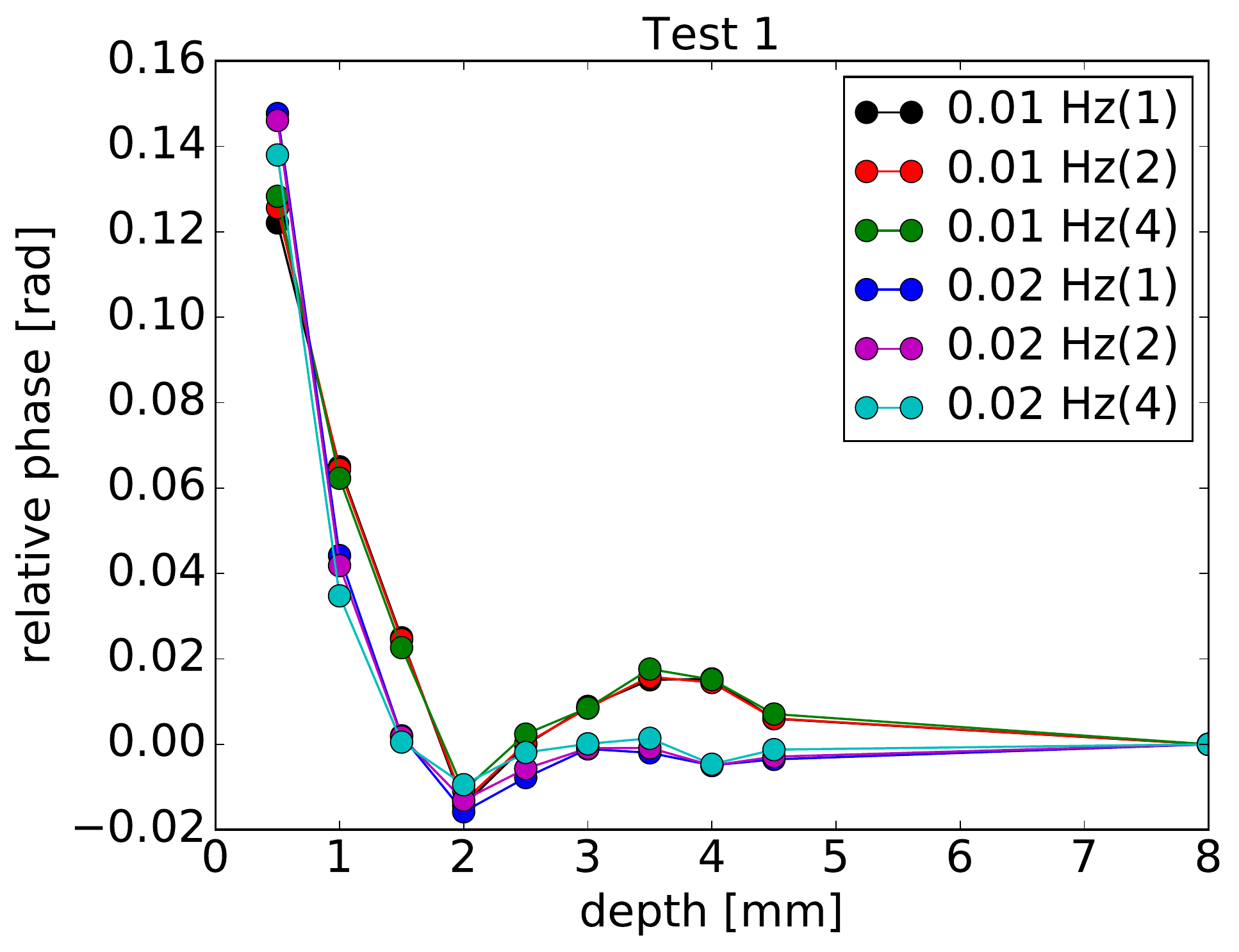}
\includegraphics[width=\figwidthhalf mm]{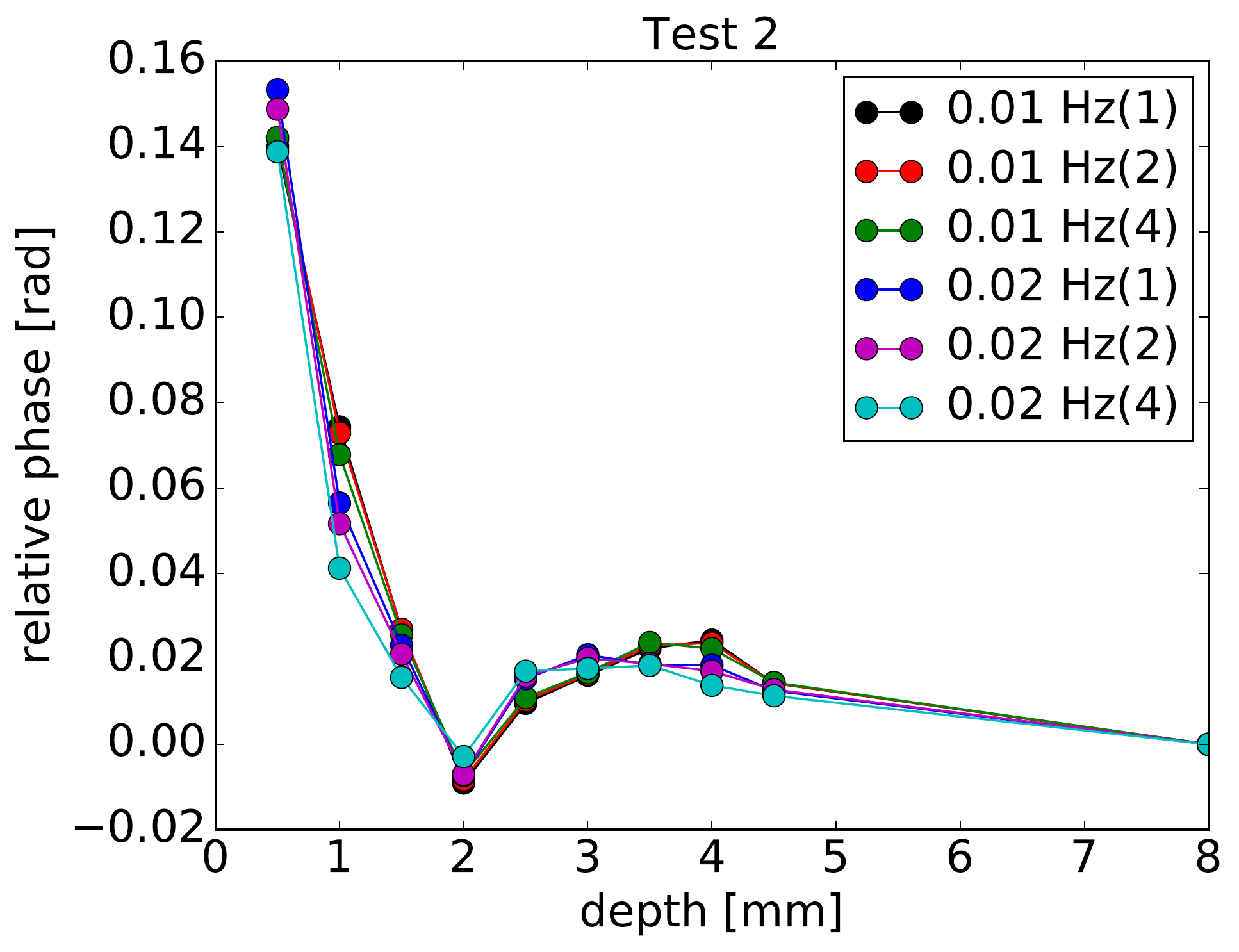}
\end{center}
\caption{Relative contrast due to depth of defects, for values of amplitude and phase using FFT, at selected frequency values with subsampling of every 2 and 4 frames (intensity).
\label{fig:ppt_difffreqsubsIrel}}
\end{figure}

\begin{figure} 
\begin{center}
\includegraphics[width=\figwidthhalf mm]{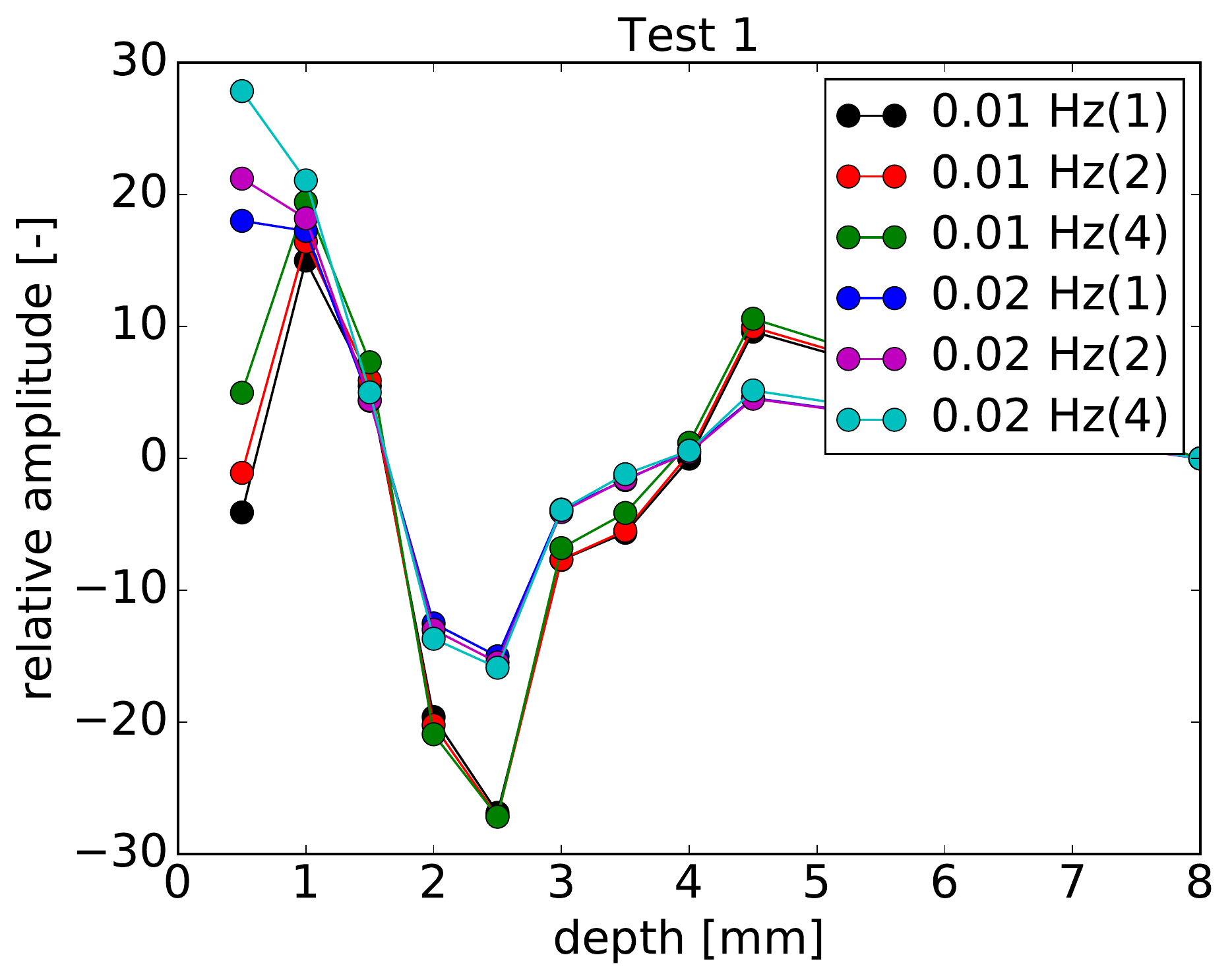}
\includegraphics[width=\figwidthhalf mm]{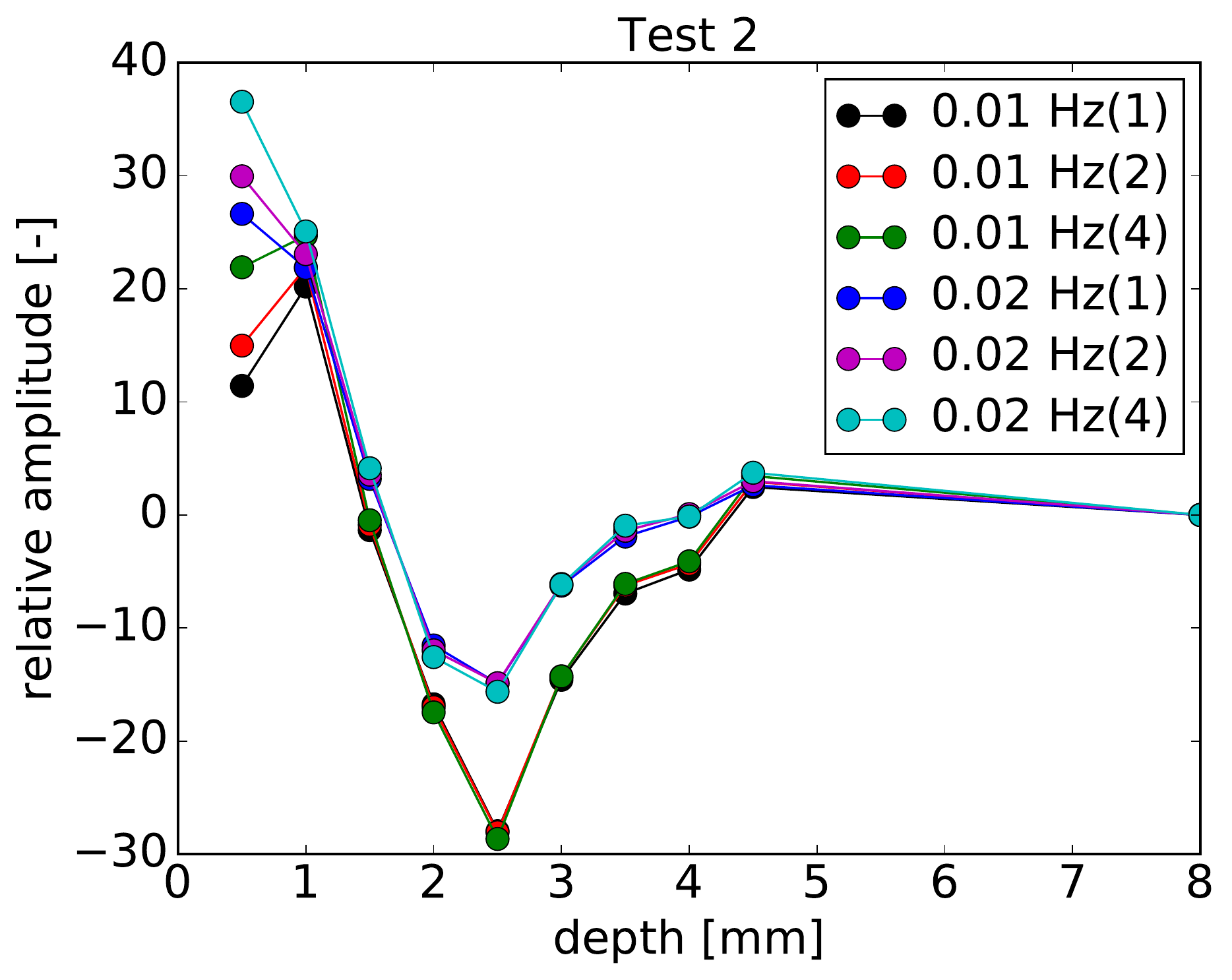}
\includegraphics[width=\figwidthhalf mm]{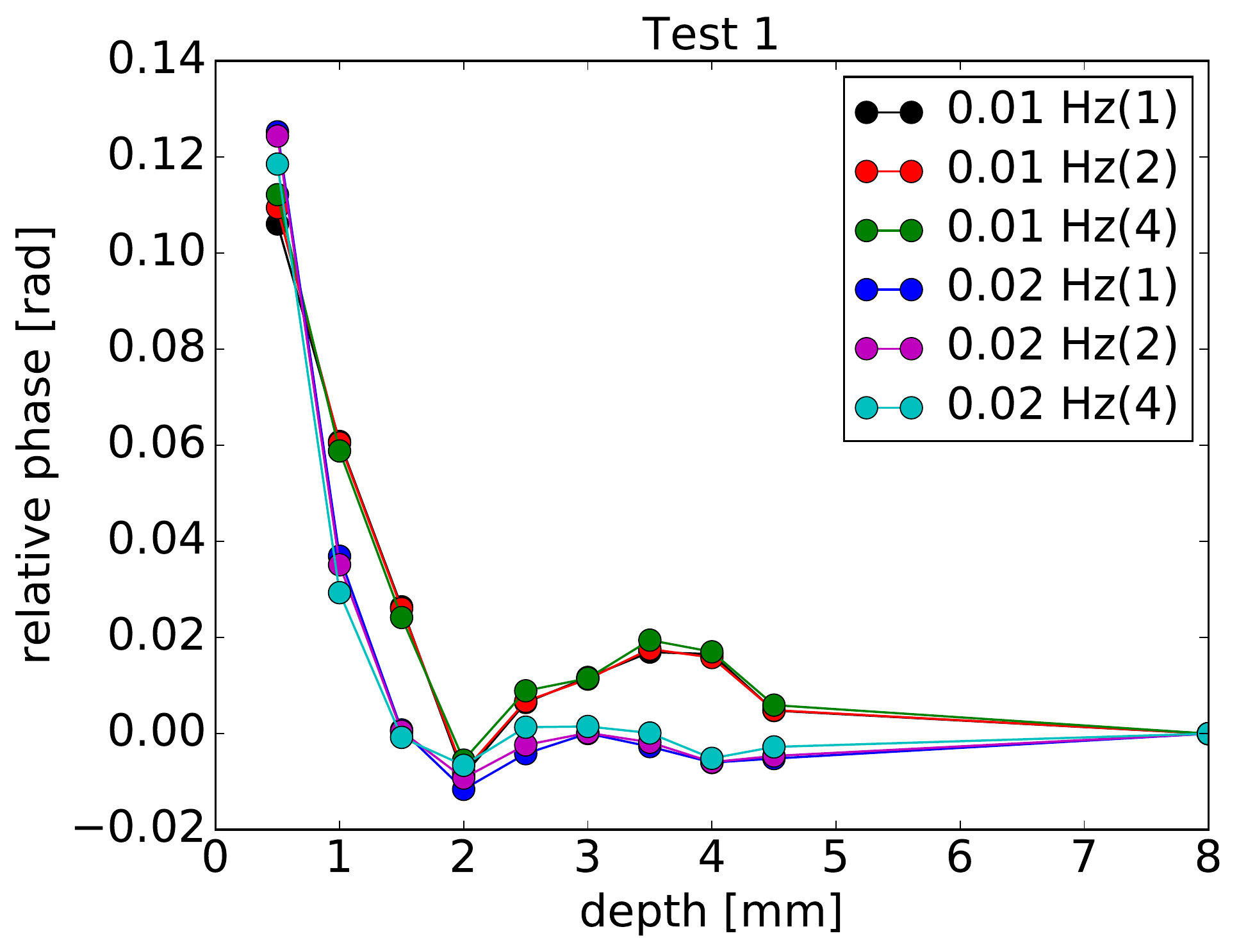}
\includegraphics[width=\figwidthhalf mm]{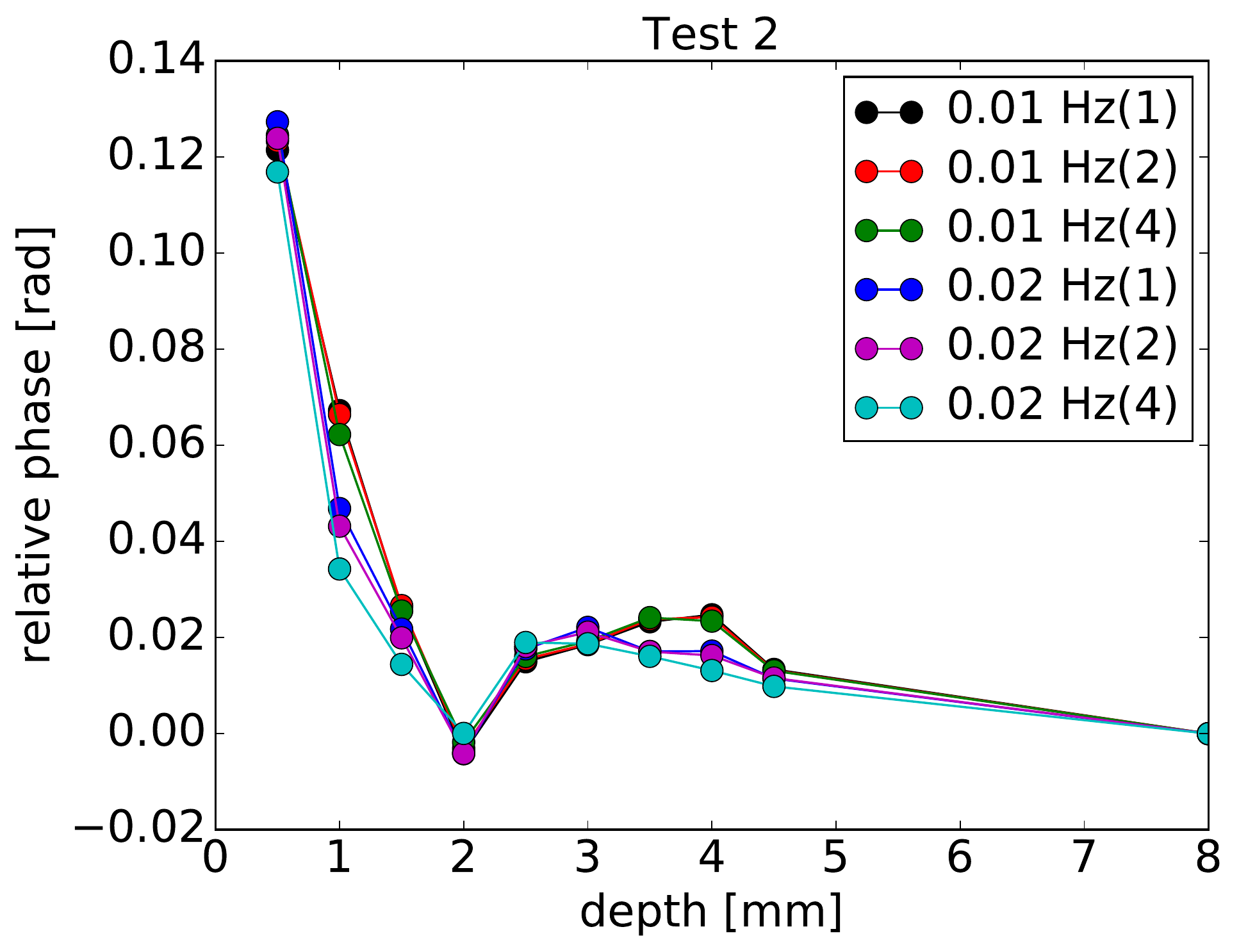}
\end{center}
\caption{Relative contrast due to depth of defects, for values of amplitude and phase using FFT, at selected frequency values with subsampling of every 2 and 4 frames (temperature).
\label{fig:ppt_difffreqsubsTrel}}
\end{figure}

In order to understand the effects of subsampling on the relative amplitude and phase contrast, these results were shifted such that the mean value of the 4 sound points was zero. The corresponding results are shown in Figs.~\ref{fig:ppt_difffreqsubsIrel} and \ref{fig:ppt_difffreqsubsTrel}. For relative amplitude, the contrast was similar despite the subsampling, except for the defect at 0.5~mm depth. For phase, the relative contrast was similar despite the subsampling, except for 0.02~Hz, where there was a clear effect of subsampling up to defect depths of 1.5~mm. These results for relative phase contrast at 0.01~Hz are similar to that of Goh~\etal~\cite{Goh2018AX}, where it was shown that the minimum number of points for the relative contrast to remain similar was around 10 points per cycle, equivalent to subsampling every 10 points in the Tests 1 and 2. The findings of Goh~\etal~\cite{Goh2018AX} are similar to those for 0.02~Hz in Tests 1 and 2 as well. This is because the phase contrast for 0.02~Hz correspond to 25,12.5 and 6.25 points per cycle. By referring to Figs.~\ref{fig:ppt_difffreqsubsIrel} and \ref{fig:ppt_difffreqsubsTrel}, it can be observed that subsampling every 4 points resulted in a more significant shift than subsampling every 2 points. This effect is because the FFT integrals are calculated as rectangles, which are approximations for a continuous function. While the relative contrast may not deviate significantly with higher sampling rates, it does not eliminate the artifacts created by FFT.

\begin{figure} 
\begin{center}
\includegraphics[width=\figwidthhalf mm]{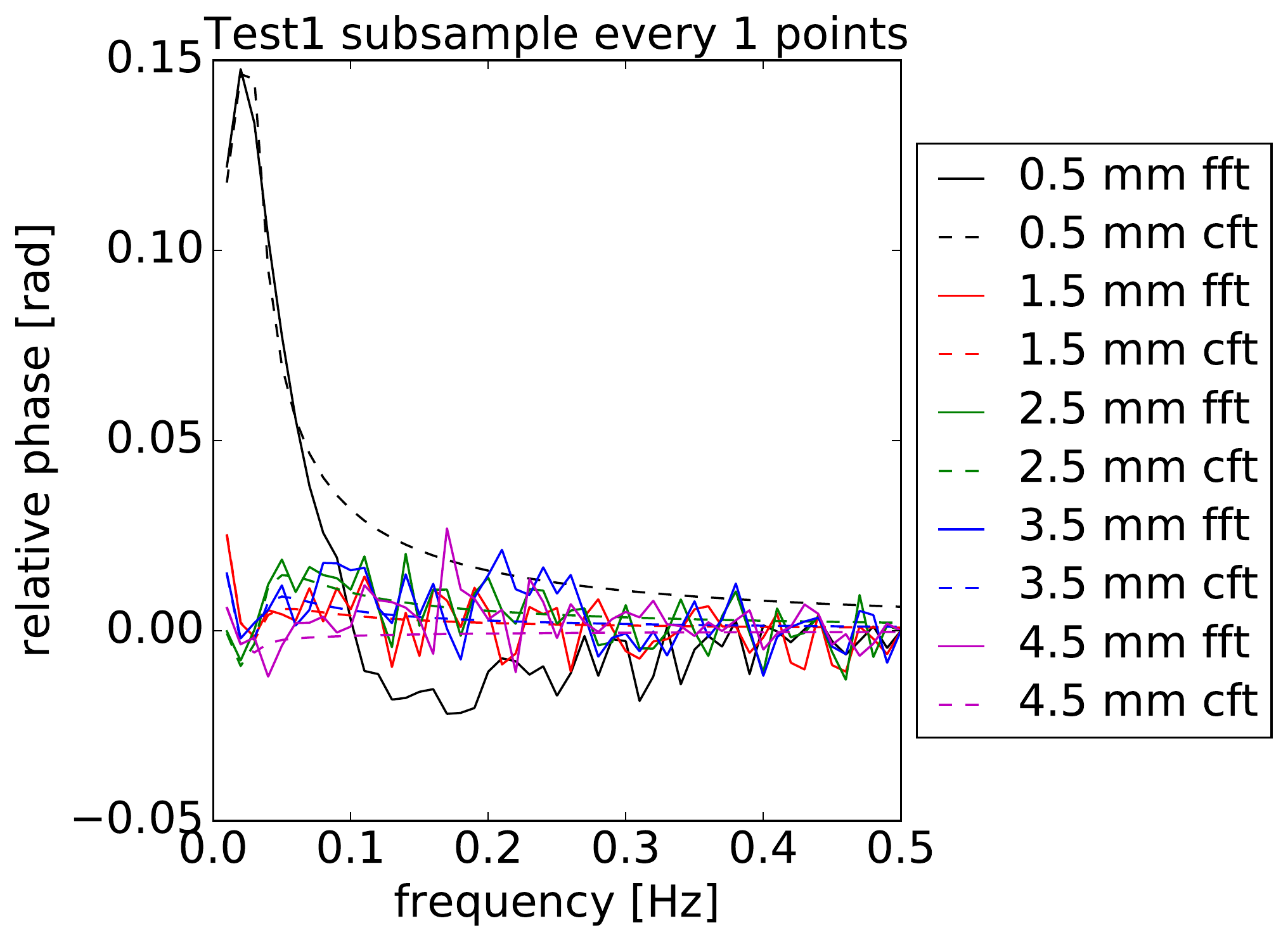}
\includegraphics[width=\figwidthhalf mm]{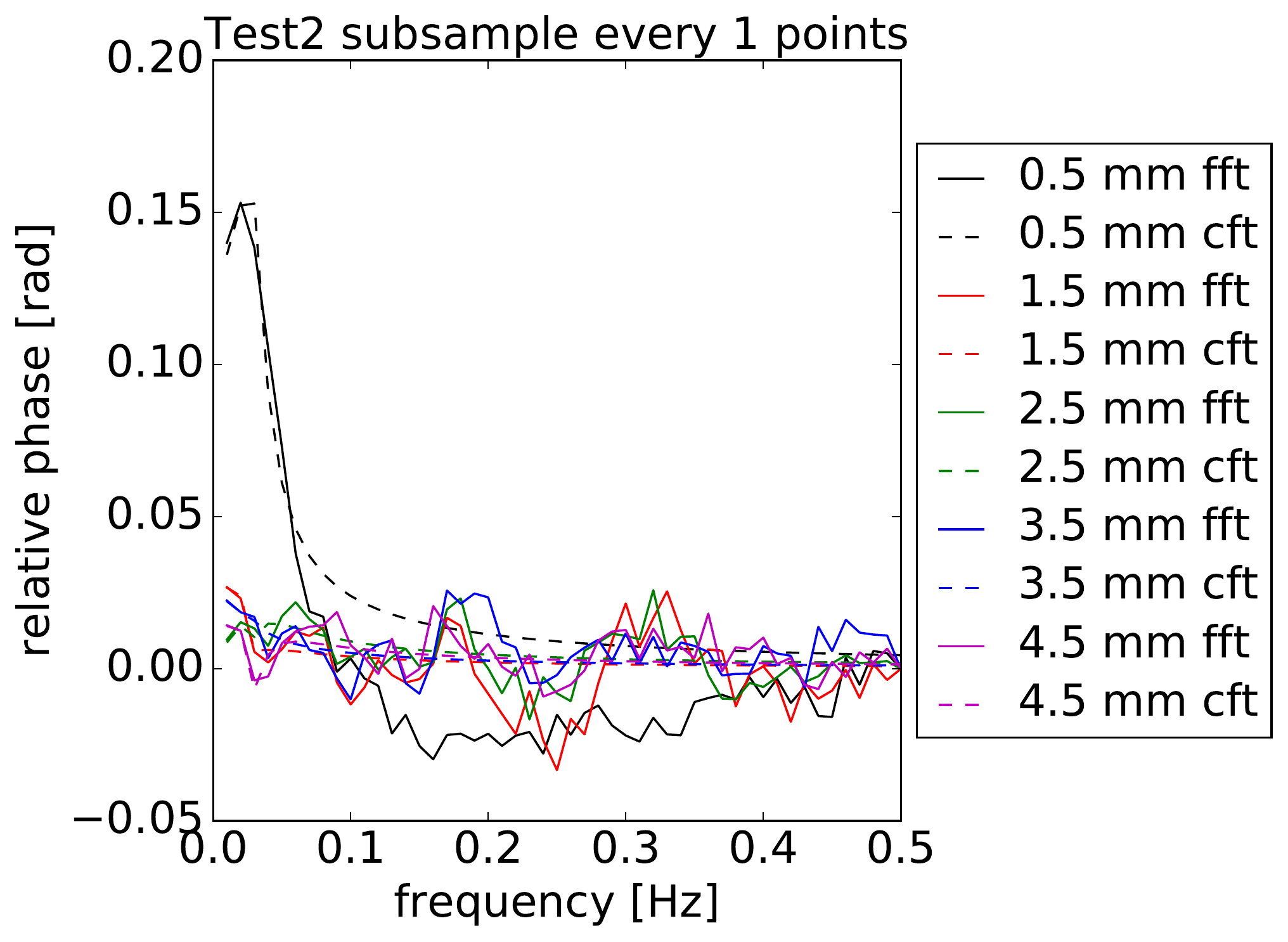}
\includegraphics[width=\figwidthhalf mm]{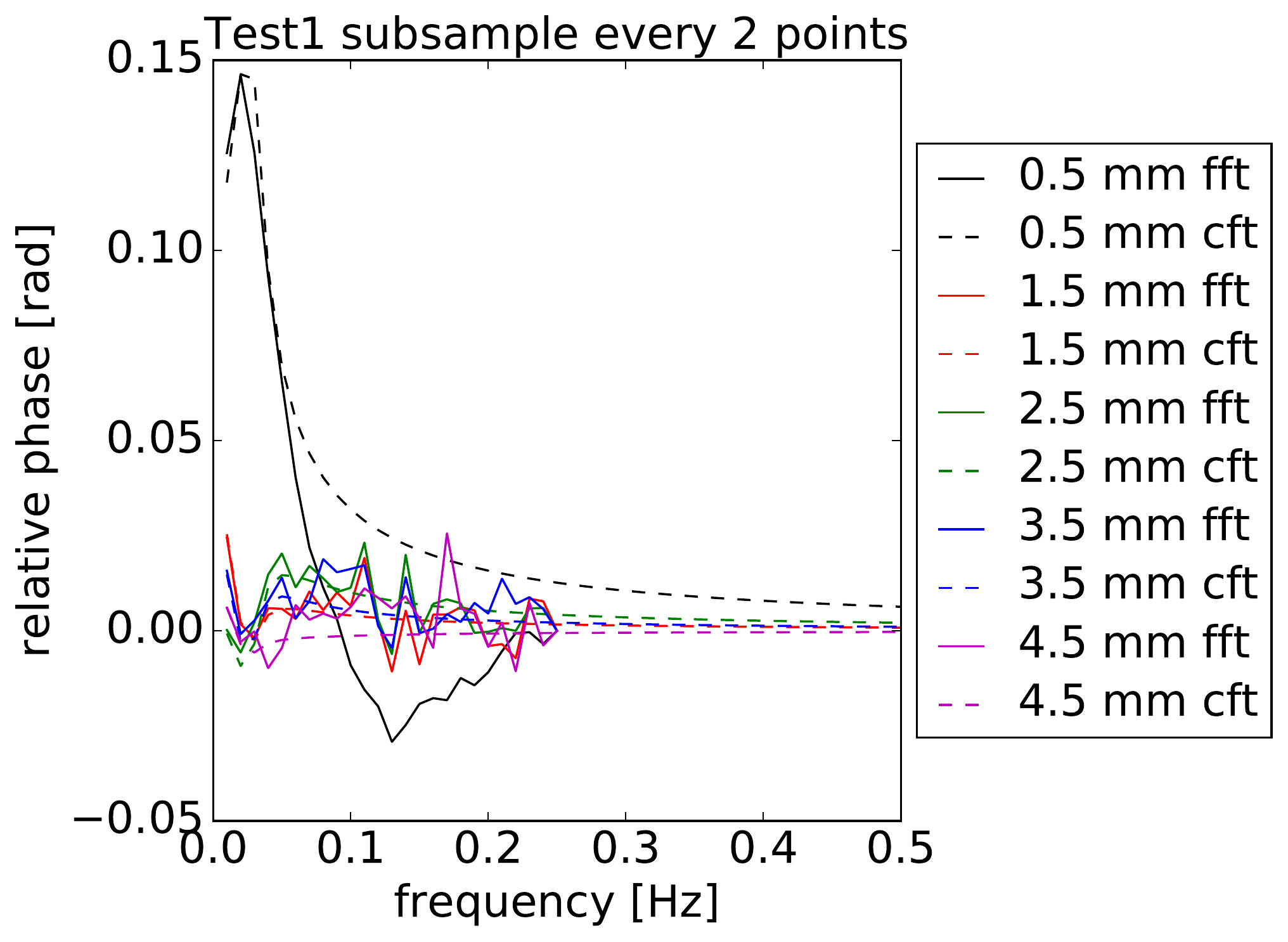}
\includegraphics[width=\figwidthhalf mm]{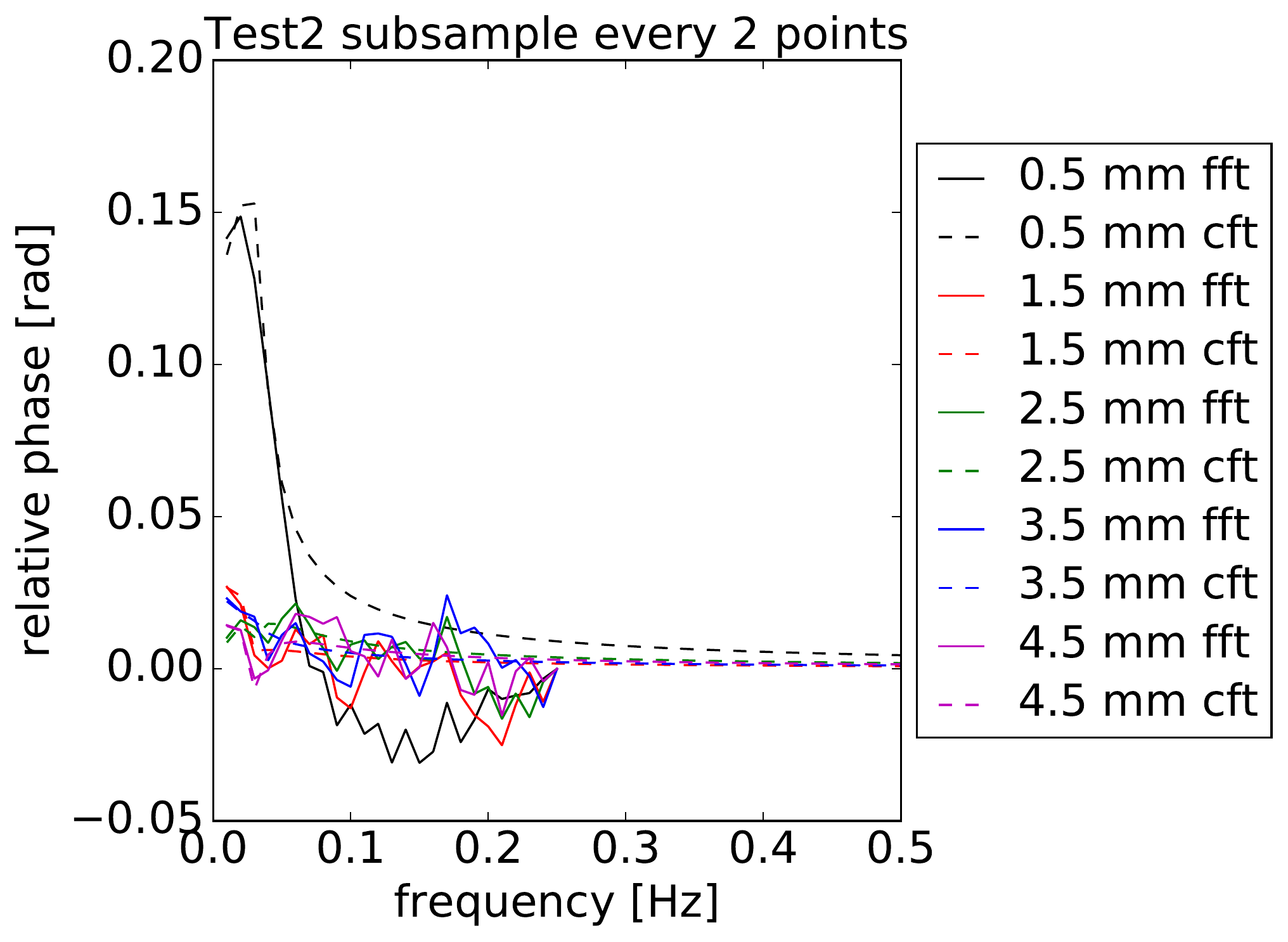}
\includegraphics[width=\figwidthhalf mm]{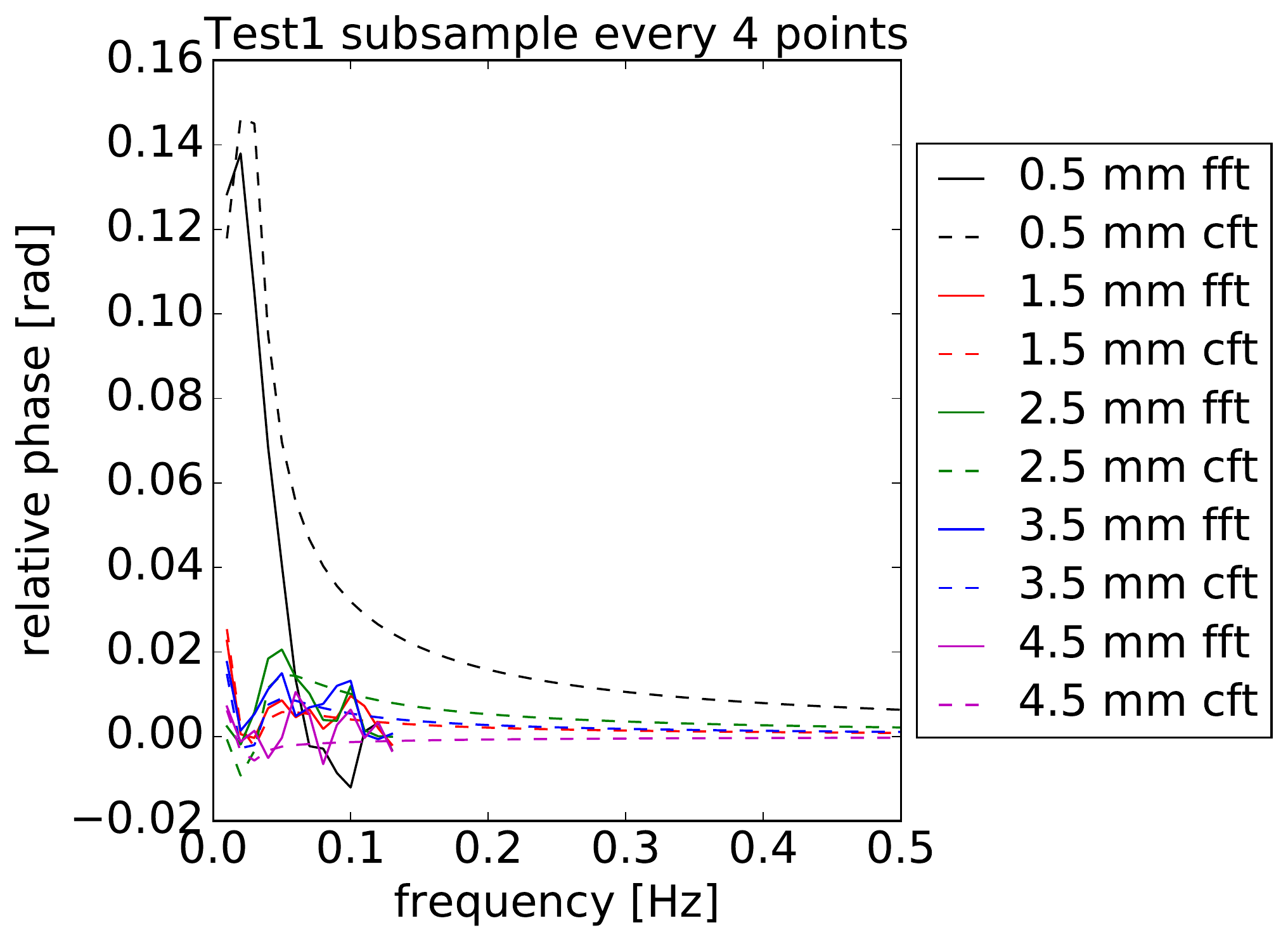}
\includegraphics[width=\figwidthhalf mm]{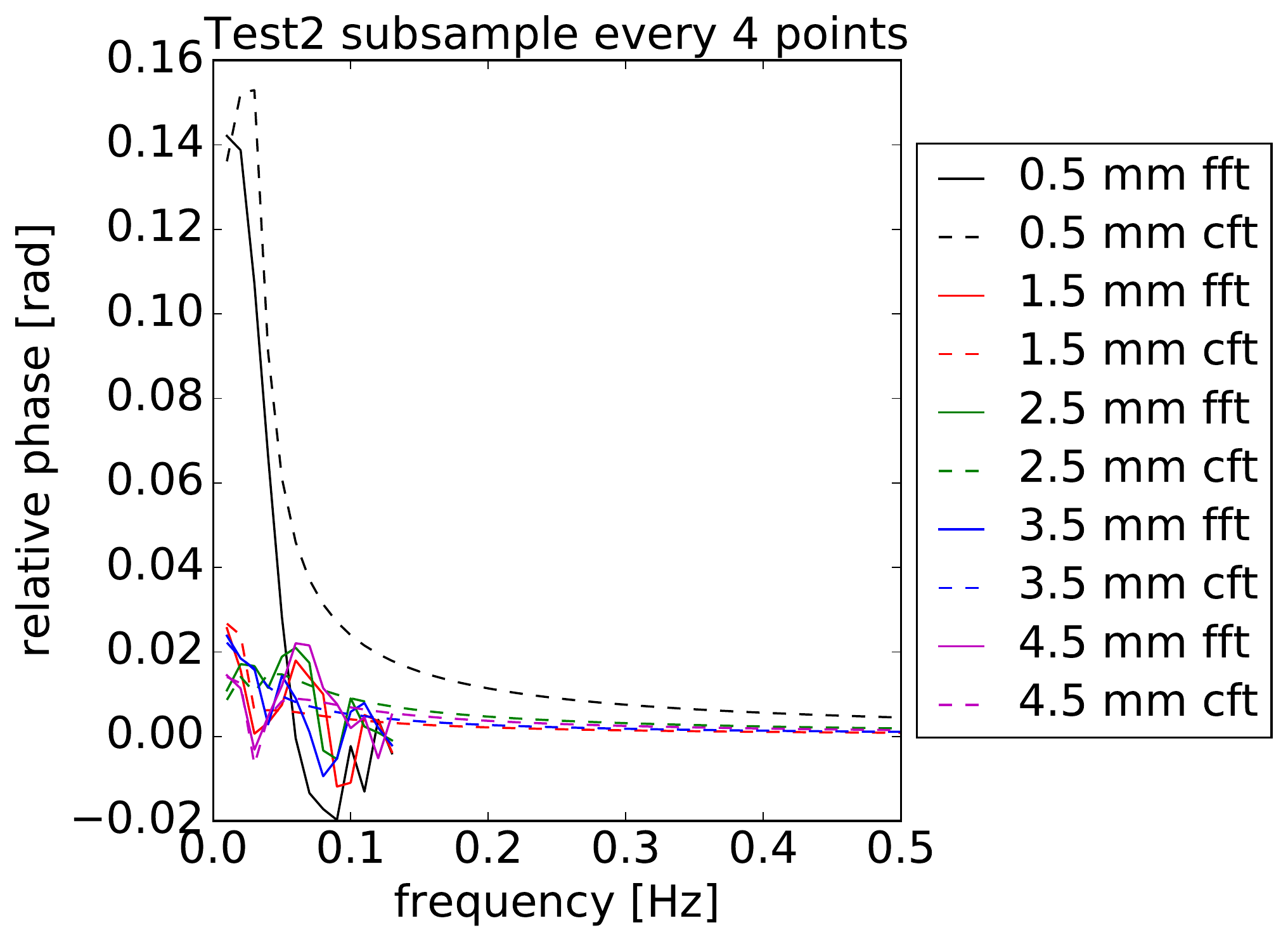}
\end{center}
\caption{Absolute phase contrast $\Delta\phi$, for FFT with subsampling of every 2 and 4 frames, versus CFT (Intensity).
\label{fig:ppt_phasecontrastI}}
\end{figure}

\begin{figure} 
\begin{center}
\includegraphics[width=\figwidthhalf mm]{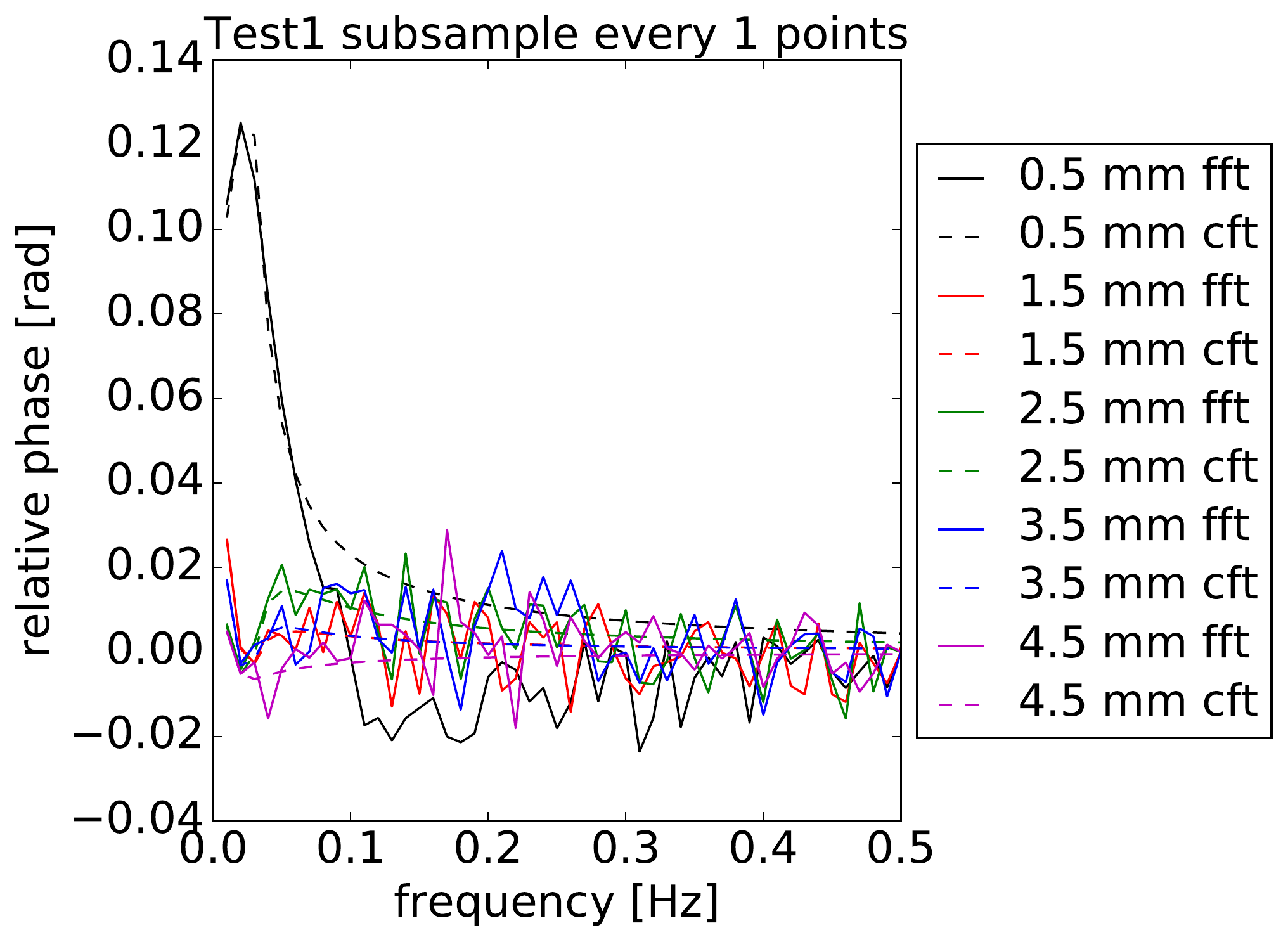}
\includegraphics[width=\figwidthhalf mm]{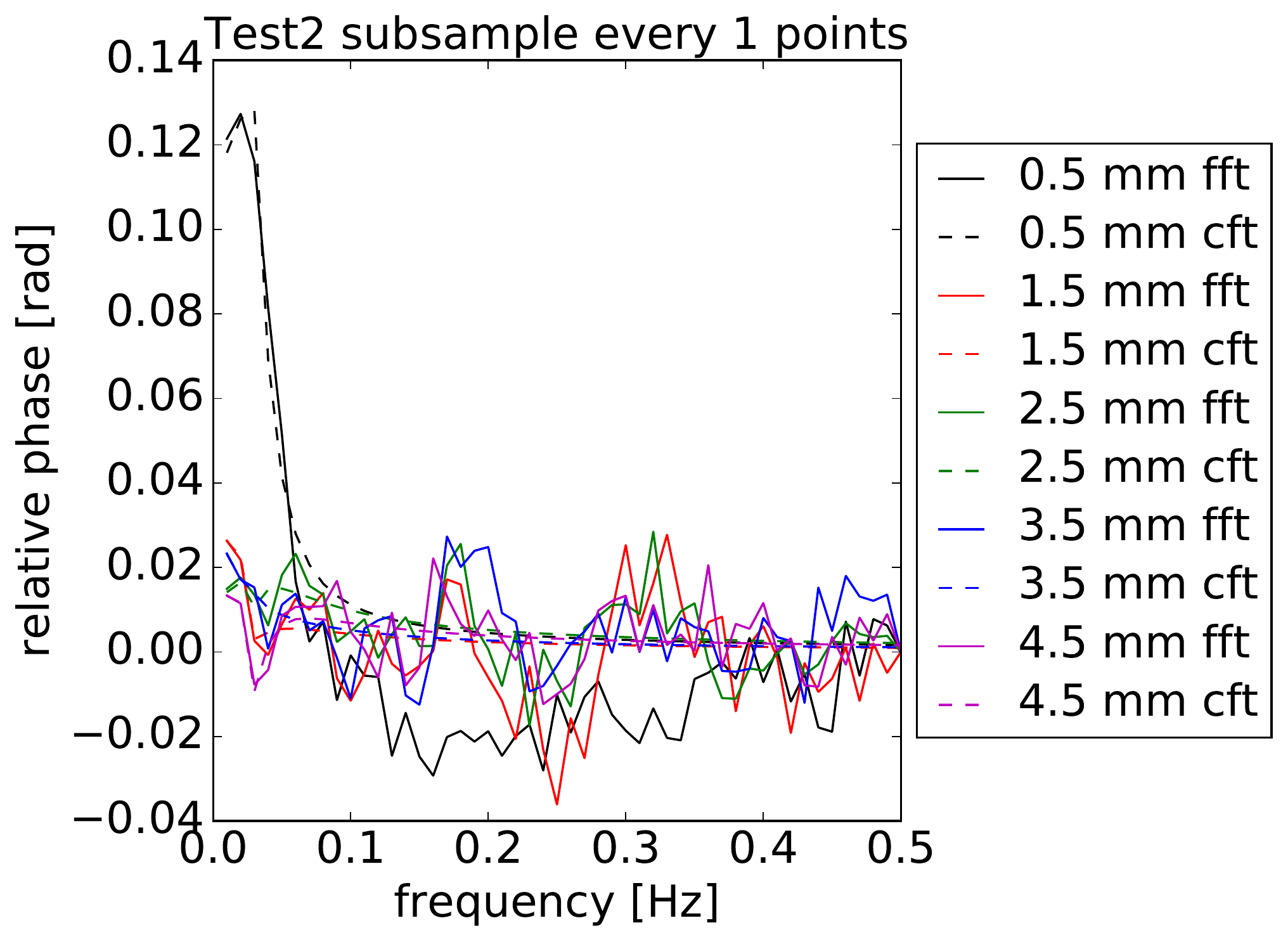}
\includegraphics[width=\figwidthhalf mm]{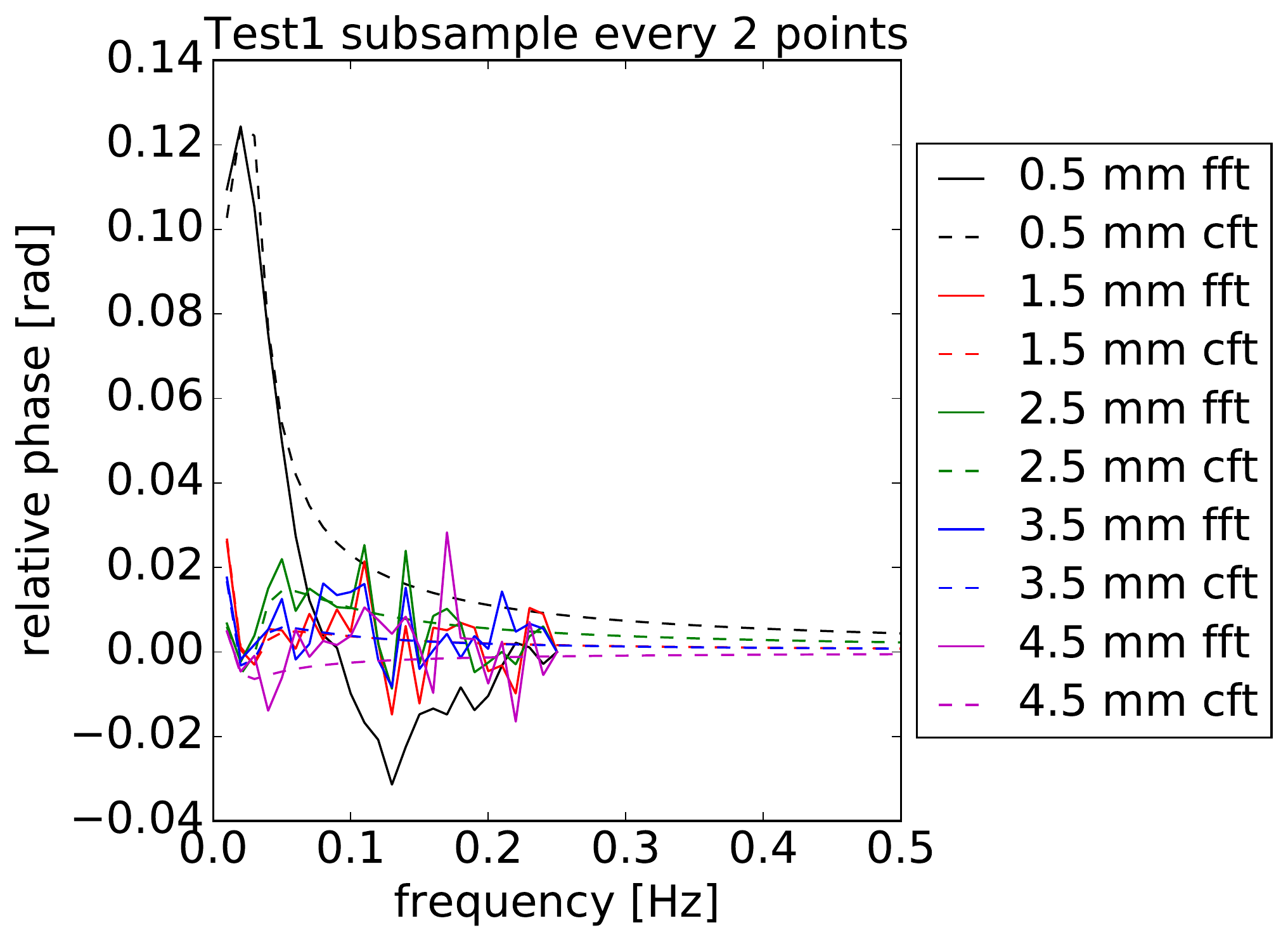}
\includegraphics[width=\figwidthhalf mm]{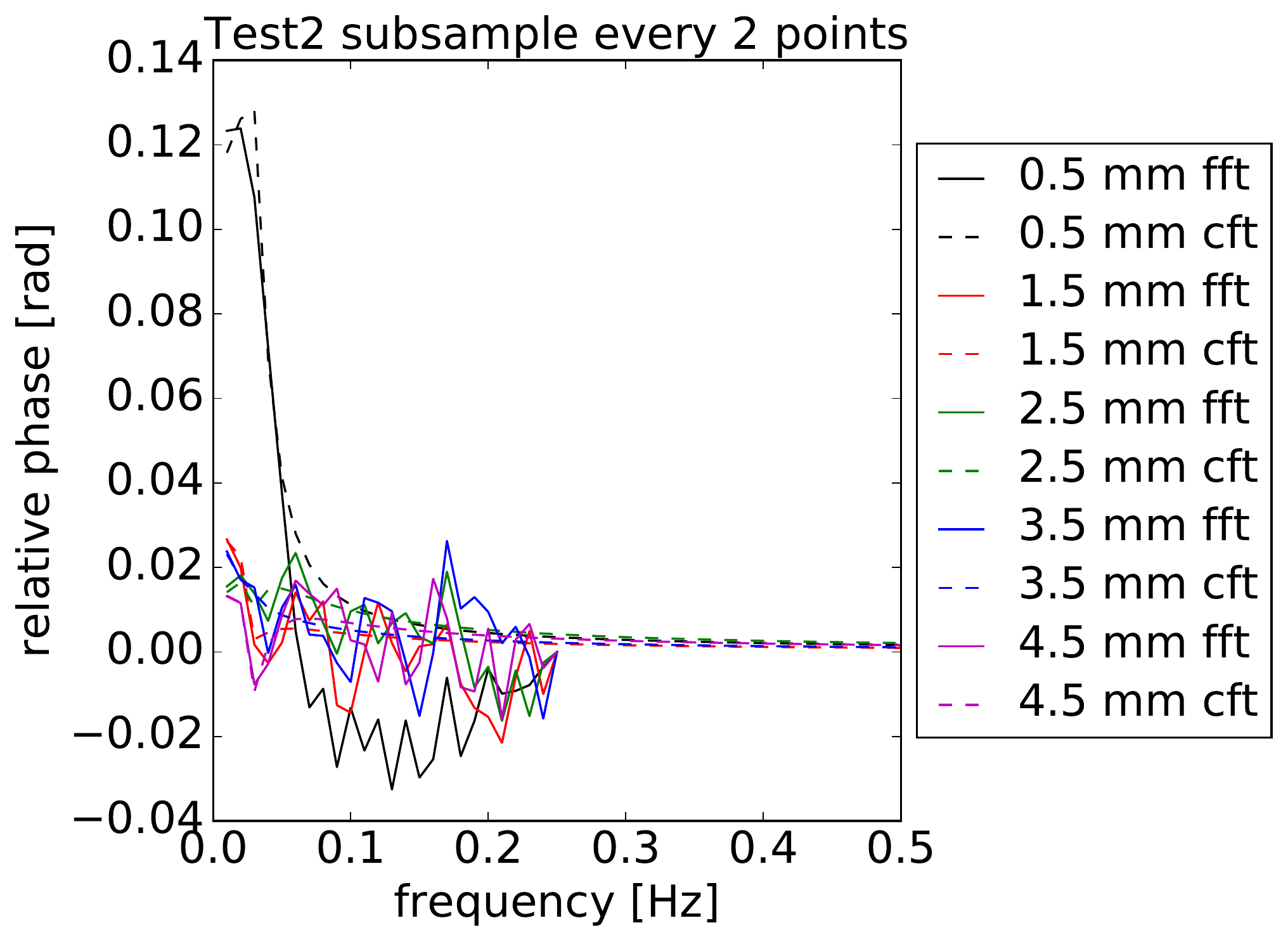}
\includegraphics[width=\figwidthhalf mm]{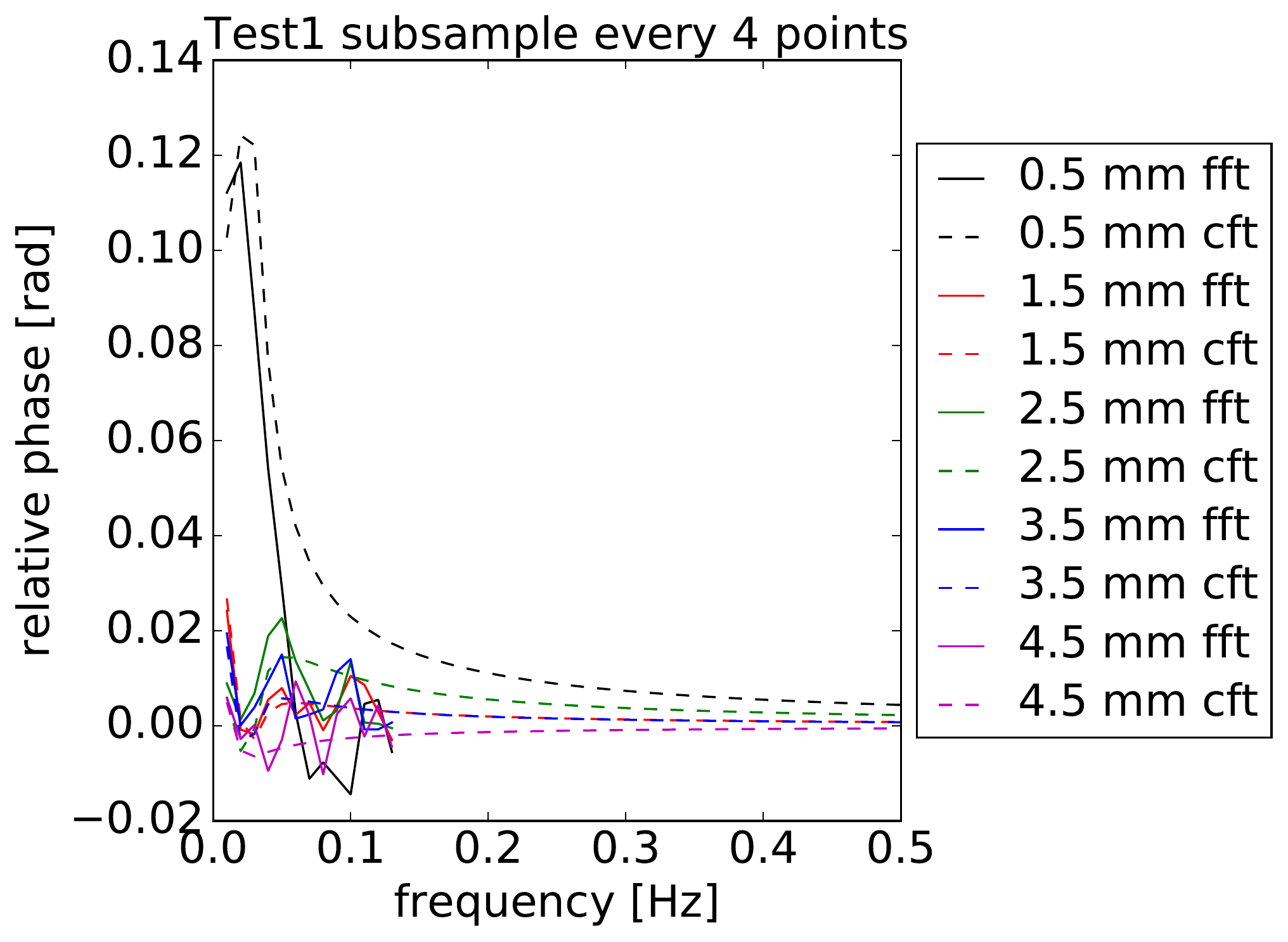}
\includegraphics[width=\figwidthhalf mm]{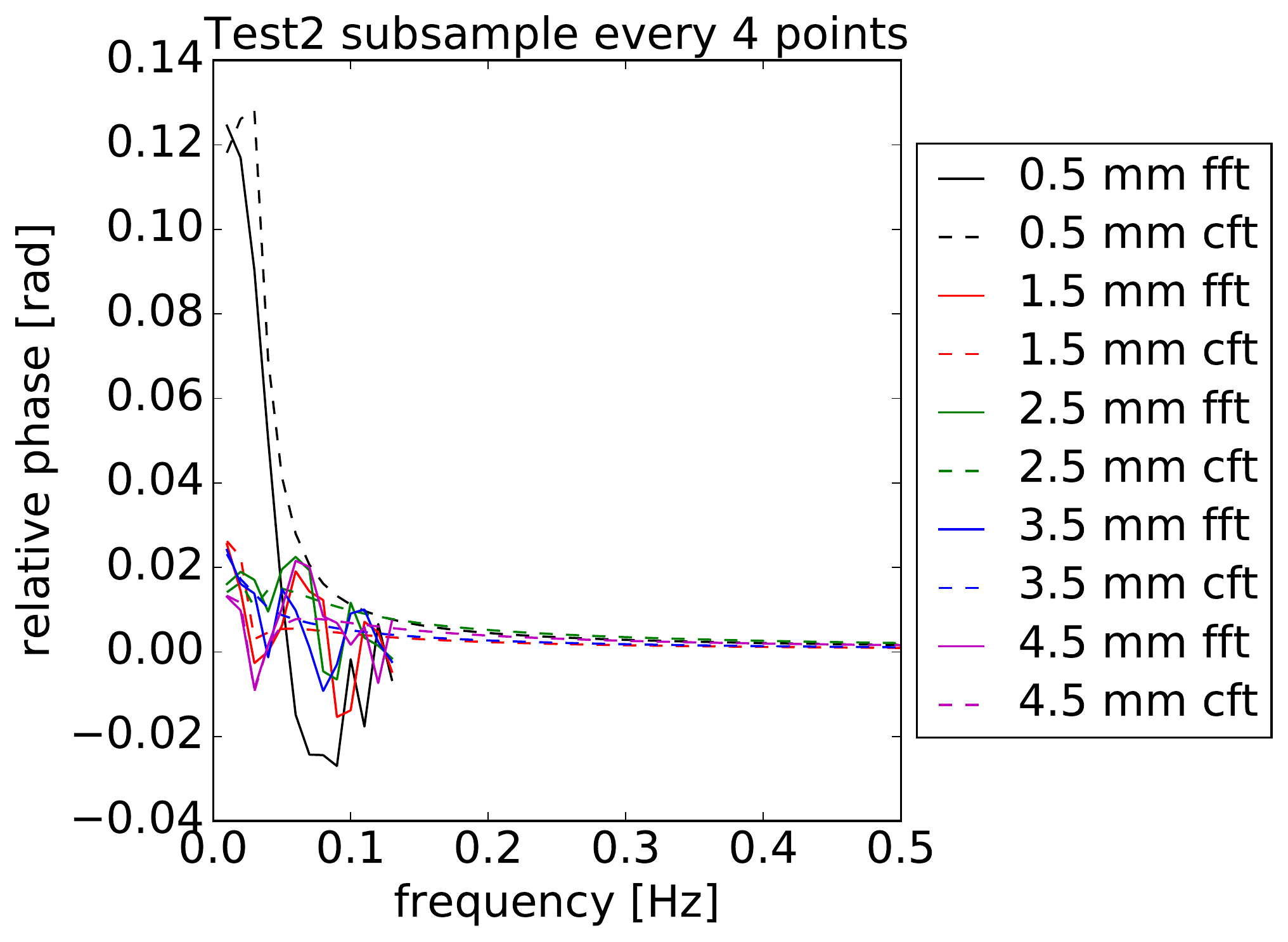}
\end{center}
\caption{Absolute phase contrast $\Delta\phi$, for FFT with subsampling of every 2 and 4 frames, versus CFT (Temperature).
\label{fig:ppt_phasecontrastT}}
\end{figure}

Next, we conduct an assessment of existing methods used to retrieve the depth of the defects with PPT. It was proposed by Ibarra-Castanedo and Maldague~\cite{Ibarra2004QIRT} that depth retrieval was possible, using absolute phase contrast as defined by

\begin{eqnarray}
\Delta\phi & = & \phi_{d} - \phi_{s}
\end{eqnarray}

where $\phi$ represents the phase value, and the subscripts $d$ and $s$ represent defective and sound areas respectively. A similar approach is adopted, using FFT and proposed CFT on the curves, to ascertain the differences. The phase values obtained at sound areas (N, S, E, W) in the current work were averaged to obtain the equivalent phase contrast. The results of FFT versus CFT are shown in Figs.~\ref{fig:ppt_phasecontrastI} and \ref{fig:ppt_phasecontrastT}, with effects of subsampling shown. Note that subsampling was not applied to polynomial fitting in these figures, in order for clarity. There are certain features of interest in these results. For FFT, it is obvious that the absolute phase contrast will converge to zero at frequencies corresponding to half the sampling frequency. This is not true for CFT, where the defined absolute phase contrast tends towards non zero asymptotic values, as expected. Between the full sample and subsampling every 2 frames, the absolute phase contrast at higher frequencies are also very different. This indicates a strong dependence on sampling rates. This presents a clear obstacle for depth retrieval, as the absolute phase contrast is not independent of measurement parameters.

\begin{figure} 
\begin{center}
\includegraphics[width=\figwidthhalf mm]{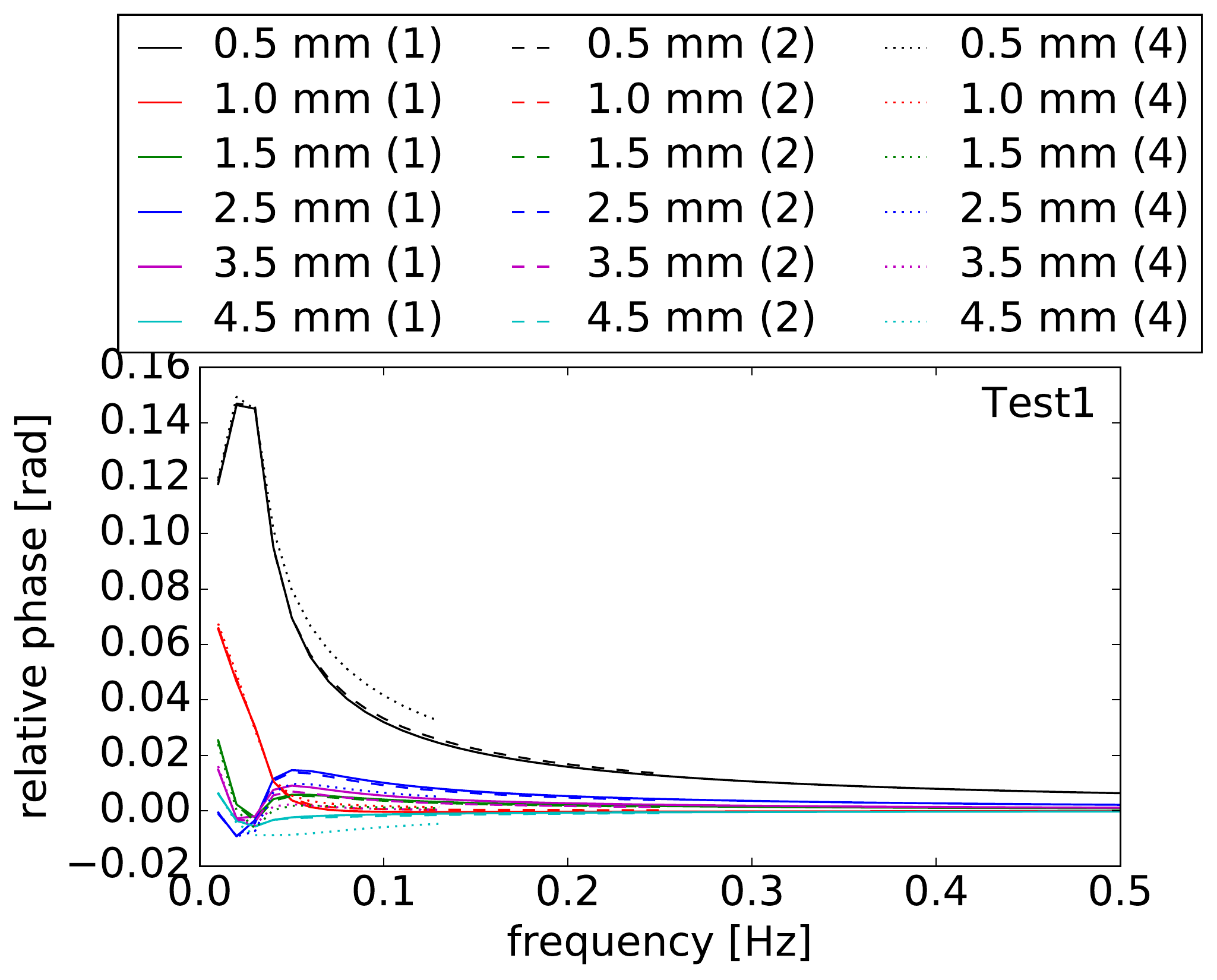}
\includegraphics[width=\figwidthhalf mm]{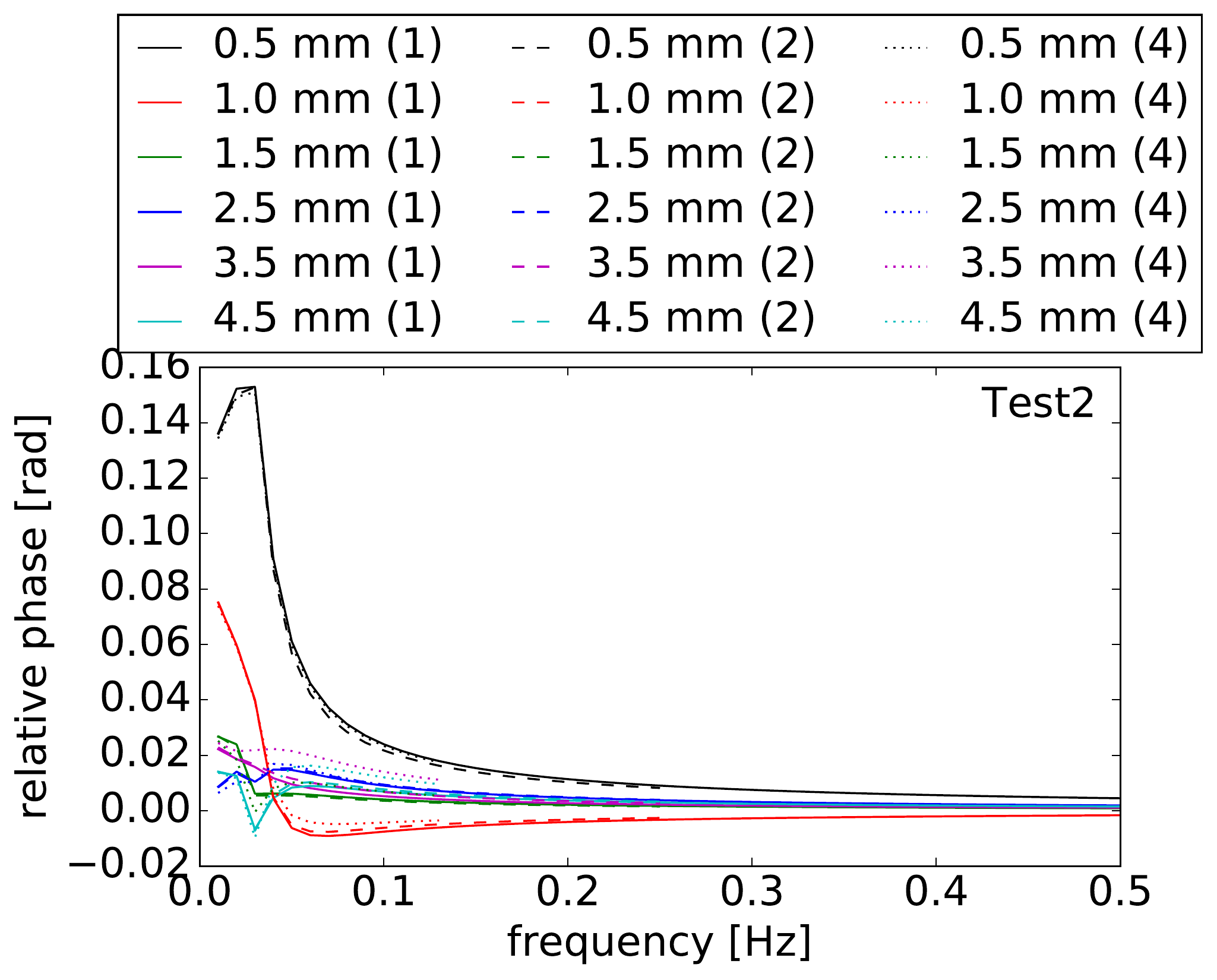}
\includegraphics[width=\figwidthhalf mm]{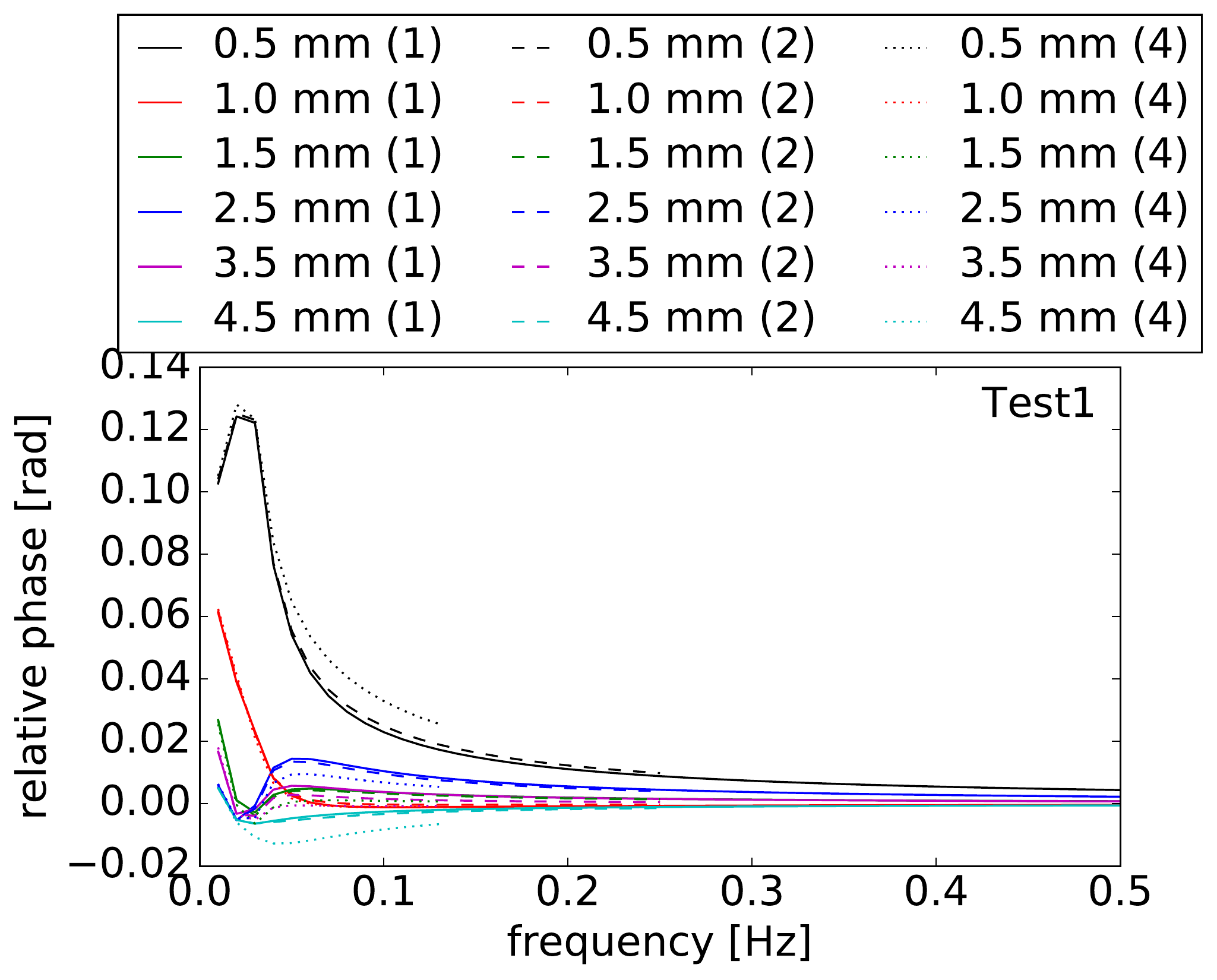}
\includegraphics[width=\figwidthhalf mm]{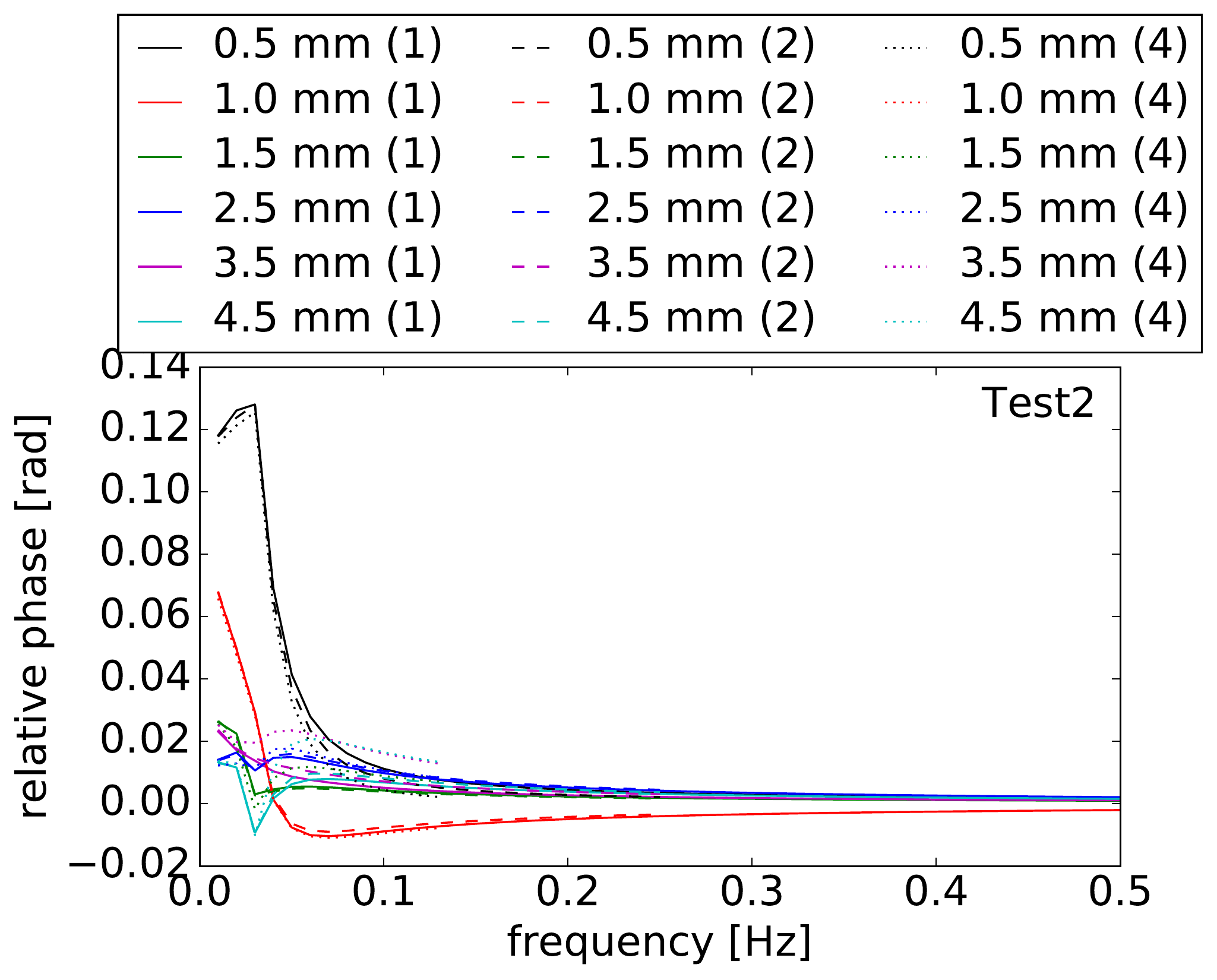}
\end{center}
\caption{Absolute phase contrast $\Delta\phi$, for CFT with subsampling of every 2 and 4 frames, using intensity (top row) and temperature (bottom row).
\label{fig:ppt_phasecontrast_cftonly}}
\end{figure}

The relative effects of subsampling on absolute phase contrast using polynomial fitting and CFT, are shown in Fig.~\ref{fig:ppt_phasecontrast_cftonly}. The difference is less prominent when CFT is used, when only using half the number of points. However, significant deviations can be observed when subsampling every 4 points. This could be due to the lack of points near the start of the cooling curve, which affected the polynomial fitting algorithm. This is further amplified as the start of the cooling curve affects the spectra to a much larger extent, especially at lower frequencies. With these results, it is recommended that the proposed polynomial fitting method with CFT be adopted for PPT measurements, to minimise the uncertainties in the data. As long as the sampling rate is not low to the extent that the start of the cooling curve is not captured sufficiently, this method should suffice for standardisation of current and future PPT data. Furthermore, it is relatively simple to implement on existing data. A collaborative effort will certainly move the field forward in this aspect, especially for depth retrieval and automated detection of defects with machine learning. In a sense, this method acts somewhat to curate data for machines learning models to succeed where they may have failed.

\section{Conclusion}
\label{sect:conclude}

A proposed polynomial fitting method with Continuous Fourier Transform (CFT) was benchmarked using Thermography measurements, to find consistent amplitude and phase information from truncated signals. It was shown that existing Fast Fourier Transform (FFT) methods implemented on Pulse Phase Thermography (PPT) measurements may not be ideal for a comprehensive understanding of phenomena due to inherent artifacts introduced by FFT. The proposed method was benchmarked for defects at different depths, as well as sound areas, with subsampling effects investigated. It was shown that the proposed method is consistent for both amplitude, phase and absolute phase contrast, as long as the sampling rates were sufficient to capture the start of the cooling curves. It is hoped that a collaborative effort across different labs will move the field forward, via a clearer understanding of Thermography, towards machine learning for Advanced Thermography.







\bibliographystyle{model1-num-names}
\bibliography{thermographybib}







\newpage

\begin{table}[!h]
\begin{center}
\caption{Sample tolerance values for polynomial fit of a certain order, for 13 sample points
\label{tab:polyfit}}
\begin{tabular}{|c|c|c|c|c|}
\hline
Tolerance & 5th order & 7th order & 9th order & 11th order\\
\hline
Intensity (max) [-] & 600 & 300 & 150 & 90\\
Intensity (mean) [-] & 90 & 40 & 20 & 12\\
Temperature (max) [K] & 2.5 & 1.5 & 0.6 & 0.38\\
Temperature (mean) [K] & 0.3 & 0.15 & 0.1 & 0.0675\\
\hline
\end{tabular}
\end{center}
\end{table}

\newpage

\end{document}